\documentclass[useAMS,usegraphicx,usenatbib]{mn2e}
\usepackage{times}   
\usepackage{mathptm} 
\usepackage{color}
\usepackage{xcolor}
\usepackage{amsmath}
\usepackage{url}
\usepackage[normalem]{ulem}
\title[Single spin mergers]
{Mergers of double NSs with one high-spin component: brighter kilonovae and fallback accretion, weaker gravitational waves}
\author[Rosswog et al.]
{S. Rosswog\thanks{E-mail: stephan.rosswog@astro.su.se}$^{1,2}$, P.  Diener$^{3,4}$, F. Torsello$^2$, T.~M. Tauris$^5$,
N. Sarin$^{6, 7}$\\
$^1$ University of Hamburg, Hamburger Sternwarte, Gojenbergsweg 112, 21029, Hamburg, Germany\\
$^2$ The Oskar Klein Centre, Department of Astronomy, AlbaNova, Stockholm University, SE-106 91 Stockholm, Sweden\\
$^3$ Center for Computation \& Technology, Louisiana State University, 70803, Baton Rouge, LA, USA\\
$^4$ Department of Physics \& Astronomy, Louisiana State University, 70803, Baton Rouge, LA, USA\\
$^5$ Department of Materials and Production, Aalborg University, Skjernvej 4A, DK-9220~Aalborg {\O}st, Denmark\\
$^6$ Nordita, Stockholm University and KTH Royal Institute of Technology Hannes Alfvéns väg 12, SE-106 91 Stockholm, Sweden \\
$^7$ The Oskar Klein Centre, Department of Physics, AlbaNova, Stockholm University, SE-106 91 Stockholm, Sweden}

\def\p{\partial}
\def\msun{M$_{\odot}$}
\def\Msun{M$_{\odot}$ }
\def\be{\begin{equation}}
\def\ee{\end{equation}}
\def\bi{\begin{itemize}}
\def\i{\item}
\def\ei{\end{itemize}}
\def\ben{\begin{enumerate}}
\def\een{\end{enumerate}}
\def\bea{\begin{eqnarray}}
\def\eea{\end{eqnarray}}
\def\bt{\begin{tabbing}}
\def\et{\end{tabbing}}
\def\gcc{gcm$^{-3}$}

\def\edo{\end{document}}

\def\M4{M$_4$}
\def\M6{M$_6$}
\def\lcm6{LCM$_6$}
\def\qcm6{QCM$_6$}

\def\wh3{W$_{\rm H,3}$}
\def\wh4{W$_{\rm H,4}$}
\def\wh5{W$_{\rm H,5}$}
\def\wh6{W$_{\rm H,6}$}
\def\wh7{W$_{\rm H,7}$}

\newcommand{\SpB}{\texttt{SPHINCS\_BSSN}\; }
\newcommand{\spB}{\texttt{SPHINCS\_BSSN}}
\newcommand{\spid}{\texttt{SPHINCS\_ID}}
\newcommand{\spiD}{\texttt{SPHINCS\_ID}\; }
\newcommand{\Ma}{\texttt{MAGMA2}\;}

\newcommand{\Lo}{\texttt{LORENE}\;}

\newcommand{\fu}{\texttt{FUKA}}

\newcommand{\McL}{\texttt{McLachlan}\;}

\def\fu{\texttt{FUKA }}
\def\Lo{\texttt{LORENE }}

\begin{document}
\date{Draft version}

\pagerange{\pageref{firstpage}--\pageref{lastpage}} \pubyear{2020}

\maketitle

\label{firstpage}

\begin{abstract}
Neutron star (NS) mergers where both stars have negligible spins are commonly considered
as the most likely ``standard'' case. In globular clusters, however, the majority of 
NSs have been spun up to millisecond (ms) periods and, based on observed systems,
we estimate that a non-negligible fraction of all double NS mergers
($\sim 4\pm2\;\%$) contains one component with a spin of a (few) ms. We 
use the Lagrangian numerical relativity code \SpB to simulate mergers where one
star has no spin and the other has a dimensionless spin parameter of $\chi=0.5$.
Such mergers exhibit several distinct signatures compared to irrotational cases.
They form only one, very pronounced spiral arm and they dynamically
eject an order of magnitude more mass of unshocked material at the original, very low
electron fraction. One can therefore expect particularly bright,
red kilonovae.
Overall, the spinning case collisions are substantially less violent and they eject
smaller amounts of shock-generated semi-relativistic material. Therefore, the ejecta
produce a weaker blue/UV kilonova {\em precursor} signal, but --- since the total amount is larger ---
brighter kilonova {\em afterglows} months after the merger. The spinning cases also have
significantly more fallback accretion and thus could power late-time X-ray flares.
Since the post-merger remnant loses energy and angular momentum significantly less efficiently  
to gravitational waves, such systems can delay a potential collapse to
a black hole  and are therefore candidates for merger-triggered
gamma-ray bursts with longer emission time scales. 
\end{abstract}

\begin{keywords}
hydrodynamics -- methods: numerical -- instabilities -- shock waves -- software: simulations -- gravitational waves - gamma-ray bursts
\end{keywords}

\section{Introduction}
Nearly exactly 100 years after their prediction \citep{einstein16}, 
gravitational waves (GWs) were detected 
for the first time in 2015: the LIGO detectors recorded the
``chirp'' signal from a merging binary black hole \citep{abbott16a}. 
The first NS merger event, GW170817, followed soon after \citep{abbott17b} 
and in its aftermath an electromagnetic firework was observed all 
across the spectrum \citep{abbott17c}. This first gravitational wave-based multi-messenger detection 
brought major leaps forward for many long-standing problems. For example, it made 
clear that NS mergers can indeed produce (short) gamma-ray bursts (sGRBs) 
\citep{goldstein17,savchenko17}, the delay between the gravitational wave peak and
the sGRB was used to very tightly constrain the propagation speed of GWs to the
speed of light \citep{abbott17b} and the event was also used for an independent
estimate of the Hubble parameter \citep{abbott17a}. It further helped
to constrain the tidal deformability 
of NSs and thereby some nuclear matter equations of state
could be ruled out \citep{abbott17c}. 
Last, but not least, it revealed the ultimate final destiny of a massive binary star
system \citep{vhdl73,tv23}, and it confirmed \citep{tanvir17,abbott17c,cowperthwaite17,smartt17,kasen17,rosswog18a} 
the long-held suspicion that NS mergers are
major sources of r-process elements in the cosmos 
\citep{lattimer74,symbalisty82,eichler89,rosswog98b,rosswog99,freiburghaus99b}.  This first
event splendidly demonstrated the power and promise of multi-messenger astrophysics
for the near future.\\
At the end of  the LVC science run O3, $\sim\!90$ gravitational wave events have been
observed \citep{abbott21a} and this number  will  increase rapidly with future
science runs at improved sensitivity. While the initially most conservative expectations for NS
mergers were component masses near 1.4~\msun, nearly exactly circular orbits 
\citep{peters63,peters64} and irrotational NS binaries \citep{bildsten92,kochanek92}, 
 observations during approximately the last decade revealed a much broader diversity.
In terms of masses, NSs with masses down to 1.17~\Msun were observed \citep{martinez15} 
while the highest secure NS masses exceed 2 \Msun 
\citep{antoniadis13,fonseca21}. Even more extreme objects that could potentially  
(but don't have to) be NSs have been detected by the LIGO/Virgo collaboration
(GW190814 and GW200210\_092254; \cite{GW190814,GWTC-3}). In both cases a $\sim23$ ~\Msun 
black hole (BH) merged with a $\sim 2.6$~\Msun object. The latter objects could 
themselves be the result of a binary NS merger \citep{bartos23}.
Most recently, another peculiar binary system, PSR J0514-4002E, has been observed that is of 
particular interest for our study here \citep{bdf+24}: it contains a millisecond pulsar 
(period $\sim5.6$~ms) in an eccentric orbit ($e=0.71$) around a massive compact 
object with a mass between 2.1 and 2.7~\msun, yet another 
candidate for a NSs with extremely large mass.
Apart from its very large overall mass, this binary 
system is remarkable, because it could be an example of a 
double NS (DNS) binary that contains a millisecond pulsar (MSP), i.e. it could be an example of the 
systems whose mergers we are studying here. 
This binary system is located in the globular cluster NGC~1851 and, in general,
more than half of the NSs observed in globular clusters
are spinning with periods of a few milliseconds\footnote{For a list of further MSPs 
in compact binary systems, we refer to P. Freire's website 
(https://www3.mpifr-bonn.mpg.de/staff/pfreire/NS\_masses.html).}.
As we will discuss in
Sect.~\ref{sec:discussion_fraction}, mergers that involve a MSP are not just a merely academic possibility,
but observations suggest that they should account for a noticeable fraction ($\sim 4\%$) 
of the  merging NS binaries.\\
Despite the preference for irrotational systems, the effect of NS 
spins on the merger process has been explored since the late 1990s, at that 
time in Newtonian-plus-GW-backreaction simulations. For example \cite{ruffert96,zhuge96,ruffert97a,rosswog99} 
explored, in addition to irrotational binaries, also corotating binaries and cases with both NS spins being 
anti-alligned with the orbital angular momentum. In \cite{rosswog00} 
binaries systems with only one spinning star were explored, though
not with a clear motivation based on stellar evolution grounds. 
More recently, the effects of spins were also explored in full numerical relativity
simulations. For example, \cite{kastaun13} used approximate initial conditions 
to study the prompt collapse to a BH and its resulting spin while 
\cite{bernuzzi14a} started their general relativistic simulations from 
constraint satisfying initial conditions mergers, each time both NSs had the same spin. 
Motivated by NSs in globular clusters, 
\citet{east16} modeled dynamical capture binaries with non-zero eccentricities 
and spins and thereby also considered cases with only one spinning star. 
More recently, several studies have further explored NS mergers 
with non-zero spins, see for example \cite{dietrich17b,ruiz19,tsokaros19,east19,most19,most21,papenfort22} and \cite{dudi22}.\\
Motivated by the insight from binary stellar evolution that mergers 
that contain a MSP occur at a non-negligible rate, see Sect.~\ref{sec:discussion_fraction},
and by the observation of PSR J0514-4002E \citep{bdf+24}, we study here how
 mergers with one highly spinning component  differ in various potentially
observable signatures from irrotational systems of the same mass.
Our paper is organized as follows. In Sec.~\ref{sec:formation} we discuss 
how DNS systems with only one rapidly spinning stellar component
may form. In Sec.~\ref{sec:method} we summarize our computational methodology, in particular
the Lagrangian numerical relativity code \SpB and how we construct constraint-satisfying
initial conditions with the code \spiD that can be linked to either the \Lo \citep{gourgoulhon01,lorene} or  the \fu  \citep{papenfort21,kadath}
initial data solver library. In Sec.~\ref{sec:results} we discuss the simulation setup
(\ref{sec:sim_setup}), the merger morphology (\ref{sec:morph}), the gravitational wave emission (\ref{sec:GWs}), the dynamic
ejecta (\ref{sec:dyn_ej}) and the resulting electromagnetic emission (\ref{sec:EM}). Sec.~\ref{sec:discussion} 
is dedicated to a discussion of our results while the most salient features of NS mergers with a single, 
rapid spin are summarized in Sec.~\ref{sec:summary}.

\section{Formation of a double NS system with a rapidly spinning component}
\label{sec:formation} 
The formation of DNS systems, as expected from binary star evolution of a pair of massive stars in isolation, 
involves a long sequence of stellar interactions \citep[e.g.][]{tkf+17,tv23}.  These interactions include mass 
transfer and tides, and the binary system must survive two supernova (SN) explosions to remain bound. 
The NS spins are a crucial testimony of the origin of the progenitor binary. It has been demonstrated that for 
systems produced in the Galactic disk, the first-formed (and thus recycled) NS in DNS systems can only be 
spun-up to spin periods longer than $\sim\!10-12\;{\rm ms}$ \citep{tlp15} because of the short-lasting and
 inefficient mass-transfer stages of the progenitor high-mass X-ray binary.\\ 
 However, the situation is quite different for DNS systems located in dense stellar environments like globular 
 clusters. Here, because of the possibility of close encounters among stars and other binaries, NSs can be 
 assembled in pairs that include a fully recycled MSP via exchange reactions. An example of a DNS candidate 
 system formed via such dynamical interactions is PSR~J1807$-$2500B \citep{lfrj12}, which is a 
 binary 4.2~ms pulsar found in the globular cluster NGC~6544. 

\begin{figure}
\vspace{0.5cm}
  \begin{center}
  \includegraphics[width=0.50\textwidth, angle=0]{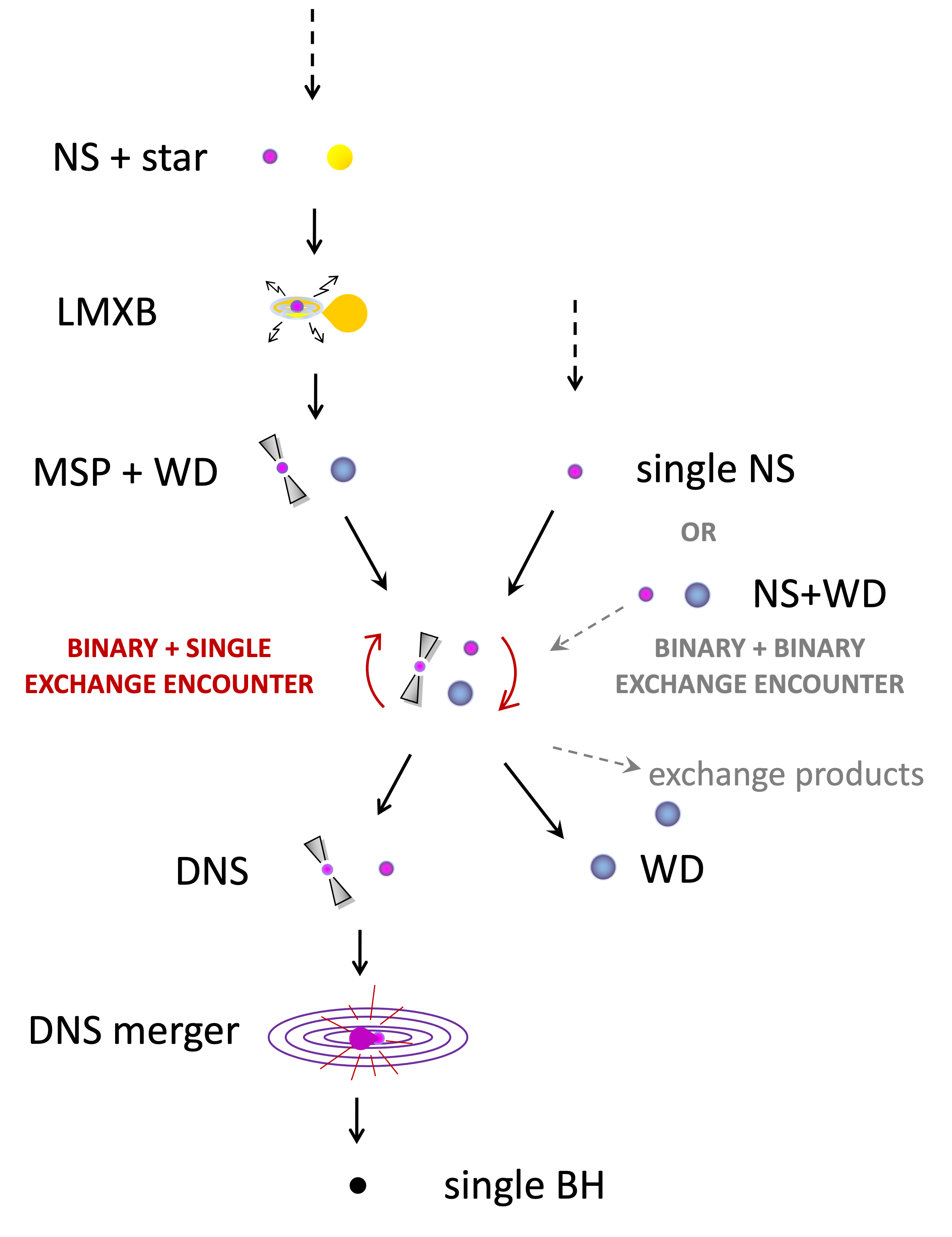}
\caption{Formation scenario of a DNS merger, with one component being a rapidly spinning MSP, produced in a dense stellar environment via exchange encounters.
Acronyms used in this figure: NS: neutron star; LMXB: low-mass X-ray binary; MSP: millisecond pulsar; WD: white dwarf; DNS: double NS; BH: black hole.}
  \label{fig:cartoon}
  \end{center}
\end{figure}

There are currently about 23 DNSs known in our Galaxy, about half of which will merge within a 
Hubble time \citep{tv23}. It is noteworthy that two out of the three globular cluster DNS pulsars, 
J1807$-$2500B and J0514$-$4002A, have been fully recycled to spin periods of 4.2~ms and 5.0~ms, 
respectively. This is most likely a result of long-term recycling in a low-mass X-ray binary (LMXB) 
system \citep{acrs82,rs82}, later followed by an encounter exchange of the MSP into the present 
DNS system, see Fig.~\ref{fig:cartoon}. The bottleneck, in terms of temporal evolution, of the 
formation channel depicted in Fig.~\ref{fig:cartoon} is the nuclear fusion timescale of the low-mass 
star that evolves to become the donor star in the LMXB system in which the NS is recycled to ms 
spin period. This donor star should have a mass of less than $2\;M_\odot$ to secure efficient 
recycling \citep{ts99,tlk12}. The nuclear evolution timescale is thus at least 1.3~Gyr. However, this 
timescale is often significantly less than the Hubble time and thus plays no big role for the detection 
statistics of DNS mergers with a fast-spinning NS component given that the in-spiral timescale of 
the DNS system due to GWs can take any value between a few Myr and up to a Hubble time.

A recent discovery in the Galactic globular cluster NGC~1851 \citep{bdf+24} provides further 
evidence for exchange interactions producing a MSP orbited by another NS or BH, and may 
thus hint at the possibility for producing GW source components in the BH lower-mass gap.
These components themselves may originate from DNS mergers in dense clusters \citep{bdf+24} 
or in hierarchical triple systems \citep{sra+14}. To summarize, it is evident that nature may 
produce DNS systems (and mergers) in which one component is a MSP with very rapid spin.
We note that both globular clusters, nuclear star clusters, and hierarchical triple systems have 
been suggested in relation to the origin of some recently detected GW source components 
\citep{rzp+16,ll21,brg+23}.

The fastest spinning MSP known to date has a spin period of 1.40~ms \citep[PSR~J1748$-$2446ad,][]{hessels06}. Recycling an MSP to a spin period of e.g. $1.0\;{\rm ms}$, requires accretion of, at least, $0.24\;M_\odot$ of material \citep[see Eq.~(14) in][]{tlk12}, and it is uncertain whether disk-magnetospheric conditions in the vicinity of the accreting NS in an LMXB allows for such fast spin \citep[see discussions in][]{tv23}. 
Nevertheless, in the following we study the merger process of a DNS 
system in which one NS component has a spin parameter $\chi\equiv S/M_{\mathrm{ADM}}^2 = 0.5$, which, somewhat depending on
the EOS, corresponds to a spin period of 
$\sim 1.2$~ms (see Table~\ref{tab:runs}),
while the other component has a negligible spin (in our simulations $\chi_2 = 0$).
The latter approximates  the typical spin period $\ga 1.0\;{\rm s}$ of an old non-recycled NS which has spun down due to significant magneto-dipole radiation from an (initially) strong B-field of order $\sim\!10^{13}\;{\rm G}$.

There can be delay times of several Gyr from the formation of the DNS system until it merges, but this is strongly depending on the orbital period ($P_{\rm orb}$) and eccentricity of the system, and also depending on the chirp mass ($\mathcal{M}$); see \citet{pet64}. For a binary with relatively small eccentricity $e\la 0.6$, this delay time is approximately given by \citep{maggiore08}:
\begin{equation}
  \tau_{\rm GW}\simeq 10\;{\rm Myr}\;\;\left(\frac{M_\odot}{\mathcal{M}} \right)^{5/3}\left(\frac{P_{\rm orb}}{{\rm hr}}\right)^{8/3} \left( 1 - e^2\right)^{7/2}\,.
\end{equation}

As an example, a circular DNS system with an orbital period of 8~hr and two NS component masses of $1.3\;M_\odot$ will merge after a time interval of about 2.0~Gyr, but the same system with eccentricity $e=0.9$ would merge within about 6~Myr. For the circular case,  the non-recycled NS would have spun down to a very slow spin period (justifying the use of $\chi =0$), whereas a recycled MSP with $P\simeq 1.2\;{\rm ms}$ and a low B-field ($\sim\! 10^7\;{\rm G}$) has such a small value of $\dot{P}\sim\! 10^{-21}$ that its position would remain ``frozen'' in the ($P,\,\dot{P}$)-diagram  and thus it would retain its rapid spin for several Gyr, therefore justifying the use of $\chi = 0.5$. 

\section{Methodology}
\label{sec:method}
Here we concisely summarize our simulation methodology as implemented 
in the current ``version 1.'' of our fully general relativistic, Lagrangian hydrodynamics
code \SpB  \citep{rosswog21a,diener22a,rosswog22b,rosswog23a}. This novel numerical
relativity code evolves the spacetime by solving the full set of Einstein equations
via a standard BSSN-formulation \citep{shibata95,baumgarte99}. The spacetime evolution
part is performed with finite difference methods and with fixed mesh-refinement in a very similar 
way as in Eulerian numerical relativity codes. Differently from them, we
evolve matter with freely moving particles according 
to a modern formulation of relativistic Smooth  Particle Hydrodynamics (SPH).
Our approach with a mesh for the spacetime 
and particles for the matter evolution requires a continuous back-and-forth
mapping from the particles to the grid (the mapped property is the energy-momentum tensor 
$T_{\mu \nu}$) and from the grid to the particle positions (mapped are the metric properties). 
Thus, our evolution code consists of three main sectors: matter evolution, spacetime evolution
and the coupling between the  two. These different elements will be briefly summarized below.\\
\subsection{Hydrodynamics}
\label{sec:hydro}
We solve the relativistic hydrodynamics equations with a modern version of Smooth Particle
Hydrodynamics, see \cite{monaghan05,rosswog09b,price12a,rosswog15c} for general reviews
of the SPH method. A step-by-step derivation of the general relativistic SPH equations can be found 
in Sec. 4.2 of \cite{rosswog09b}, a concise summary is provided in \cite{rosswog10a} and \cite{rosswog15c} 
and the exact version of the equations that we use together with all technical details can be found in our 
recent code paper devoted to ``version 1.0'' of \SpB \citep{rosswog23a}.\\
We distinguish between a ``computing frame'' in which the simulations are performed and a local
fluid rest frame in which, for example, thermodynamic quantities are measured. Our equations can be 
derived from a discretized relativistic Lagrangian\footnote{Note that we follow the convention that energies 
are measured in units of $m_0 c^2$, where $m_0$ is the baryon mass. This is discussed in more detail
in Sec.~2.1 of \cite{diener22a}.}
\be
L_{\rm SPH, GR}= - \sum_b \frac{\nu_b}{\Theta_b} [ 1 + u(n_b,s_b)],
\ee
where $\nu_b$ is the (conserved) baryon number of SPH particle $b$, $\Theta_b= (-g_{\mu\nu} v^\mu v^\nu)^{-1/2}_b$
is a generalized Lorentz factor with $v$ being the coordinate velocity, $u$ is the internal energy per baryon and 
$n$ and $s$ are the local rest-frame baryon number density and specific entropy. We measure a computing frame
baryon number density, $N= \sqrt{-g} \Theta n$, $g$ being the determinant of the spacetime metric, in a similar 
way to Newtonian SPH
\be
N_a= \sum_b \nu_b\, W(|\vec{r_a} - \vec{r}_b|,h_a),
\label{eq:N_sum}
\ee
where the smoothing length $h_a$ characterizes the support size 
of the SPH smoothing kernel $W$, see below. The equations we evolve in time
are  the canonical energy per baryon (see \cite{rosswog09b} for details),
\be
e_a= \left(S_i v^i + \frac{1 + u}{\Theta}\right)_a = \left(\Theta \mathcal{E} v_i v^i + \frac{1 + u}{\Theta}\right)_a
\label{eq:can_en}
\ee
and  the canonical momentum per baryon
\be
(S_i)_a = (\Theta \mathcal{E} v_i)_a,
\label{eq:can_mom}
\ee
where $\mathcal{E}= 1 + u + P/n$ is the relativistic enthalpy per baryon with $P$ being the gas pressure.
The corresponding evolution equations are 
\be 
\left(\frac{d e_a}{dt}\right) = -\sum_b \nu_b \left\{ \frac{P_a}{N_a^2}  \;  v_b^i   \; D^a_i +  
\frac{P_b}{N_b^2} \;  v_a^i \; D^b_i \right\}   -\left(\frac{\sqrt{-g}}{2N} T^{\mu \nu} \frac{\p g_{\mu \nu}}{\p t}\right)_a
\label{eq:dedt_full}
\ee
and
\be
\frac{d(S_i)_a}{dt}  =  -\sum_b \nu_b \left\{ \frac{P_a}{N_a^2}  D^a_i  +  
\frac{P_b}{N_b^2} D^b_i \right\}  + \left(\frac{\sqrt{-g}}{2N} T^{\mu \nu} \frac{\p g_{\mu \nu}}{\p x^i}\right)_a
\label{eq:dSdt_full},
\ee
where 
\be
D^a_i \equiv   \sqrt{-g_a} \;  \frac{\p W_{ab}(h_a)}{\p x_a^i} \quad {\rm and} \quad 
D^b_i \equiv    \sqrt{-g_b} \; \frac{\p W_{ab}(h_b)}{\p x_a^i}
\label{eq:kernel_grad}
\ee
and $W_{ab}(h_k)$ is a shorthand for $W(|\vec{r}_a - \vec{r}_b|/h_k)$. We have implemented 
a large variety of different kernel functions, but our preferred ones are a C6-smooth Wendland kernel
\citep{wendland95} for which we use exactly 300 contributing neighbours, which we had scrutinized in \cite{rosswog15b},  and a member of the 
harmonic-like kernels \citep{cabezon08}, $W^{\rm H}_n$, with $n=8$ for which we use 220 
contributing neighbour particles. For the explicit expressions 
and the motivations behind these choices, we refer to our recent detailed code paper 
\citep{rosswog23a}.\\
The hydrodynamic terms
are enhanced by dissipative terms to allow for a robust treatment of shocks. See, for example, Fig. ~4 in \cite{rosswog21a} for
a relativistic 3D shock tube test and Sec.~2.1.1--2.1.3 in \cite{rosswog22b} for the explicit expressions for the dissipative
terms that we are using. In experiments with the Newtonian \Ma code \citep{rosswog20a} 
we had found that {\em slope-limited reconstruction in artificial dissipation terms}
massively suppresses unwanted dissipation effects. We therefore  apply this technique
also in \spB.
 In addition, we also evolve the dissipation parameters in time as described in detail in 
Sec.2.1.3 in  \cite{rosswog22b}. 

\subsection{Spacetime evolution}
\label{sec:ST}
We evolve the spacetime according to the (``$\Phi$-version'' of the) BSSN
equations \citep{shibata95,baumgarte99}. We have written wrappers around code extracted
from the \McL thorn of the Einstein Toolkit \cite{ETK:web,loeffler12}, see Sec.~2.3 of \cite{rosswog23a} for the explicit expressions.
The derivatives on the right-hand side of the BSSN equations are evaluated via standard Finite Differencing
techniques where  we use sixth order differencing as a default. We have recently implemented
a fixed mesh refinement for the spacetime evolution, which is described in detail in \cite{diener22a}, to which
we refer the interested reader. For the gauge choices, we use a variant of ``1+log-slicing'',
together with a variant of the ``$\Gamma$-driver'' shift condition  \citep{alcubierre02,alcubierre08,baumgarte10}.

\subsection{Coupling between the particles and the mesh}
\label{sec:M_ST_coupling}
The \SpB approach of evolving the spacetime on a mesh and the matter fluid via particles
requires a continuous information exchange between the two entities: the gravity part of the particle
evolution is driven by derivatives of the metric, see Eqs.~(\ref{eq:dedt_full}) and (\ref{eq:dSdt_full}),
which are known on the mesh, while the energy momentum tensor that is needed as a source in the BSSN equations 
is known on the particles. This bi-directional information exchange is needed at every Runge-Kutta substep or,
in our case with an optimal 3$^{\rm rd}$ order Runge-Kutta algorithm \citep{gottlieb98}, three times per numerical time step.\\
The mesh-to-particle step is performed via 5$^{th}$ order Hermite interpolation, that we have developed \citep{rosswog21a} 
in extension of the work of \cite{timmes00a}. Contrary to a standard Lagrange polynomial interpolation, the Hermite 
interpolation guarantees that the metric remains twice differentiable as particles pass from one grid cell to another
and thus avoids the introduction of additional  noise. Our approach is explained in detail in Sec. 2.4 of 
\cite{rosswog21a} to which we refer the interested reader.\\
For the more challenging particle-to mesh mapping, we use 
in the latest version of \SpB a mapping technique
that is based on a ``local regression estimate'' (LRE). This new approach is explained in  detail in Sec.~2.4 of \cite{rosswog23a}. 
Here we only summarize the main ideas and refer to our original paper for details.
The major idea is to assume that the function to be mapped 
is known at the particle positions and that this function can be expanded in a Taylor series around any given grid point.
This Taylor expansion can be interpreted as an expansion in a polynomial basis with to-be-determined ``optimal'' coefficients.
These coefficients are obtained by minimizing an error functional similar to common least square approaches. \\
In principle,
such expansions could be performed up to arbitrarily high order, but (i) the size of the matrices to be inverted and therefore
the computational effort increases rapidly with polynomial order, (ii) with higher order the matrices can get closer 
to being singular and (iii) high order is not everywhere the best approach. For example, when encountering a sharp 
transition such as the NS surface, high-order polynomials can lead to spurious, ``Gibbs-phenomenon''-like 
oscillations and in such a situation a lower polynomial order delivers a better result. Therefore, we apply a ``Multi-dimensional 
Optimal Order Detection'' (MOOD) approach. The idea is to perform the LRE-mapping for different polynomial orders 
(in practice we use up to quartic polynomials which requires a $35 \times 35$ matrix to be inverted), discard those
results that are considered ``unphysical'' (e.g. negative energy density, $T_{00} < 0$), and choose out of the remaining options
the one which is the best representation of  the surrounding particles according to an error measure. This method
works very well in practice and we refer the interested reader to Sec.~2.4 and Appendix D of \cite{rosswog23a} for more details and tests.
 
\subsection{Constructing initial conditions}
\label{sec:ID}
We set up our initial SPH configurations with our initial data code \spid, see \cite{rosswog23a}, Sec. 3, for the description of the latest version. 
This code can be linked to either the \Lo \citep{gourgoulhon01,lorene} or  the \fu  \citep{papenfort21,kadath}
initial data solver library. These initial data solvers provide constraint satisfying solutions for the matter distribution and 
the corresponding spacetime for a given physical system. This is the first step towards initial conditions
for \spB, but more needs to be done: since we want to use equal mass particles (for purely numerical reasons), the particle
distribution needs to accurately reflect the matter density solution (found by the initial data solver) and the
distribution of the particles should also ensure  a good SPH interpolation accuracy. To achieve this, we use a general relativistic version
of the ``Artificial Pressure Method'' (APM) that was originally originally developed in a Newtonian context \citep{rosswog20b}.
The main idea is to start from a guess distribution of particles, measure their local error compared to the
result found by the initial data libraries and translate this error into an ``artificial pressure''. This latter pressure is then
used in an equation very similar to a hydrodynamic momentum equation to push the particle into a position where
it minimizes its error. Our earlier versions of the APM \citep{rosswog21a,diener22a,rosswog22b} were straight 
forward translations of the Newtonian method where the ``artificial pressure'' is calculated from the density error.
Recently \citep{rosswog23a}, we realized that we get --for the same computational effort-- slightly more acurate
results if we use the error in the physical pressure (rather than in the density) to calculate the ``artificial pressure''.
For details of this latest version we refer to Sec.~3 in \cite{rosswog23a}.\\
While \Lo has been a major work horse for many groups in setting up initial conditions, it struggles in 
achieving more extreme mass ratios and it does not allow for the construction of binaries with arbitrary spins. In this study, we therefore
use initial conditions exclusively based on the \fu library.

\begin{table*}
	\centering
	\caption{Summary of the performed simulations. The masses are always $2 \times 1.3$~\msun, the dimensionless
    spin parameter of star one is always $\chi_1=0$
	and we always use 2 million SPH particles, an initial separation of 45~km, 7 initial grid refinement levels
	with a grid spacing $\Delta x= 369$~m on the finest refinement level. The columns are, in order, 1) the name of the run, 2) the equation of state, 3) spin parameter
    $\chi_2$ of star 2, 4) angular velocity, 5) spin period, 6) total initial ADM mass, 7) total initial
    ADM angular momentum, 8) the total
    radiated energy, 9) the total radiated
    energy in percentage of the initial ADM mass, 10) the total
    radiated angular momentum and 11)
    the total radiated angular momentum in percentage of the
    initial ADM angular momentum. The units are given in the column headings.}
	\label{tab:runs}
	\begin{tabular}{lrrcrrrrrrr} 
		\hline		
run & EOS & $\chi_2$ & $\Omega_2$ [\msun$^{-1}$] & $P_2$ [ms] & $M_{\mathrm{ADM}}$ [\msun] & $J_{\mathrm{ADM}}$ [\msun$^2$] & $\Delta E_{\mathrm{rad}}$ [\msun] & $\Delta E_{\mathrm{rad}}$ $[\%]$ & $\Delta J_{\mathrm{rad}}$ [\msun$^2$] & $\Delta J_{\mathrm{rad}} [\%]$\\
		\hline
		\texttt{SLy\_irr} & SLy & 0 & 0 & --- &2.577 & 6.874 & $>0.078$ & $>3.01$ & $>1.93$ & $>28.1$\\
		\texttt{SLy\_sspin} & SLy &   +0.5 & 0.027 & 1.145 & 2.577 & 7.652 & $>0.045$ & $>1.73$ & $>1.32$ & $>17.3$ \\		
		\texttt{APR3\_irr} & APR3 & 0 & 0 & --- & 2.577 & 6.874 & $>0.058$ & $>2.24$ & $>1.66$ & $>24.2$ \\
		\texttt{APR3\_sspin} & APR3 &  +0.5 & 0.026 & 1.184 & 2.577 & 7.653 & $>0.027$ & $>1.03$ & $>0.949$ & $>12.4$ \\
		\texttt{MS1b\_irr} & MS1b & 0 & 0 & --- & 2.577 & 6.878 & $>0.023$ & $>0.87$ & $>0.907$ & $>13.2$\\
		\texttt{MS1b\_sspin} & MS1b &  +0.5 & 0.019 & 1.664 & 2.577 & 7.658 & $>0.014$ & $>0.55$ & $>0.682$ & $>8.9$ \\
		\hline
	\end{tabular}
\end{table*}
\begin{figure*} 
   \centering
   \includegraphics[width=\columnwidth]{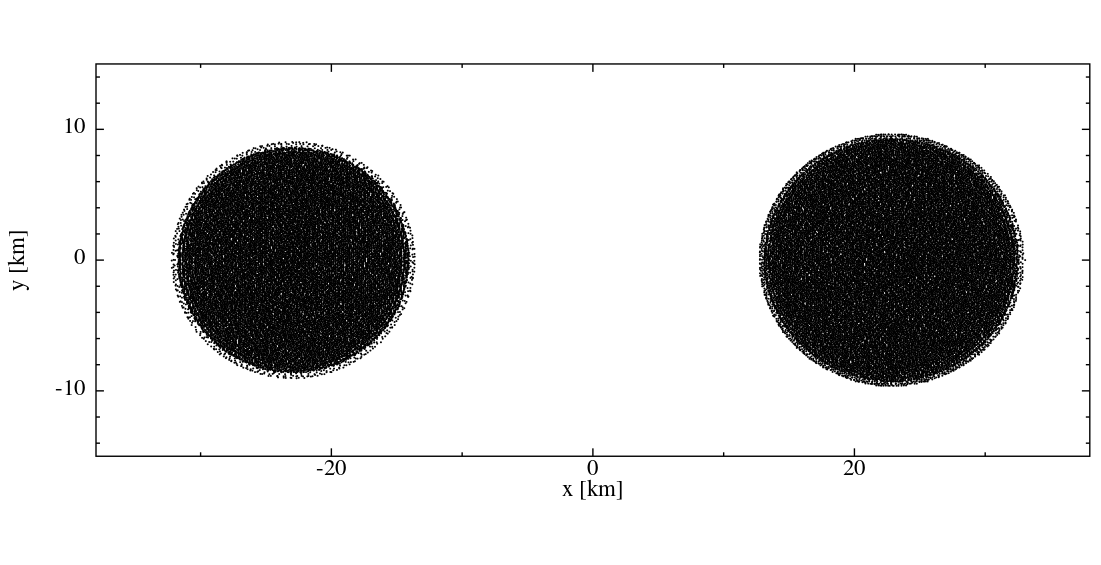} 
   \includegraphics[width=\columnwidth]{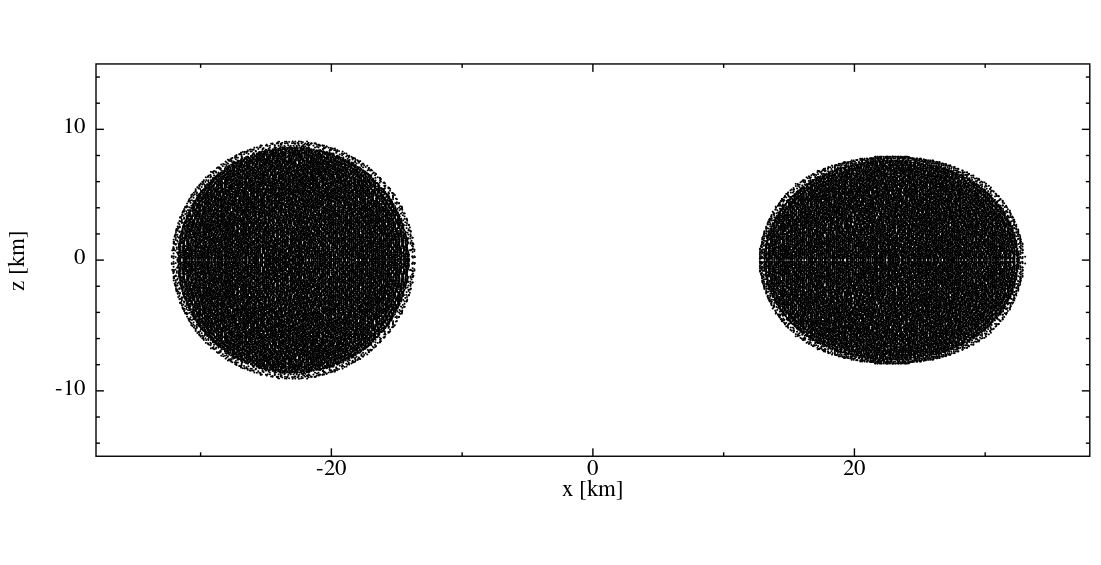}
   \caption{Initial configuration of simulation  \texttt{APR3\_sspin} with $\chi_2=0.5$.
   Shown are the SPH particle positions, projected on the $XY$ (left) and the $XZ$ plane (right).
   The spinning star (each time the one to the right), is substantially larger in the orbital plane 
   (left panel) and substantially rotationally flattened (right panel).}
   \label{fig:ID_APR3}
\end{figure*}

\subsection{Equations of state (EOS)}
\label{sec:EOS}
Our hydrodynamic equations as described in Sec.~\ref{sec:hydro} still need to be closed by an equation of state.
Currently, we are using 
piecewise polytropic approximations to cold nuclear matter equations of state \citep{read09}, that 
are enhanced with an ideal gas-type thermal contribution to both pressure and specific internal energy, a 
common approach in numerical relativity simulations. For explicit expressions please see Appendix A of \cite{rosswog22b}. 
To date, we have implemented 14 piecewise polytropic equations of state, but for our purposes here, we restrict ourselves to
\bi
\i  SLy \citep{SLY_eos}: maximum TOV mass  $M_{\rm TOV}^{\rm max}= 2.05$~\msun, tidal deformability of a 1.4~\Msun star 
$\Lambda_{1.4}= 297$
\i APR3 \citep{akmal98}: $M_{\rm TOV}^{\rm max}= 2.39$~\msun, $\Lambda_{1.4}= 390$
\i MS1b \citep{MS1_EOS}: $M_{\rm TOV}^{\rm max}= 2.78$~\msun, $\Lambda_{1.4}= 1250$.
\ei
For the tidal deformabilities we have quoted the numbers from Table 1 of \cite{pacilio22}.
For all cases we use the piecewise polytropic fit according to Table III in \cite{read09} and we use a  thermal polytropic  exponent
$\gamma_{\rm th}= 1.75$ as a default.
Given the observed  mass of 2.08$^{+0.07}_{-0.07}$~\Msun for J0740+6620 \citep{cromartie20} the SLy EOS is 
still above the $2\sigma$ lower bound of 1.94~\msun, but probably too soft and we consider it as a limiting case
on the soft side.
Concerning the currently ``best educated guess'' of the maximum NS mass, a number of
indirect arguments point to  values of $\sim 2.2-2.4$~\Msun \citep{fryer15,margalit17,bauswein17,shibata17c,rezzolla18,sarin20a,ai23}, 
and a recent Bayesian study \citep{biswas22} suggests a maximum TOV mass of 2.52$^{+0.33}_{-0.29}$~\msun\, \citep[see also introduction in][and references therein]{grb21}, close to the 
values of e.g. the APR3. From our investigated EOSs, we therefore consider APR3 as the most realistic one 
which is also consistent with the findings of \cite{pacilio22,Biswas:2021pvm}. The MS1b
EOS with its very high maximum mass of 2.78~\Msun and its large tidal deformability 
is disfavored by the observation of GW170817 \citep{abbott17b}, we therefore consider it as a limiting case on the stiff side.

\section{Results}
\label{sec:results}
%
\begin{figure*} 
   \centerline{
   \includegraphics[width=2.3\columnwidth]{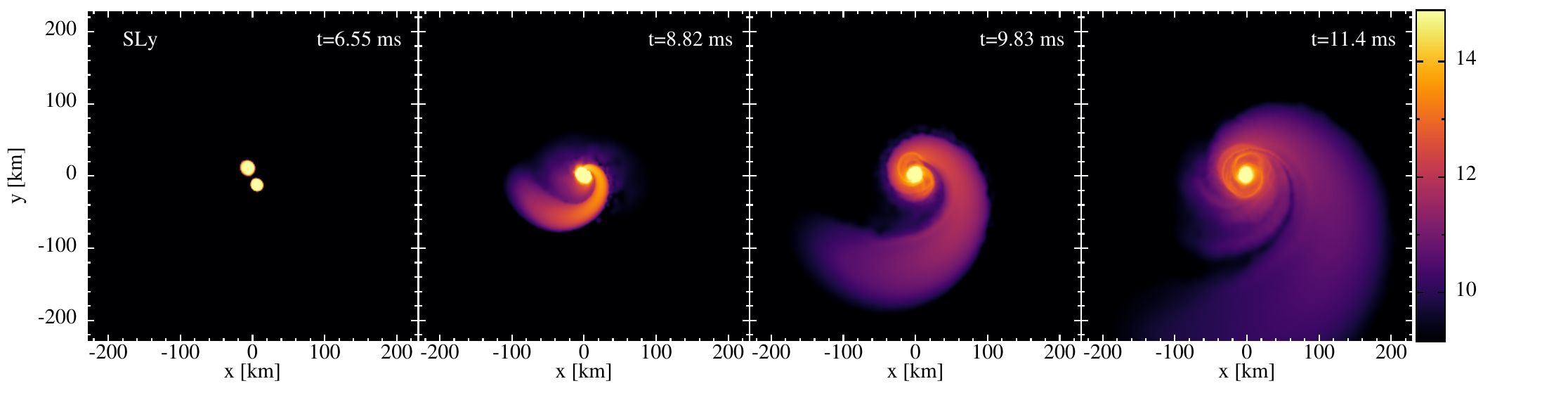}  
   }   
  \centerline{
   \includegraphics[width=2.3\columnwidth]{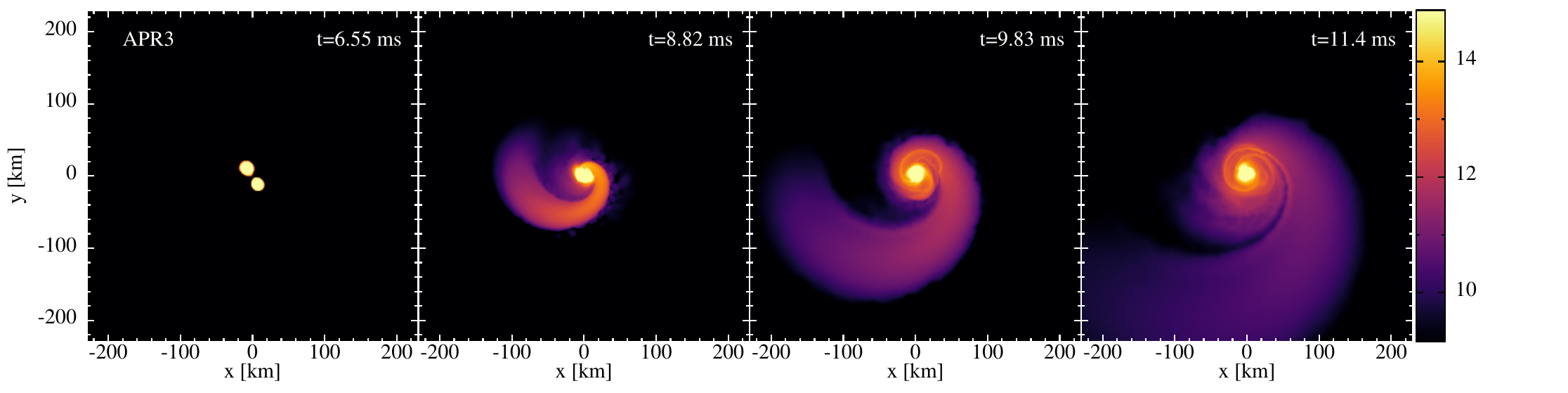} 
   }    
   \centerline{
   \includegraphics[width=2.3\columnwidth]{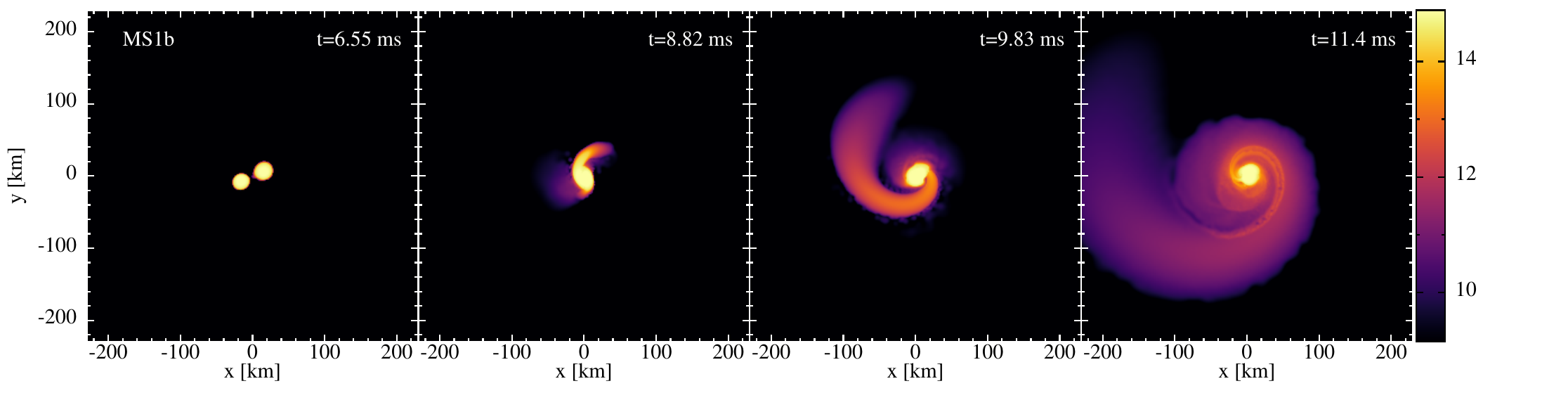} 
   }    
   \caption{Matter density in the orbital plane for the spinning cases ($\chi_1=0$ and $\chi_2=0.5$) for the SLy (top row), APR3 (middle row) and
   the MS1b (bottom row) equations of state.}
   \label{fig:dens_evol}
\end{figure*}
%
\begin{figure*} 
   \centerline{
   \includegraphics[width=\columnwidth]{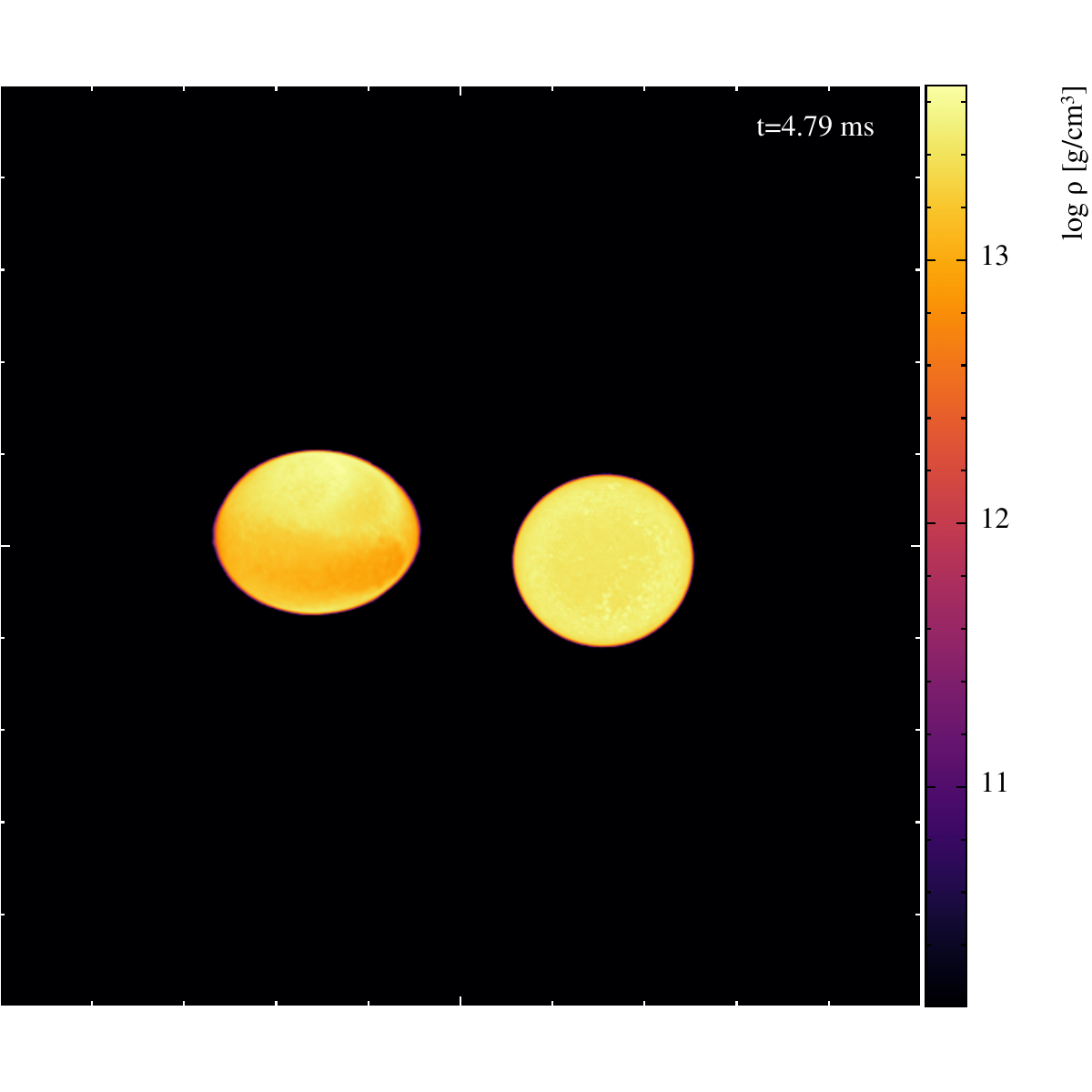} 
   \includegraphics[width=\columnwidth]{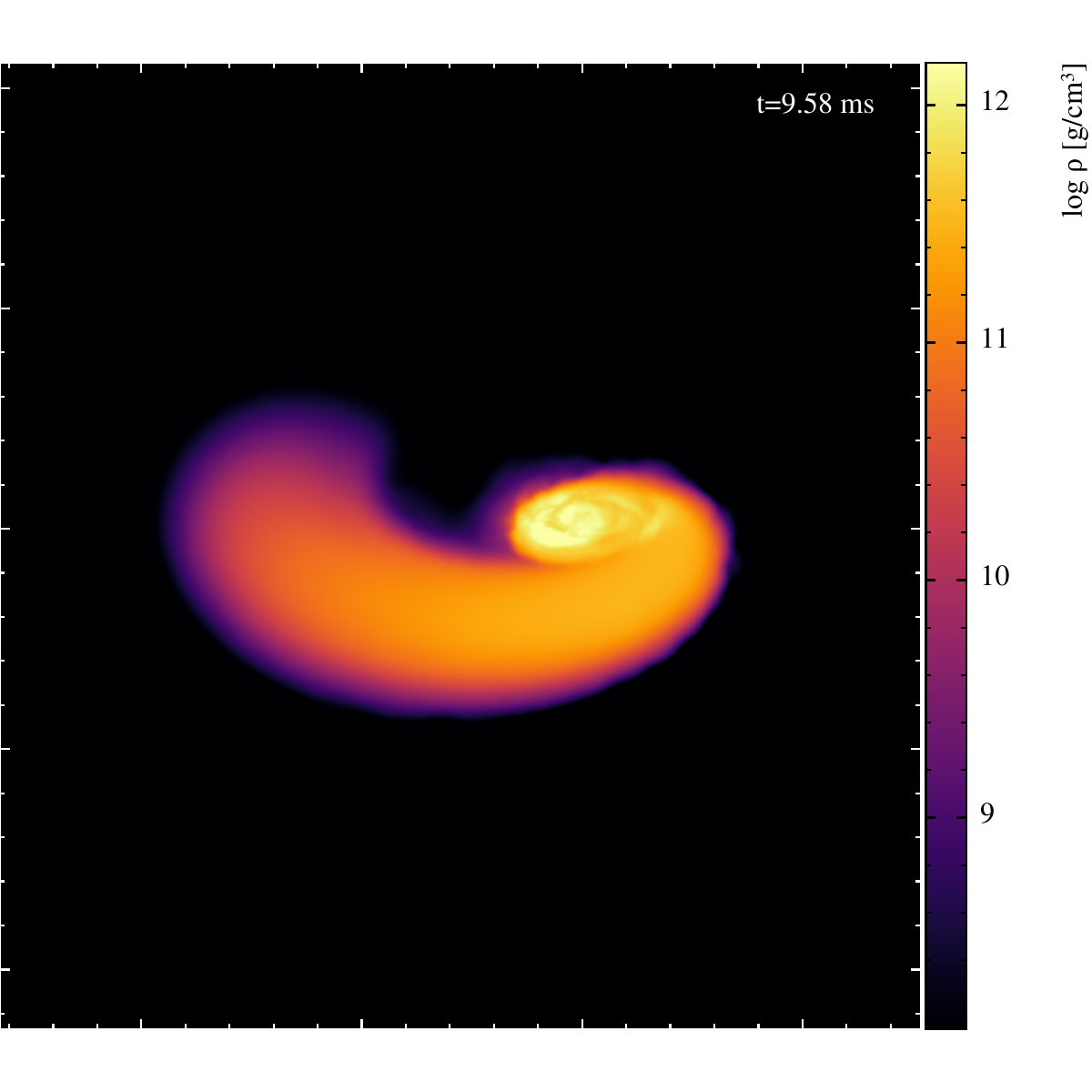}}
    \centerline{
   \includegraphics[width=\columnwidth]{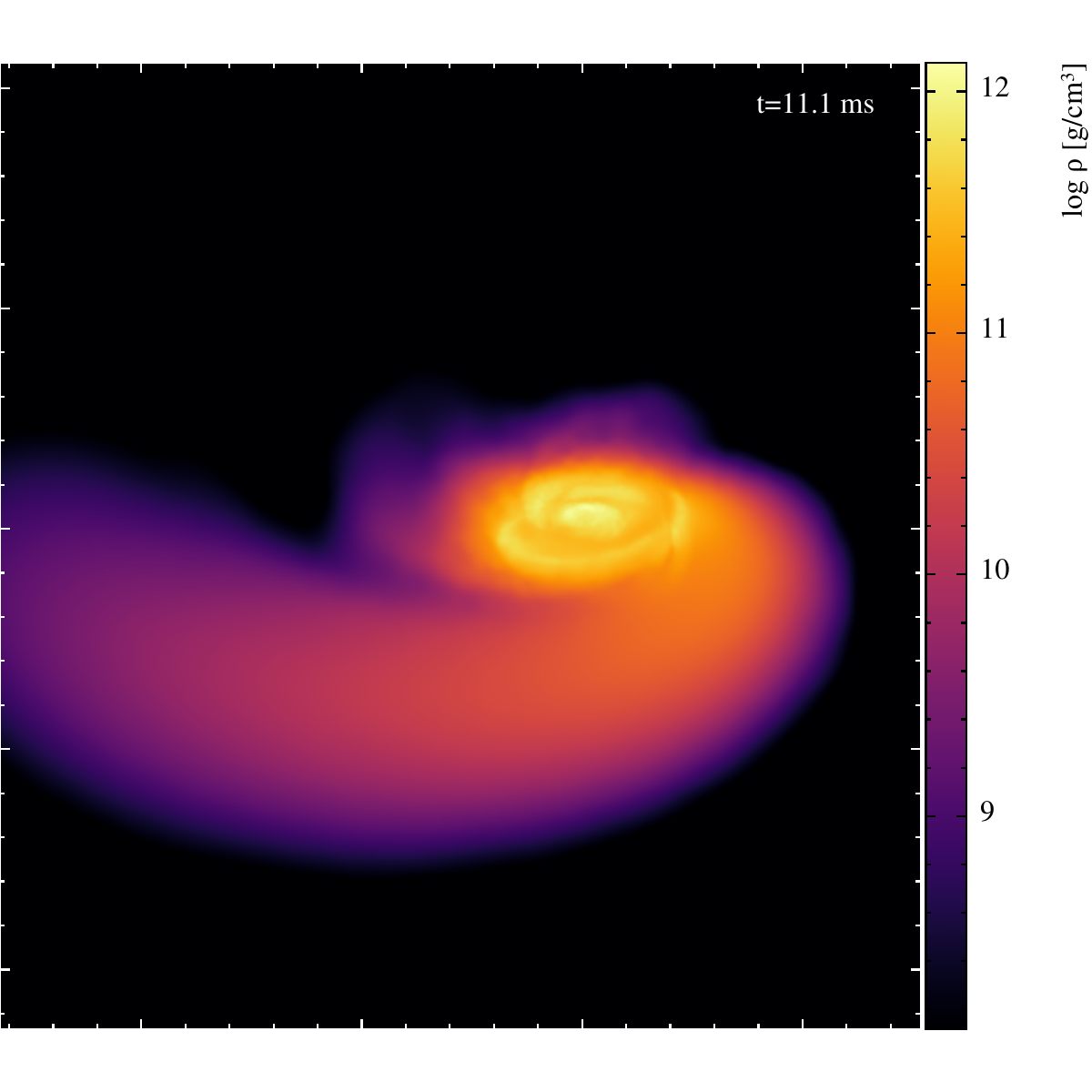} 
   \includegraphics[width=\columnwidth]{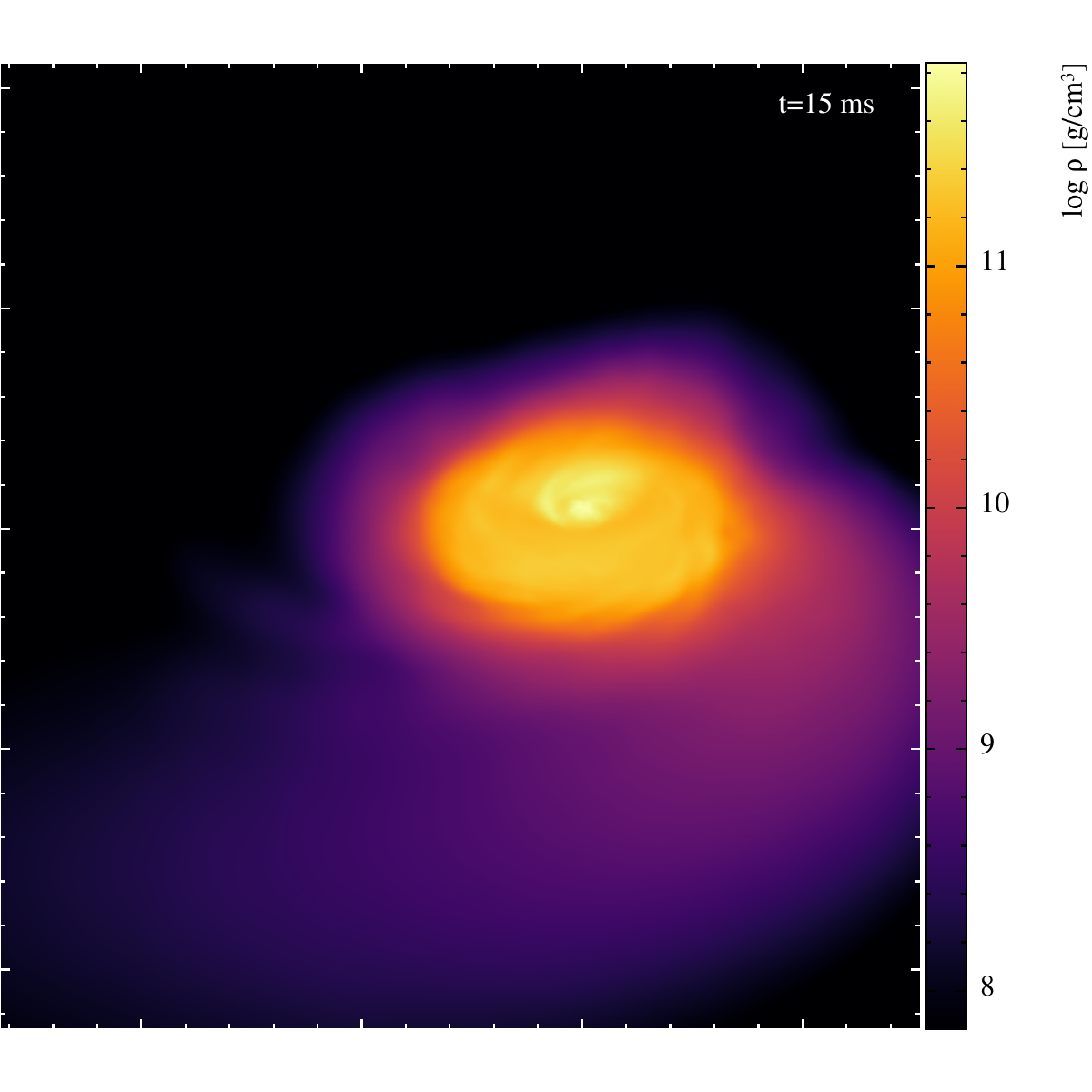}}
      
   \caption{Volume rendering of simulation APR3\_sspin ($\chi_1=0$ and $\chi_2=0.5$). White ticks on the axes refer to
   multiples of 10~km. }
   \label{fig:volren_APR3}
\end{figure*}
\subsection{Simulation setup}
\label{sec:sim_setup}
In this exploratory study, we restrict ourselves to equal mass binary systems with 1.3~\Msun
for each star and use (in order of increasing stiffness) the SLy, the APR3 and the MS1b
EOS. For each case, we run a reference simulation of an irrotational binary and
one where one of the stars has a spin parameter $\chi_2=0.5$ which, depending on the EOS, corresponds to  
spin periods between 1.15 and 1.66~ms, close to the 1.40~ms of PSR~J1748$-$2446ad \citep{hessels06}. 
Both spinning and non-spinning cases are set up by using the \fu initial data solver  \citep{papenfort21}. 
The simulations start ($t=0$) from an initial separation of 45~km and are performed with 2 million 
SPH-particles, and initially 7 mesh refinement levels out to $\approx 2268$~km 
in each coordinate direction. The initial minimum
grid spacing is $\Delta x= 369$~m, but as laid out in detail in \cite{rosswog23a},
new refinement levels are added dynamically, when a time step criterion is met.
These simulations are summarized in Table~\ref{tab:runs}.
To illustrate how large the effect of the spin $\chi_2= 0.5$ is on the stellar structure, we show
in Fig.~\ref{fig:ID_APR3} the  particle positions of the initial configuration of 
our reference run \texttt{APR3\_sspin} projected on both the $XY$  and $XZ$ plane.
The spinning star is  significantly rotationally flattened with the extent in the $Z$-direction
being only about 0.8 of the extent in the $XY$-plane.

\subsection{Morphology}
\label{sec:morph}
The spin has a significant impact on the morphological evolution. Since
the spinning star is substantially enlarged in the orbital plane, it
is more vulnerable to the tidal field from the more compact 
non-spinning companion and becomes disrupted into a large
tidal tail, see Fig.~\ref{fig:dens_evol}, so that the evolution resembles 
one of an unequal mass binary system. The rapidly rotating high-density core
keeps punching into the forming accretion torus, thereby shock-heating the
torus, see the fourth columns (t= 11.4~ms) in Fig.~\ref{fig:dens_evol}.
A volume rendering of simulation \texttt{APR3\_sspin} is shown
in Fig.~\ref{fig:volren_APR3}. As can be seen from panels two to four, 
the shocks driven into the forming torus heat up the debris which,
impeded by matter in the orbital plane, is expanding vertically and 
cause a puffing up of the torus.\\
\begin{figure*} 
   \centerline{
   \includegraphics[width=2\columnwidth]{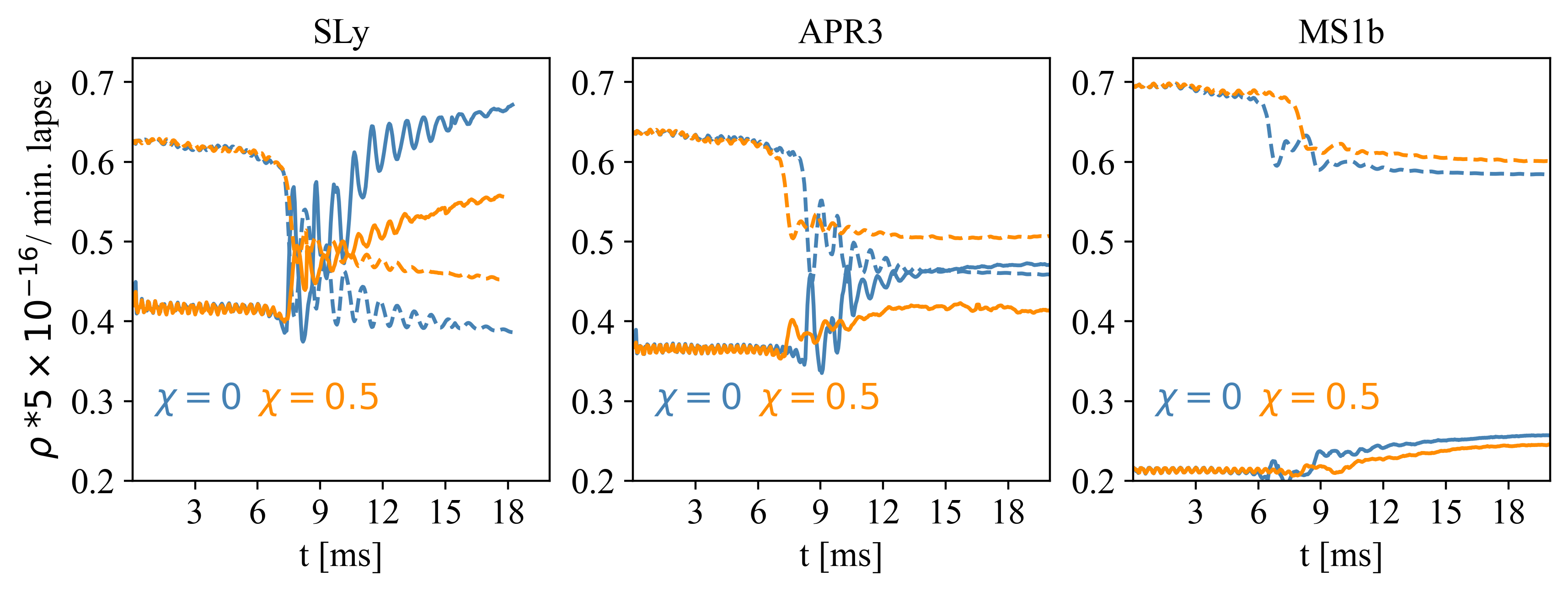}}
   \caption{Maximum densities and minimum lapse values (dashed lines)
   for the different EOSs. The initial peak densities are $8.41 \times 10^{14}$~g~cm$^{-3}$  for SLy, $7.38 \times 10^{14}$~g~cm$^{-3}$  for APR3 and $4.32 \times 10^{14}$~g~cm$^{-3}$  for MS1b), 
   the density values in the plot are scaled by a factor of $5 \times 10^{-16}$. The non-spinning cases are each time shown in blue, the spinning cases in orange.}
   \label{fig:rhomax_lapse}
\end{figure*}
We show in Fig.~\ref{fig:rhomax_lapse} the evolution of the peak densities (solid lines) and  the minimum value
of the lapse function (dashed lines), the non-spinning case is shown in blue, the spinning one in orange. The maximum density for the softest EOS (SLy) is about twice as large as for the stiffest EOS (MS1b). During the inspiral,
the peak density and lapse are practically identical between spinning and non-spinning binaries for all cases.  The two quantities show a clear anti-correlation in the sense that the lapse reaches a minimum at maximum 
compression and vice versa. For all EOSs, the collision is {\em substantially more violent in the non-spinning case}, where  
larger densities and lower lapse values are reached and the post-merger oscillations are of larger amplitude and persist
for longer in the non-spinning case. As expected, these effects are more pronounced for softer EOSs.

\subsection{Gravitational wave emission}
\label{sec:GWs}
\begin{figure*} 
   \centerline{
   \includegraphics[width=2\columnwidth]{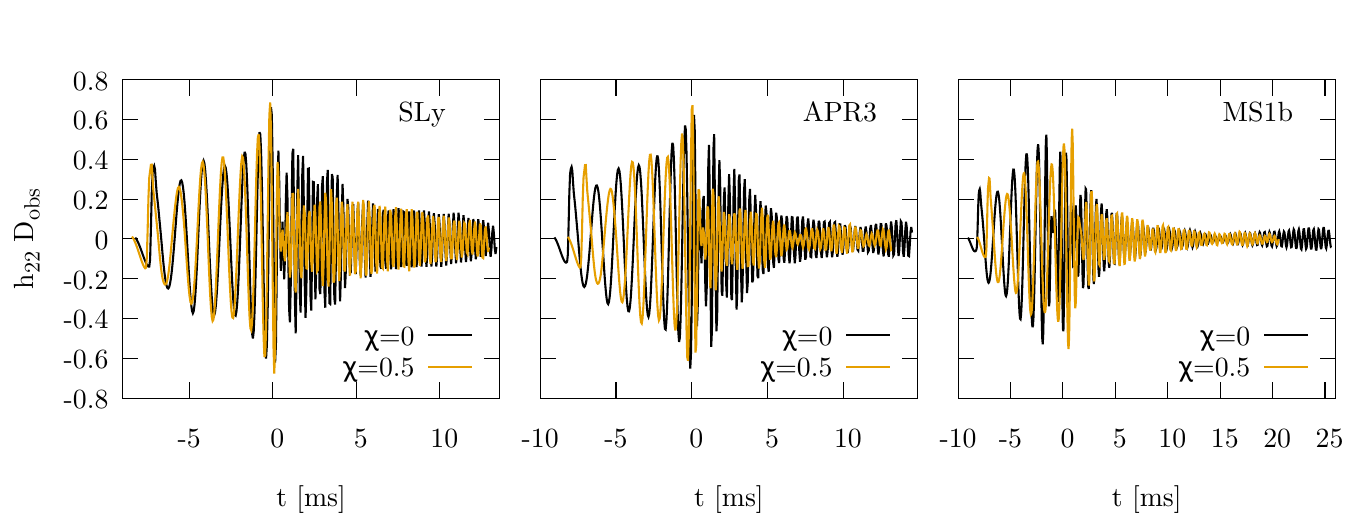} }
   \caption{The $\ell=2$, $m=2$ modes of the gravitational wave strain
   extracted via the Weyl scalar $\Psi_4$. The left panel is for the SLy
   equation of state, the middle panel is for APR3 and the right panel is
   for MS1b. In all cases the waveforms are plotted with black lines for
   the non-spinning case and with orange lines for the spinning case.
   For ease of comparison the waveforms have been shifted in time so
   that the peak amplitude is at $t=0$.}
   \label{fig:GWs_psi4}
\end{figure*}

\begin{figure*} 
   \centerline{
   \includegraphics[width=2\columnwidth]{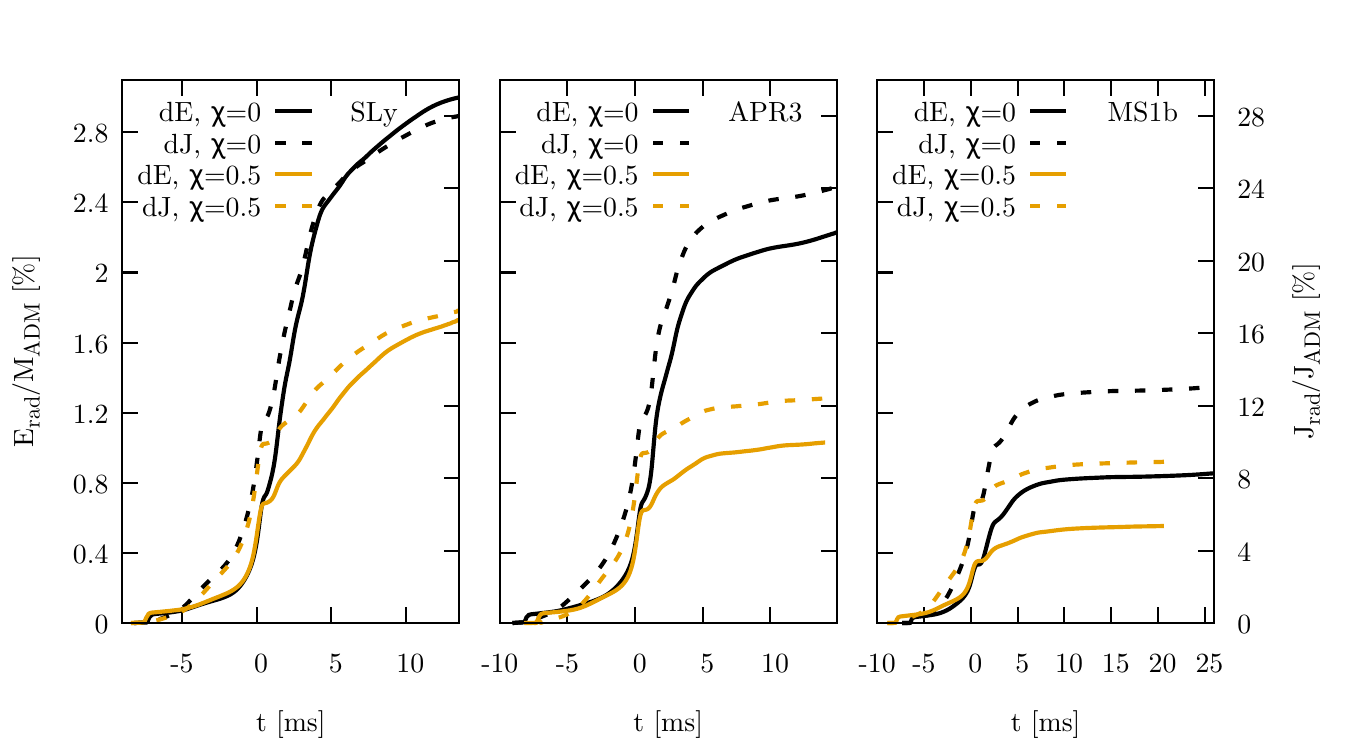} }
   \caption{The radiated energy (left axis, solid lines) and angular momentum (right axis, dashed lines) in percentage of the initial ADM values. The non-spinning cases are plotted with black lines and the
   spinning cases with orange lines. The left panel is for the SLy
   equation of state, the middle panel is for APR3 and the right panel
   is for MS1b. For ease of comparison the waveforms have been shifted
   in time so that the peak amplitude is at $t=0$.}
   \label{fig:GWs_dEdJ}
\end{figure*}

\begin{figure*} 
   \centerline{
   \includegraphics[width=2\columnwidth]{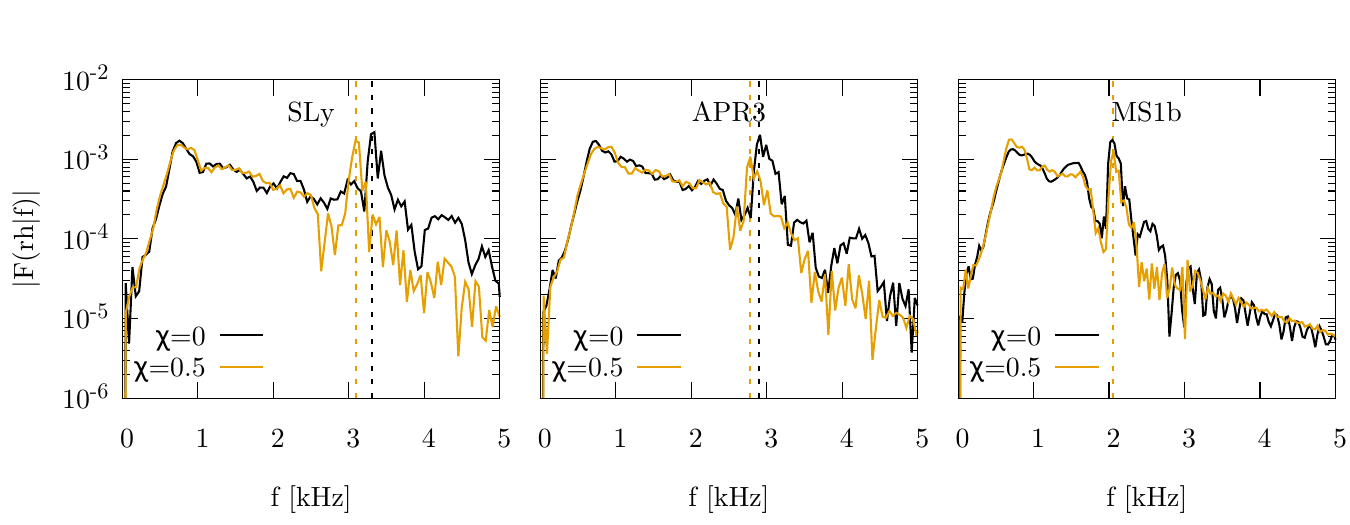} }
   \caption{The spectra of the gravitational wave signal. The left
    panel is for the SLy equation of state, the middle panel for APR3 and
    the right panel for MS1b. In all case that black line is for the
    non-spinning case and the orange line is for the spinning case.
    The positions of the main peaks of the spectra after merger are indicated with vertical dashed lines.}
   \label{fig:GWs_spectra}
\end{figure*}

We have used the \texttt{WeylScal4} thorn of the Einstein Toolkit 
\citep{loeffler12} to postprocess our simulation data and then analyzed 
the resulting waveforms using \texttt{Kuibit} \citep{kuibit}. In
Fig.~\ref{fig:GWs_psi4} we show the dominant $\ell=2$, $m=2$ mode for all
EOSs. In all cases, we see a small initial transient (the first, slightly too large peak)
which is due to imperfect initial conditions
and subsequently the expected ``chirp signal''.
The waveforms are aligned
in time so that the peak of the amplitude coincides at $t=0$.  Directly thereafter, the merged object is strongly compressed, with the bulk of matter being close to axial symmetry and therefore the GW-amplitude at this stage is minimal. On bouncing back, a bar-like structure forms and the amplitude increases again sharply.\\
For all EOSs, the non-spinning case is shown in black while the 
spinning case is shown in orange. The gravitational wave amplitudes are 
largest for the softest EOS and smallest for the stiffest EOS. In all
cases the spinning binary has a weaker signal just after merger than the
non-spinning binary. This is consistent with the observation that the
collision is generally more violent in the non-spinning case (as seen in the density/lapse evolution, Fig.~\ref{fig:rhomax_lapse}, and also in the ejecta properties, see below). Note, however, that
in the case of SLy and MS1b, the GW amplitudes become similar about 
5 ms after merger. Also observe that in some cases the amplitude decays,
but then starts to increase again. This happens at about 7~ms for APR3
and at 15~ms for MS1b.
In Fig.~\ref{fig:GWs_dEdJ}, we show the radiated energy (solid curves)
and angular momentum (dashed curves) in percentage of the initial ADM
values. For all EOSs the non-spinning cases are shown in black while the
spinning case is shown in orange. Consistent with the observation that
the wave amplitude decreases for stiffer EOS, the overall radiated
energy and angular momentum is largest for SLy and smallest for MS1b and
consistent with the observation that the non-spinning collision is
more violent, the {\em radiated energy and angular momentum is significantly
smaller for the spinning case}. It is also worth noting that much more
energy and angular momentum are radiated after the merger than during the
simulated inspiral. The SLy EOS cases also still radiate significantly at the end of
the simulations, while in the other  cases the loss of angular momentum and energy has largely calmed down.
 For massive enough binaries, the less efficient GW-emission for spinning cases will increase the lifetime of a central remnant before it collapses 
to a BH \citep{east19,papenfort22}.\\
In Fig.~\ref{fig:GWs_spectra} we show the spectra of the full waveforms
for all the EOSs. For the softer EOSs (SLy and APR3), not only are the 
peak amplitudes smaller but it also shifts  moderately ($\Delta f< 200$ Hz) to lower frequencies for the
spinning case compared with the non-spinning case.  This is consistent with the findings of \cite{bernuzzi14a,dietrich17} and \cite{east19}. The shift in frequency 
is largest for the softest EOS and, in contrast, essentially zero for
the stiffest EOS. This is consistent with the density evolution shown Fig.\ref{fig:rhomax_lapse}:
for the softer EOSs the density difference between spinning and non-spinning case is much more pronounced
than for stiffer EOSs and the larger peak densities result in higher post-merger GW-frequencies.
It is also worth noting that the secondary peaks,
visible in the spectra of the non-spinning cases, are suppressed in the
spinning case for all EOSs.\\
Observationally, such differences in the post-merger phase are likely too small to be measurable at current sensitivities. However, they may be observable with future gravitational-wave instruments~\citep{CosmicExplorer, EinsteinTelescope, NEMO}. We note that the shift in peak frequency, could also be an important consideration for studies that link the tidal deformability to the post-merger frequency to infer the presence of a hadron-quark phase transition~\citep[e.g.,][]{Bauswein2019}.

\subsection{Dynamic ejecta}
\label{sec:dyn_ej}
Since we only perform simulations for $\sim 20$~ms and neither include magnetic fields
nor neutrino transport, we focus here  on the dynamic ejecta. As a criterion to identify 
them, we use the ``Bernoulli criterion'', see e.g. \cite{rezzolla13a},
\be
- \mathcal{E} U_0 > 1, 
\label{eq:Bernoulli}
\ee
where $U_0$ is the time component of a particle's four-velocity and $\mathcal{E}$
is  the specific enthalpy. To avoid falsely identifying hot matter near the centre as unbound,
we apply the Bernoulli criterion only to matter outside of a coordinate radius
of 100 ($\approx 150$~km) together with the additional condition of the
radial velocity being positive. In a previous study \citep{rosswog22b}
we had found good agreement between the Bernoulli and the ``geodesic criterion'', $-U_0 > 1$.

\subsubsection{Masses}
\begin{figure} 
   \centerline{
   \includegraphics[width=1.35\columnwidth]{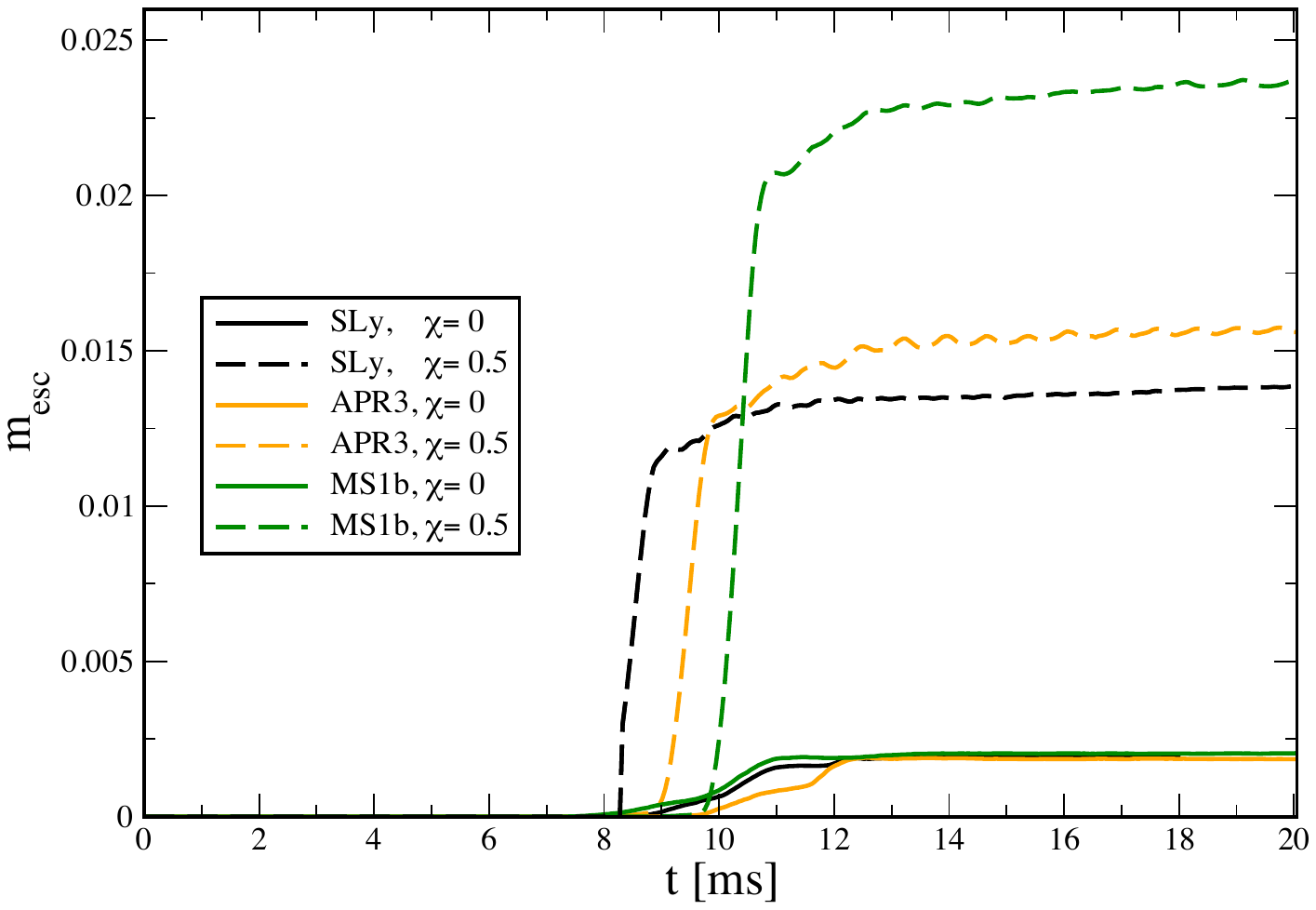} }
   \caption{Mass identified as unbound as a function of time (masses in units of \msun; see main text for more information).}
   \label{fig:mesc_t}
\end{figure}

In Fig.~\ref{fig:mesc_t} we show the evolution of the mass identified as unbound by the
above described criterion.
Clearly, the NS spin has a major effect and increases 
the dynamic ejecta masses by roughly an order of magnitude. While
the irrotational cases yield practically the same ejecta masses for
all EOSs, for the cases with spin there is a significant difference between
ejecta masses correlated with the EOS stiffness. This is because
the bulk of the ejecta is launched via tidal torques that are much larger
for less compact NSs. As a corollary this implies that the
bulk of the dynamical ejecta is very close to the original NS 
$\beta$-equilibrium value and will be dominated by electron fractions
of $\sim 0.1$. This will yield heavy r-process ($A>130$) which, in turn,
will power a red kilonova component.\\
\begin{table*}
	\centering
	\caption{Dynamic ejecta: columns 2 to 7; masses are in units of $10^{-3}$~\msun, the numbers in brackets indicate which percentage is in ``polar''/``equatorial'' ejecta. All velocities given in units of $c$.  Secular ejecta: columns 8 and 9;  column shows 8 the disk masses at the end of each simulation, these numbers should be considered as lower limits.
    The last column shows an estimate of the secular ejecta msses ($= 0.3 \; M_{\rm disk}$).}
	\label{tab:ejecta_masses}
	\begin{tabular}{lrrrrrrrcc} 
		\hline
		run &  $m_{\rm ej,all}$  & $m_{\rm ej, pol}$ (\%) & $m_{\rm ej, equ}$  (\%)& $\langle v \rangle_{\rm all}$  & $\langle v \rangle_{\rm pol}$ & $\langle v \rangle_{\rm equ}$ & $M_{\rm disk}$ [\msun] & $m_{\rm sec}$ [\msun]\\
		\hline
		SLy\_irr &  1.95 & 0.114 (5.8) & 1.84 (94.2) & 0.205 & 0.249 & 0.203 & 0.14 & 0.04\\
		SLy\_sspin & 13.71 & 0.102 (0.7) & 13.6 (99.3) & 0.163 & 0.151 & 0.163 & 0.27 & 0.08\\		
		APR3\_irr &  1.85  & 0.145 (7.8) & 1.70  (92.2)& 0.209 & 0.227 & 0.207 & 0.20 & 0.06\\
		APR3\_sspin & 15.5 & 0.343 (2.2) & 15.1 (97.8) & 0.172 & 0.137 & 0.173 & 0.25 & 0.08\\
		MS1b\_irr & 2.03 & 0.088 (4.3) & 1.95 (95.7) & 0.162 & 0.138 & 0.164 & 0.27 & 0.08\\
		MS1b\_sspin & 23.72 & 0.125 (0.5) & 23.6 (99.5) & 0.154 & 0.148 & 0.154  & 0.33 & 0.10\\
		\hline
	\end{tabular}
\end{table*}
The ejecta masses and velocities are summarized in Table~\ref{tab:ejecta_masses}. We have additionally
labelled the ejecta as ``polar'', defined as being within 30$^\circ$ of the rotation axis and as ``equatorial'' for 
the rest. Our spinning cases typically eject an order of magnitude more mass than the irrotational cases. 
In nature, of course, also the mass ratio will have an impact on the ejecta masses. \cite{papenfort22} 
interestingly find the largest amounts of ejecta for equal mass cases with  
large spins (rather than, say, for extreme mass ratios).

\begin{figure*}%
   \centerline{
   \includegraphics[width=2.1\columnwidth]{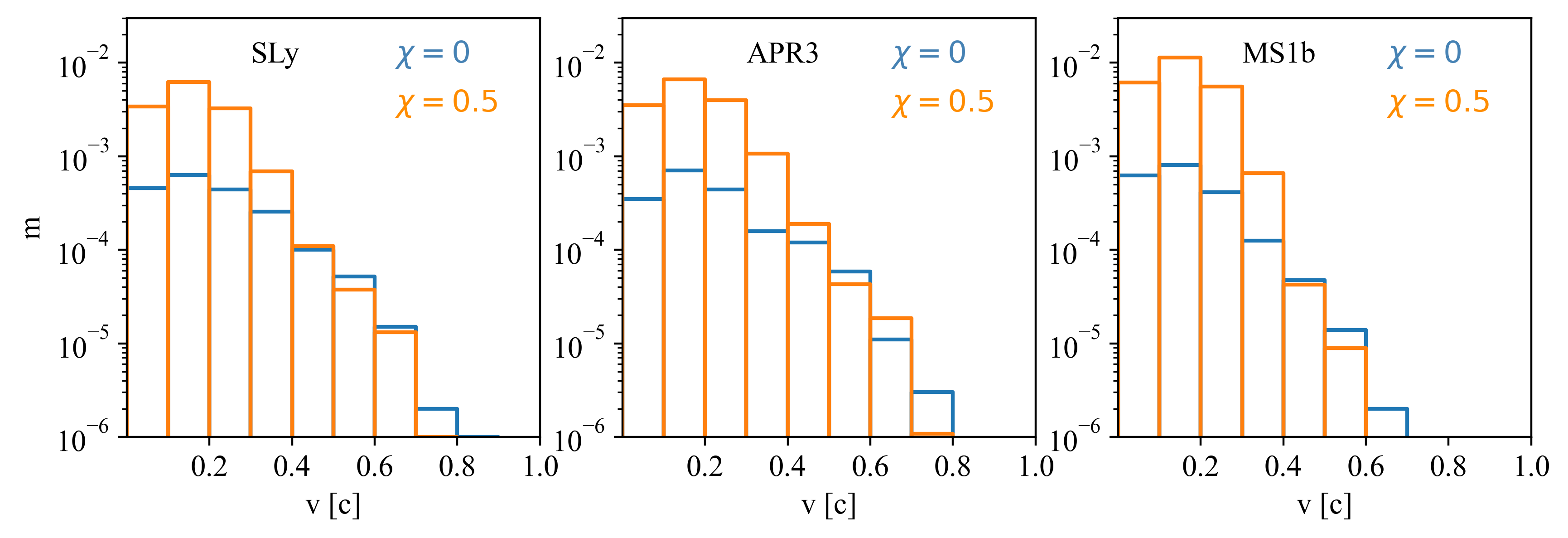} }
   \caption{Ejecta masses (in \msun) binned according to the velocities at infinity for three EOSs. Results for the irrotational cases are shown in blue, the spinning cases in orange.}
   \label{fig:binned_vel}
\end{figure*}

\subsubsection{Velocities}
In Fig.~\ref{fig:binned_vel} we show the masses in different velocity bins where the velocities refer to those asymptotically reached at 
infinity. The mass-averaged ejecta velocities hardly exceed 0.2~c, but in each of the simulations we find ejecta velocities 
exceeding 0.5~c. While this material is not very well resolved, there seems to be a robust trend of the {\em non-spinning} cases 
producing {\em more high-velocity material}.  This is a result of the less violent collision in the spinning cases
discussed in Sec.~\ref{sec:morph} where the central object avoids particularly deep compressions which, when bouncing back,
produce high-velocity ejecta.

\subsection{Secular ejecta}
\begin{figure} 
   \centerline{
   \includegraphics[width=1.1\columnwidth]{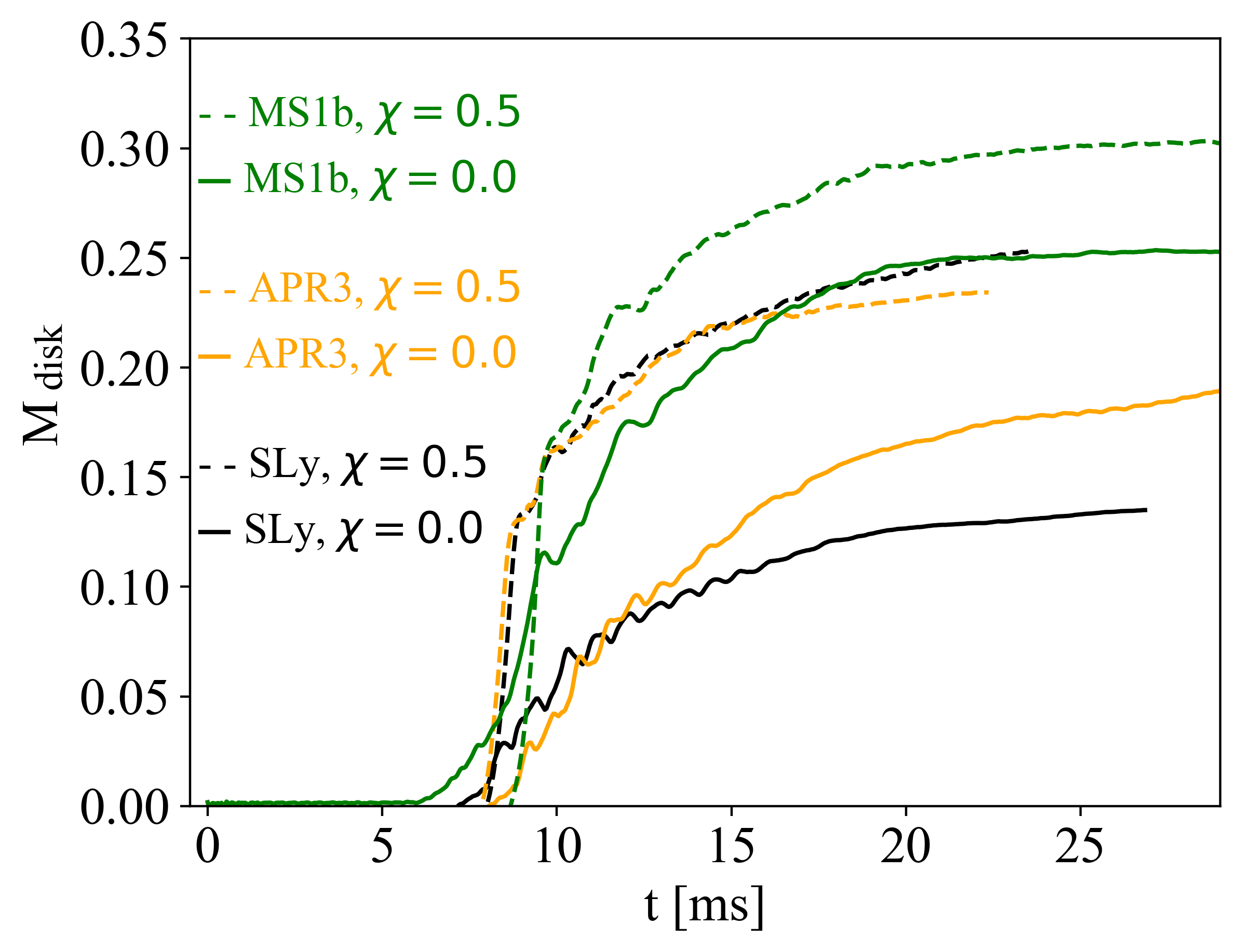}}
   \caption{Evolution of the disk masses (masses in units of \msun), where we use the mass with $\rho<10^{13}$ \gcc $\;$ that is bound to the central object as an estimate
   of the disk mass.}
   \label{fig:mdisk_t}
\end{figure}
Apart from the dynamic ejecta, mergers also produce secular ejecta that
become unbound on time scales much longer than what we simulate here. 
They amount to a substantial fraction ($\sim$ 0.3 -- 0.4) of the initial disk masses and expand typically at $\sim0.1 \; c$ \citep{siegel17a,miller19,fernandez19}. 
To estimate the disk masses, we monitor the mass with $\rho < 10^{13}$ \gcc $\;$ that is still bound as a function of time, see Fig.~\ref{fig:mdisk_t}. Here we also see a substantial increase of the disk masses with spin, however, much less pronounced than for the dynamic ejecta. 
We find a factor of $\sim$ two for the softest EOS (SLy) and a substantially smaller increase  for the medium (APR3; 25 \%) and the stiffest EOS (MS1b; 22 \%). 
The estimated amount of secular ejecta, conservatively estimated as 0.3 $M_{\rm disk}$, is shown in the last column of Table~\ref{tab:ejecta_masses}. 
It is worth stressing that a) the dynamical and secular ejecta masses of the more realistic EOSs (SLy and APR3) add up for the irrotational cases to numbers that are close to but slightly larger than the estimates of $\sim$0.04 \Msun from GW170817~\citep{cowperthwaite17}. However, for the spinning cases they can be up to a factor of 2 larger, and b) the increase is the more pronounced for more softer EOS, see Table~\ref{tab:ejecta_masses}. 
A substantial disk mass increase through initial spins has also been reported by \cite{papenfort22}.

\subsection{Electromagnetic emission}
\label{sec:EM}
\subsubsection{Kilonova}
\label{sec:kilonova}
\begin{figure*} 
   \centerline{
   \includegraphics[width=0.75\columnwidth]{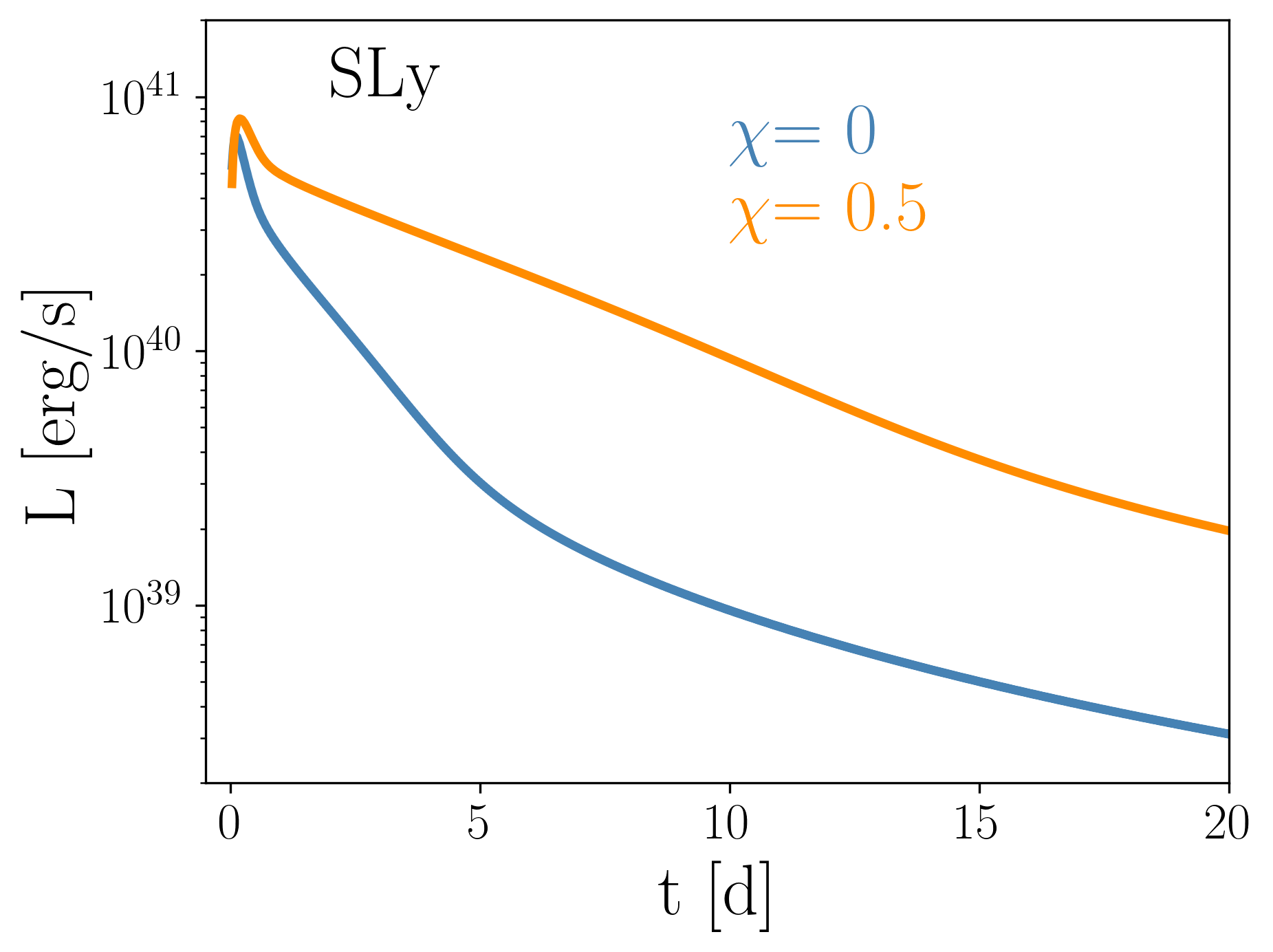} \hspace*{-0.15cm}   \includegraphics[width=0.75\columnwidth]{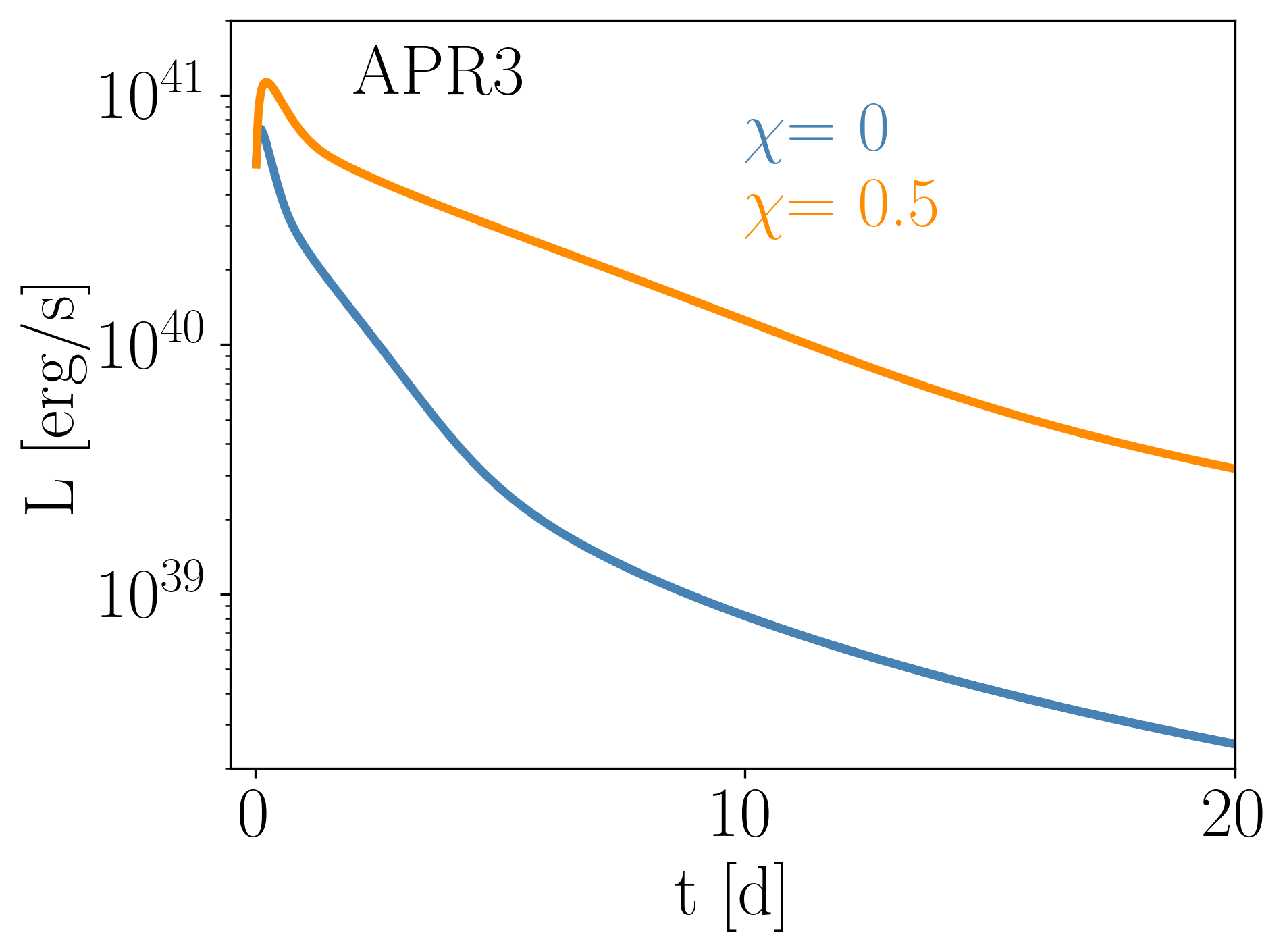}  \hspace*{-0.15cm}   \includegraphics[width=0.75\columnwidth]{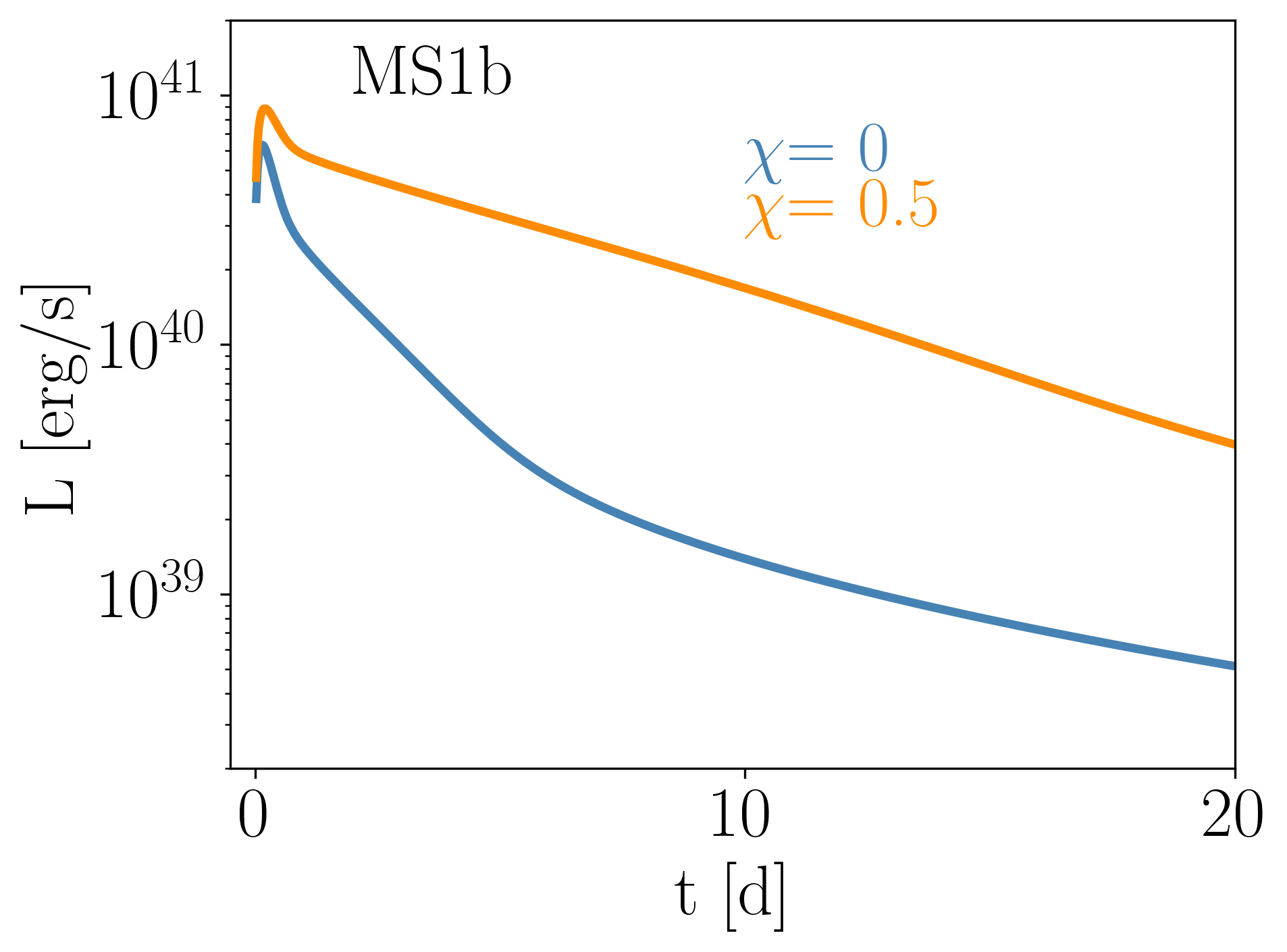}}
   \caption{Bolometric kilonova light curves  from the {\em dynamic ejecta} based on a semi-analytic eigenmode expansion model \citep{wollaeger18a,rosswog18a} for the 
   (irrotational and spinning) SLy (left), APR3 (middle) and MS1b EOS cases (right).}
   \label{fig:KN_bolometric_dynamic}
\end{figure*}
\begin{figure*} 
   \centerline{
   \includegraphics[width=0.75\columnwidth]{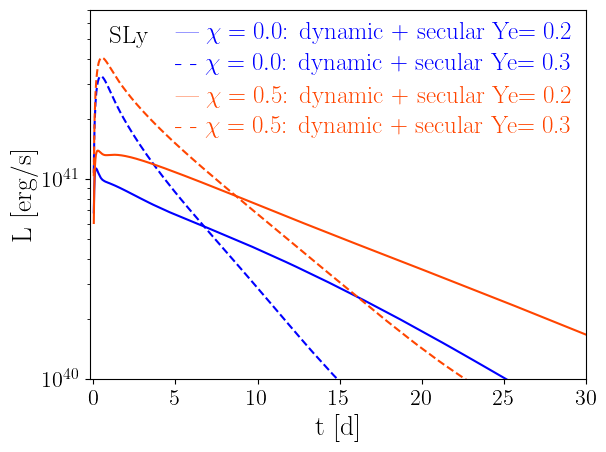} \hspace*{-0.15cm}
   \includegraphics[width=0.75\columnwidth]{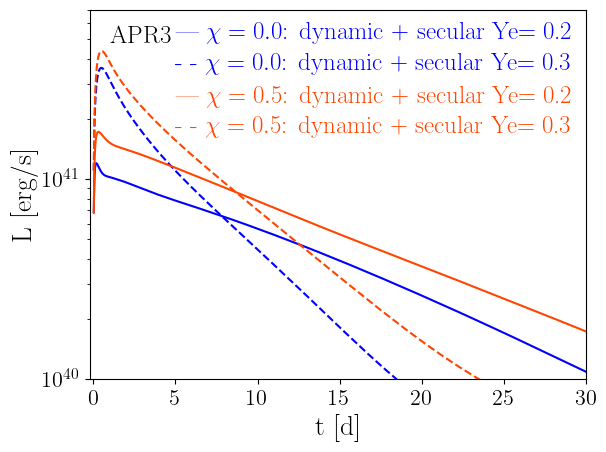}  \hspace*{-0.15cm}   \includegraphics[width=0.75\columnwidth]{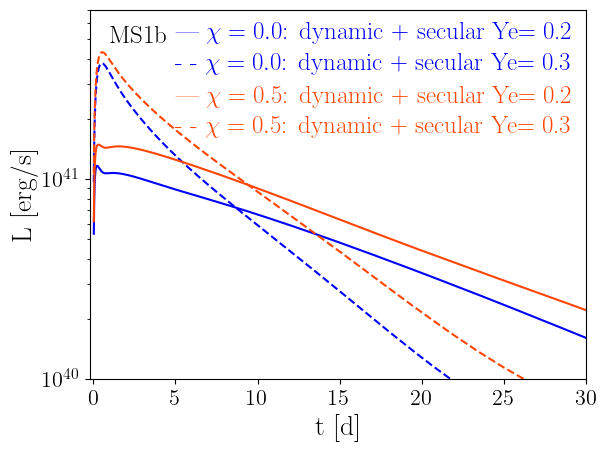}}
   \caption{ Bolometric kilonova light curves from {\em all ejecta}. i.e. dynamic
   plus secular ejecta, where the latter masses have been estimated from the disk masses. Since the electron fraction of the secular ejecta is not a settled topic, we explore a case with the secular ejecta $Y_e$ being below ($Y_e=0.2$) and another one with being above ($Y_e=0.3$) the critical electron fraction value of $Y_e^{\rm crit}\approx 0.25$.}
   \label{fig:KN_bolometric_all}
\end{figure*}

With these large differences in ejecta masses due to the spin, one can also
expect significant differences in the bolometric light curves of the resulting
kilonovae. We will begin by calculating the kilonova emission for {\em only the dynamic ejecta}, and, in a second step, we add the {\em secular ejecta} whose properties we estimate based on the disk masses. But before we do so, we want to start
with  simple order of magnitude estimates \citep{arnett80,arnett82,kasen10,metzger10b,kasen13a,piran13a,grossman14a}
to understand
the qualitative dependence on various parameters.\\
Let us assume that the ejecta are spherical
and have already reached the homologous expansion stage (this happens quickly, see
e.g. \cite{rosswog14a,neuweiler23}) and the radius evolves as $R=v t$, where $v$ is the expansion velocity and $t$ the expansion time.
For an ejecta mass $m_{\rm ej}$ we then have a density of
\be
\rho \sim \frac{m_{\rm ej}}{(4\pi/3)R^3}= \frac{m_{\rm ej}}{(4\pi/3)v^3 t^3}.
\ee
 The typical optical depth is $\tau \sim \kappa \rho R$, where $\kappa$ is a typical opacity (keep in mind that in reality this is of course wavelength-dependent). The diffusion time is then given by
\be
\tau_{\rm diff}\sim \tau \frac{R}{c}= \frac{\kappa \; m_{\rm ej}}{(4 \pi/3)c v t}.
\ee
The peak emission occurs when the expansion time, $t=R/v$, equals approximately the diffusion time $\tau_{\rm diff}$. After solving for the time, one finds 
\be
\tau_{\rm peak}\sim \sqrt{\frac{\kappa m_{\rm ej}}{(4\pi/3)c v}} \propto \kappa^{1/2} m_{\rm ej}^{1/2} v^{-1/2}.
\label{eq:t_peak}
\ee
Any initial thermal energy has at $\tau_{\rm peak}$ been lost to expansion and thereafter the energy source is radioactive decay
\be
L\sim \eta \; m_{\rm ej} \; \dot{q},
\ee
where $\eta$ is some efficiency (neutrinos, for example, are lost) and $\dot{q}$ is the nuclear heating rate per unit mass. The nuclear heating rate can often be well parametrized as a powerlaw \citep{metzger10a,korobkin12a,lippuner15,hotokezaka17a,rosswog18a,hotokezaka20,rosswog22c}
\be
\dot{q}= Q_0 \left(\frac{t}{t_0}\right)^{-\alpha}
\ee
with $\alpha \sim 1.3$, so that the luminosity
is
\be
L(t) \sim \eta Q_0 t_0^{\alpha} \;m_{\rm ej} \; t^{-\alpha}
\label{eq:lum}
\ee
and, at peak, 
\be
L_{\rm peak}= L(\tau_{\rm peak}) \propto \kappa^{-\alpha/2} \; v^{\alpha/2} \;  m^{1-\alpha/2} \approx \kappa^{-0.65} \; v^{0.65} \;  m^{0.35}.
\label{eq:L_peak}
\ee

\begin{figure*} 
   \centerline{
   \includegraphics[width=2\columnwidth]{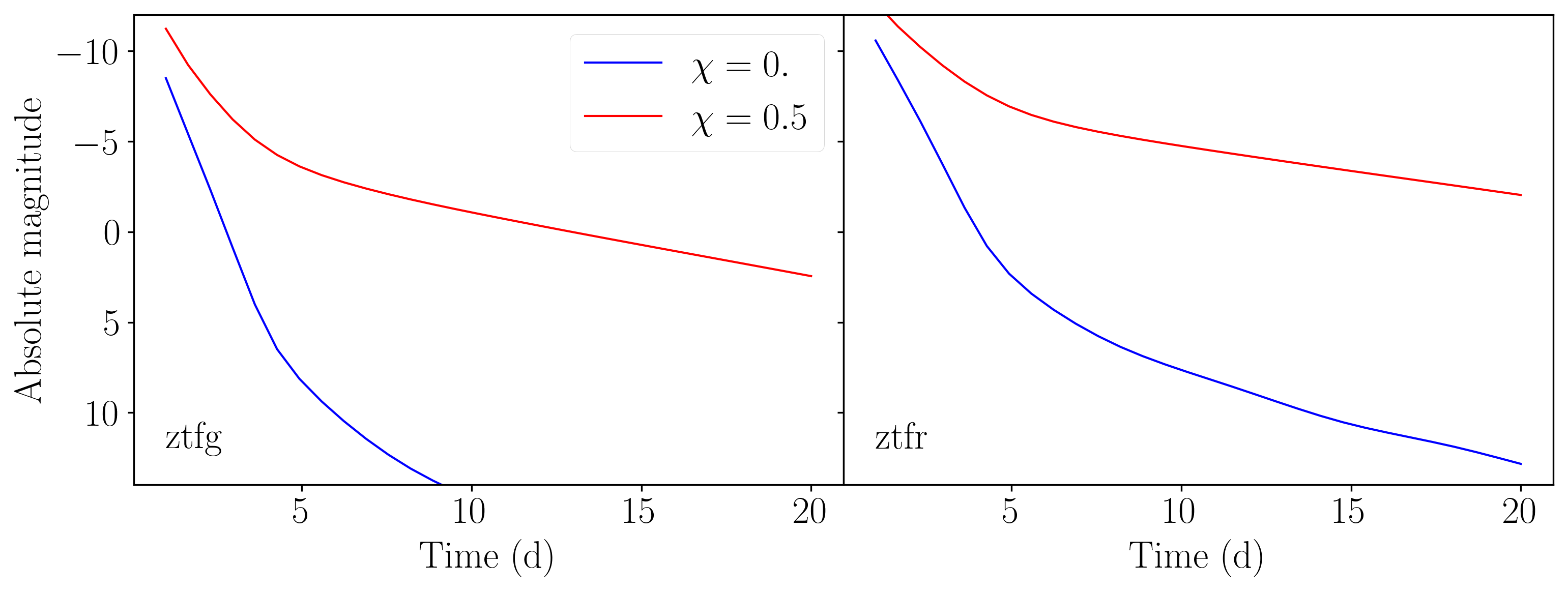}}
   \caption{Absolute magnitude lightcurves for the spinning (red) and non-spinning cases (blue) for ztfg (left) and ztfr (right) filters.}
   \label{fig:knmagnitude}
\end{figure*}

\begin{figure*} 
   \centerline{
   \includegraphics[width=2\columnwidth]{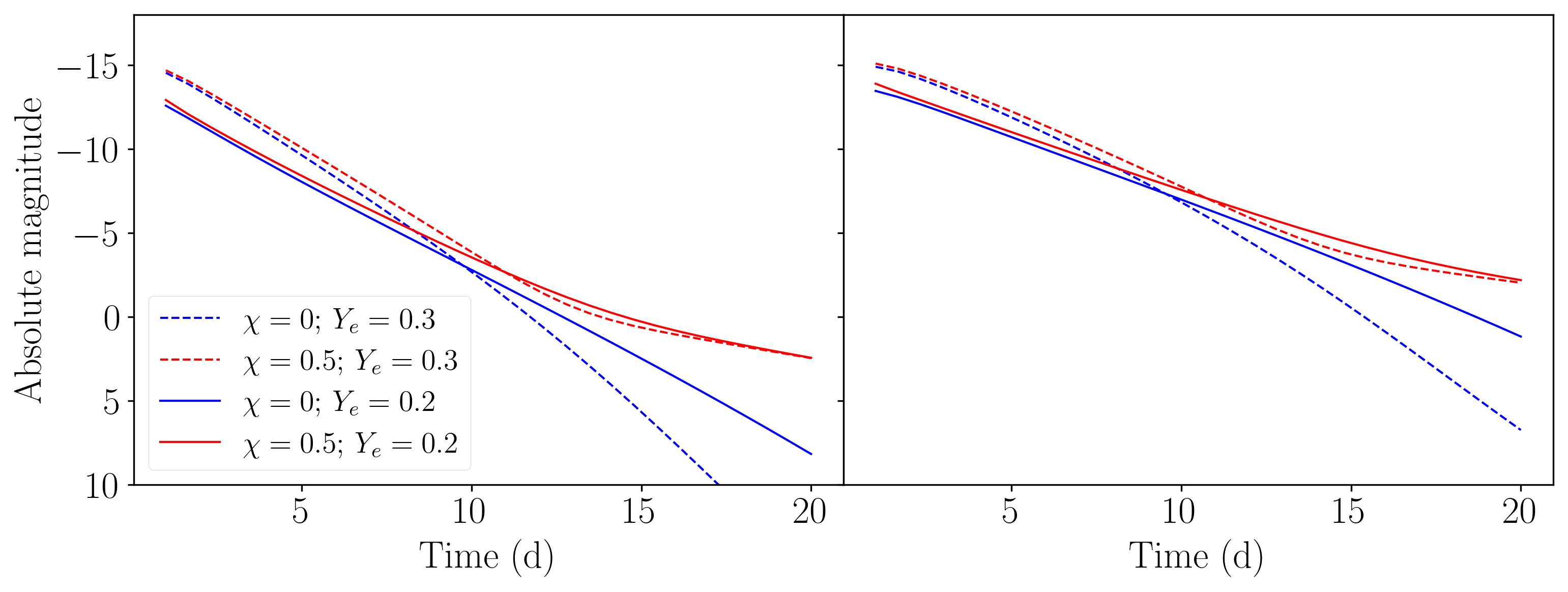}}
   \caption{Absolute magnitude lightcurves for the spinning (red) and non-spinning cases (blue) for ztfg (left) and ztfr (right) filters for the dynamical and secular ejecta.}
   \label{fig:knmagnitudeallejecta}
\end{figure*}

\noindent{\em Dynamic ejecta.}
Since our current set of simulations does not include neutrino transport
(or any approximation to it), we have no simulation-based information of the neutron-richness
of the ejecta. However, at least for the spinning cases, the ejecta are heavily dominated
by tidal ejecta that are never substantially heated and therefore are ejected with
their original, cold $\beta$-equilibrium $Y_e < 0.1$ (see, for example, Fig.~21 in \cite{farouqi22}). We take an angle of
$|\Theta| < 30^\circ$ to approximately divide the ``polar'' and ``equatorial'' ejecta and assign
them values of $Y_e^{\rm pol}= 0.3$ and $Y_e^{\rm equ}= 0.1$. 
Based on these properties, we compute  kilonova light 
curves with a semi-analytic eigenmode expansion model 
\citep{wollaeger18a,rosswog18a} which is based on \cite{pinto00}. 
We use opacities selected according to the electron fraction 
(following \cite{tanaka20a}, Table 1). For the heating due
to radioactivity, we use a fit formula \citep{rosswog22c} that yields 
the heating rate as a function of velocity, electron fraction and time.
This fit formula is based on a heating rate library that has been 
produced with the Winnet nuclear reaction network \citep{winteler12} that in turn
is based on the BasNet network \citep{thielemann11}. The network contains 5831 isotopes from the valley of $\beta$-stability to the neutron drip line, starting with nucleons and reaching up to $Z=111$. We applied a thermal efficiency that is based on \cite{barnes16a} with parameters suggested in \cite{metzger19a}. 
The resulting light curves are shown in Fig.~\ref{fig:KN_bolometric_dynamic}. 
The lightcurve peaks for the spinning cases are reached about a factor of 2 later (at $\sim 6$ hours post merger) than for the irrotational cases, broadly consistent with Eq.~(\ref{eq:t_peak}).
As expected from Eqs.~(\ref{eq:lum}) and (\ref{eq:L_peak}), the irrotational binaries produce substantially dimmer
kilonovae with luminosities at 5 days being about an order of
magnitude lower than for the single spinning star case. Among those,
the APR3 EOS case achieves the brightest peak luminosities, likely due to the largest amount of polar ejecta. In the spinning cases, the kilonova lightcurves reach their peak several hours after the non-spinning cases. This finding is broadly consistent with those of \cite{papenfort22}.\\ 
In Fig.~\ref{fig:knmagnitude}, we show the absolute magnitude lightcurves for the kilonova (only from the dynamical ejecta) for two filters, ztfg and ztfr, computed assuming a blackbody spectral energy distribution for the APR3 equation of state and the same Eigenmode expansion model as described above through {\sc{Redback}}~\citep{sarin23_redback}. These broadband lightcurves again illustrate the significant differences in the kilonova brightness and evolution for the spinning case.\\
{\em Dynamic plus secular ejecta.} We are not modelling the secular
ejecta directly, but since this channel likely
produces the largest ejecta fraction, we 
try to estimate its effects on the resulting emission.
To this end we (conservatively) assume that
30\% of the disk mass becomes unbound, see last column in Table~\ref{tab:ejecta_masses}, and
escapes at 0.1 $c$. The electron fraction of these ejecta is not fully settled yet, see \cite{rosswog22c}
Sec.~2.2.3 for a discussion and pointers to the original literature,
therefore we study one case with $Y_e$ below (0.20) and one case above (0.30) the critical
electron fraction value of $Y_e^{\rm crit}\approx 0.25$ \citep{korobkin12a,lippuner15}. The corresponding lightcurves are shown in Fig.~\ref{fig:KN_bolometric_all}. Clearly, the {\em impact of the electron fraction} (which mostly determines the opacity $\kappa$)
of the secular ejecta is significant for the peak luminosity (see Eq.~(\ref{eq:L_peak})), the peak time (see Eq.~(\ref{eq:t_peak})) and for the slope of the lightcurve decay. To fix ideas, let us concentrate on our
best guess EOS, APR3, shown in the middle panel of Fig.~\ref{fig:KN_bolometric_all}, similar statements hold for the other EOSs.
Changing the secular electron fraction from 0.2 to 0.3 increases the peak luminosity by a factor of $\sim$ 3, and also delays the time to peak by about the same factor (compare, for example, the solid blue line with the dashed blue line). The {\em impact of the spin} is qualitatively similar (compare e.g. the solid blue with the solid orange line):
it increases the peak luminosity by $\sim$ 40 \% while delaying the time to peak by nearly a factor of 2.\\
Since the secular ejecta dominate the masses by at least
factors of a few, and in some cases more than an order 
of magnitude, their properties also predominantly shape the electromagnetic emission, so that the effects of dynamical ejecta alone will be very difficult to infer from observations.\\
To again allow a more direct comparison to observations, we also show the absolute magnitude for the combination of secular and dynamical ejecta in Fig.~\ref{fig:knmagnitudeallejecta}. This follows the same trend as the dynamical ejecta only lightcurve, in so far as the spinning case produces a brighter kilonova. However as with the bolometric lightcurve the properties of the secular ejecta shape the electromagnetic emission.

\subsubsection{Kilonova precursor and afterglow}
\begin{figure*} 
   \centerline{
   \includegraphics[width=2\columnwidth]{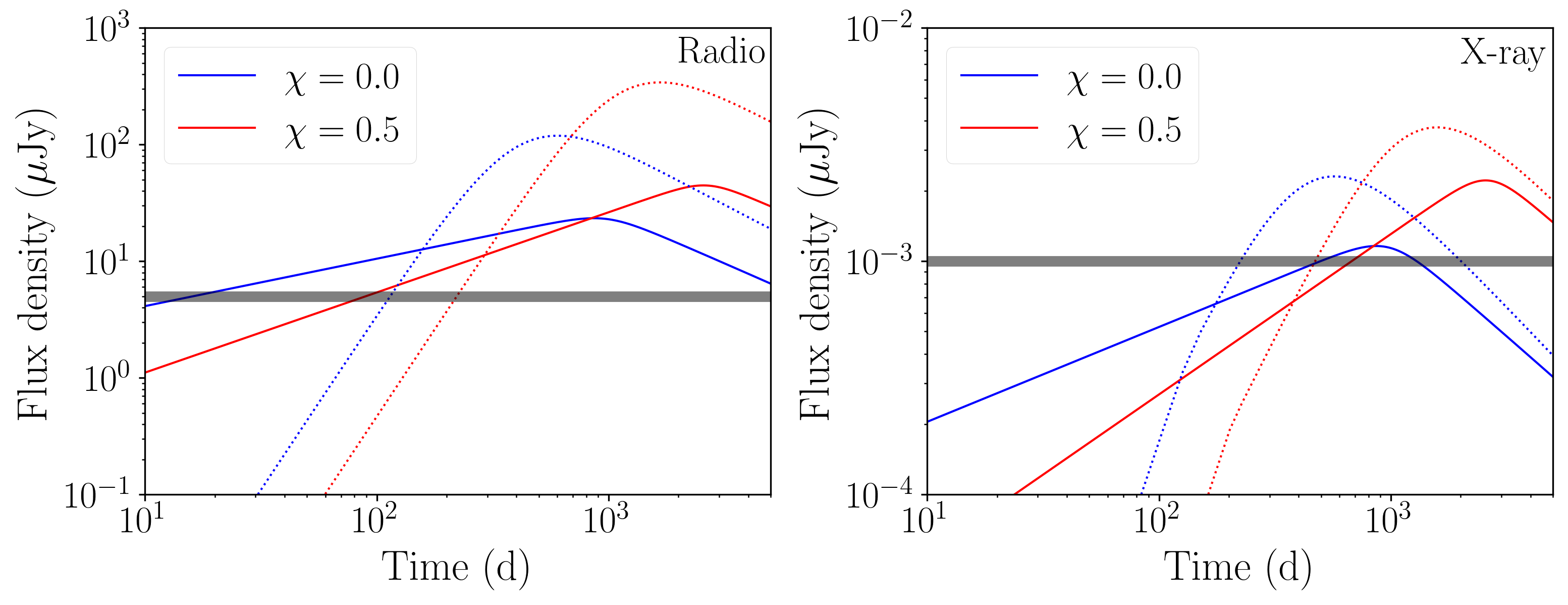}}
   \caption{Radio (left panel) and X-ray (right panel) synchrotron lightcurves at 3~{Ghz} and 5~{KeV}, respectively from the dynamical ejecta for the zero spin (blue curves) and spinning (red curves) for the APR equation of state. The solid and dashed curves represent two different modelling assumptions about the kilonova afterglow, while the gray horizontal shaded band represents the sensitivity limit of the VLA and Chandra telescopes.}
   \label{fig:KNafterglow}
\end{figure*}

We also briefly mention the fast velocity ejecta
component and its observational consequences. In each of the simulated  cases, we find dynamic ejecta velocities
above 0.5~$c$. Such a fast component
may cause additional electromagnetic signatures: in the leading, fastest ejecta, neutrons may avoid being captured
by nuclei  and their subsequent decay will trigger a blue/ultra-violet precursor to the kilonova $\sim 1$~h after the merger \citep{metzger15a}.
Additionally, these fast ejecta, once being decelerated by the 
interstellar medium, may cause a ``kilonova afterglow'' due to 
synchrotron emission \citep{nakar11a,hotokezaka15a,hotokezaka18a}. 
This may have actually been observed as a late-time increase in X-ray 
flux years after GW170817 \citep{hajela22,troja22,hajela22}. \\
To explore the effect of spin on this ``kilonova afterglow'' in detail, we calculate the kilonova afterglow signature from the dynamical ejecta using two complementary approaches to check for consistency. 
 In the first approach, we take the kinetic energy and total mass for the spinning and non-spinning case numerical simulations for the APR3 equation of state, and model the synchrotron emission following~\citet{sarin22_kne}. In particular, we model the dynamical ejecta as a one-zone spherical outflow propagating out into a constant-density interstellar medium with the observed emission a product of synchrotron radiation produced by the ejecta and external medium~\citep{nakar11a} corrected for synchrotron-self absorption and evolution into the deep-Newtonian regime~\citep{Sironi2013}. As an alternative, second approach, we take the distribution of mass and velocity from the numerical simulations described above, approximating the mass vs velocity relationship as broken power laws and calculate the synchrotron emission following~\citet{Sadeh2023}.
Our simulated kilonova afterglow for a fiducial choice of afterglow parameters; interstellar medium density, $n = 5$~cm$^{-3}$, electron power law index, $p=2.1$, magnetic and electron energy fractions of $\epsilon_{e}=0.1$, and $\epsilon_{B}=0.001$, respectively, at a luminosity distance of $40$~Mpc are shown in Fig.~\ref{fig:KNafterglow}.

For both modelling assumptions we see the same trend, that the zero spin system peaks significantly earlier than the brighter spinning case. This is consistent with physical intuition, given the latter (spinning) case has more ejected material and a larger kinetic energy. This suggests that kilonova afterglows may provide a potential late-time distinguishing feature for NSs with a spinning component, something we discuss in more detail in Sec.~\ref{sec:discussion}. 
Our numerical modelling agrees with physical intuition, peaking at $\approx 600$ and $1200$ days for the non-spinning and spinning cases respectively, broadly consistent with the deceleration timescale for our chosen parameters, i.e., the timescale where the blastwave has swept up a comparable mass to its own~\citep{nakar11a}.

\subsubsection{Fallback accretion}

\begin{figure*} 
   \centerline{
   \includegraphics[width=2.\columnwidth]{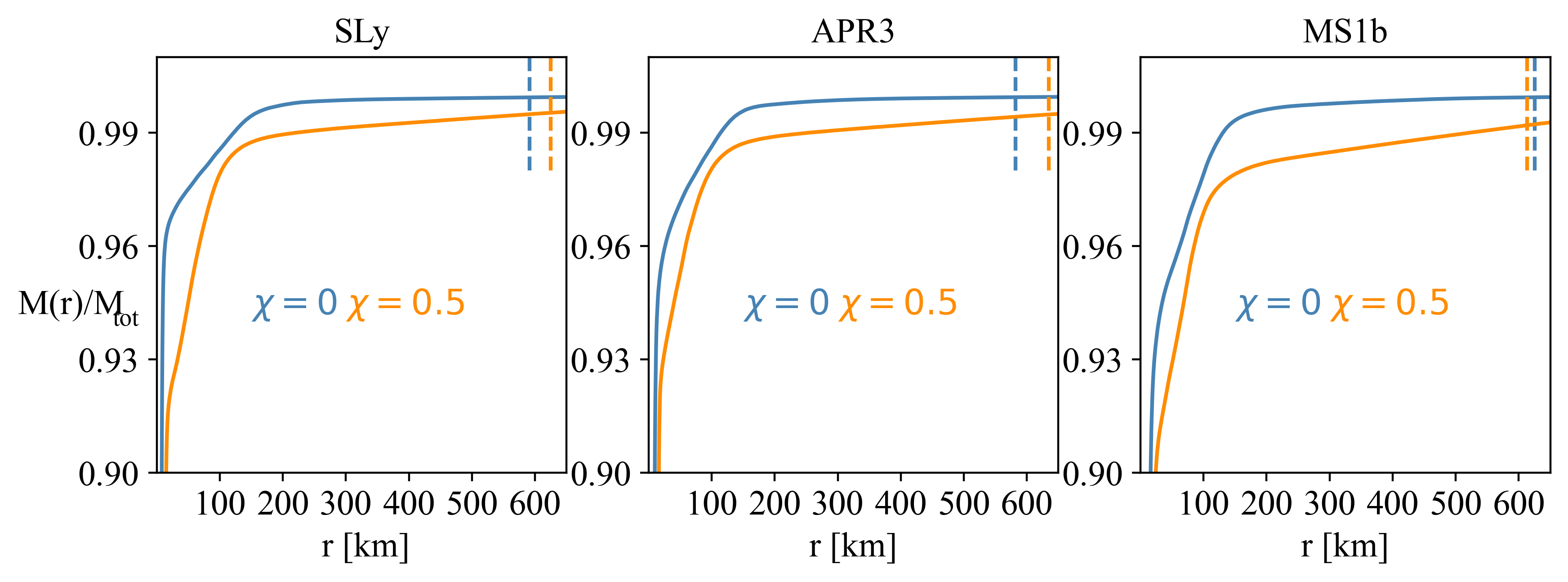} 
      }
   \caption{Baryonic mass enclosed inside a given radius ($M(r)/M_{\rm tot}$ at $t= 22.9$~ms for all considered EOSs. The dashed vertical lines indicate the approximate radii beyond which the eccentricity exceeds a value of unity, i.e. matter is unbound.}
   \label{fig:M_r}
\end{figure*}

\begin{figure*} 
   \centerline{
   \includegraphics[width=2.\columnwidth]{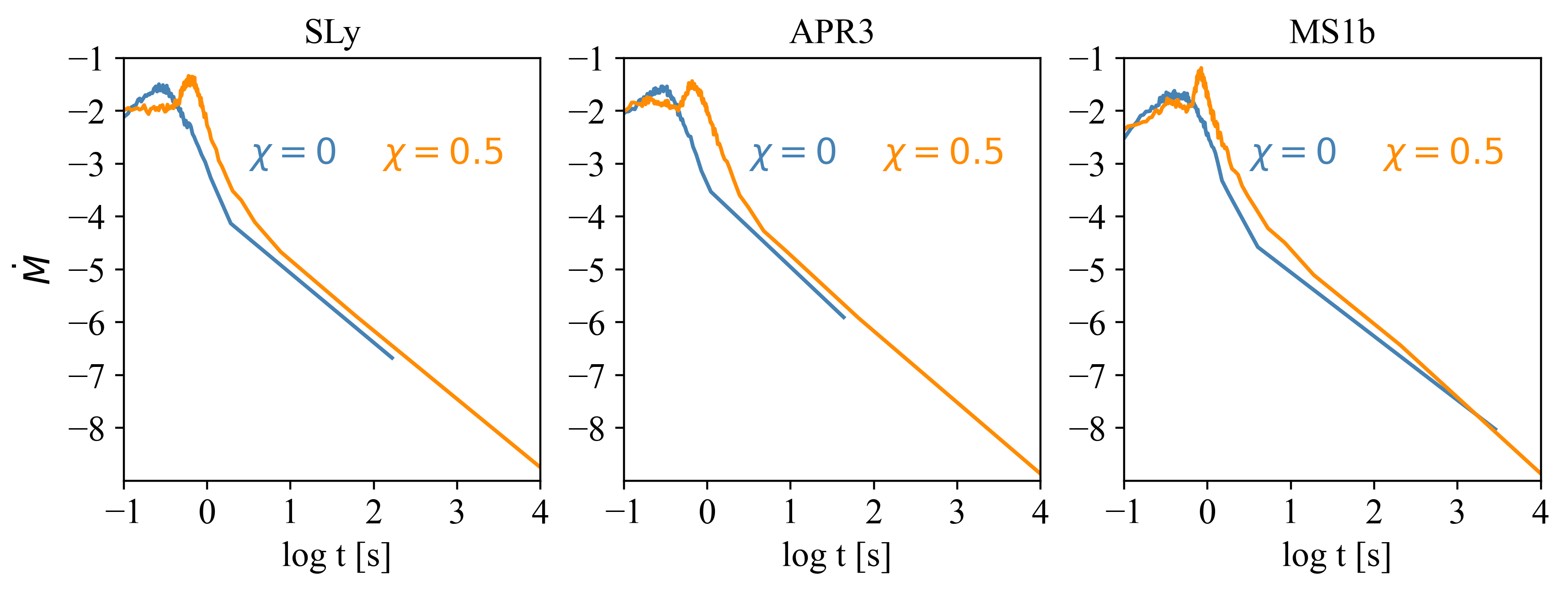}}
   \caption{Mass fallback rates (in units of \msun/s) based on the simple analytical model of Rosswog (2007) for our different simulations. }
   \label{fig:Mdot_t}
\end{figure*}

\begin{table}
	\centering
	\caption{Fallback accretion: mass estimates for fallback accretion based on either the general-relativistic Bernoulli ($m_{\rm fb,B}$) criterion and based on a Newtonian estimate of the eccentricity ($m_{\rm fb,ecc}$), both in units of $10^{-3}$~\msun. }
	\label{tab:fallback_masses}
	\begin{tabular}{lrr} 
		\hline
		run &  $m_{\rm fb,B}$  & $m_{\rm fb, ecc}$ \\
		\hline
		SLy\_irr   &  9.50  & 9.56     \\
		SLy\_sspin &  19.24 & 19.52  \\		
		APR3\_irr &  8.54  & 8.56  \\
		APR3\_sspin & 19.14 & 19,65  \\
		MS1b\_irr & 12.20 & 12.31    \\
		MS1b\_sspin & 23.58 & 23.92 \\
		\hline
	\end{tabular}
\end{table}
While the question whether fallback accretion can power some
emission on time scales substantially longer than the dynamical
time scale ($\sim$ ms) is clearly very relevant for understanding
the observed late time emission of some GRBs, it actually is very difficult
to deliver reliable estimates from full-fledged numerical relativity simulations that end at several 10~ms. We will therefore apply different
methods to ensure that our conclusions are qualitatively robust.\\
First, we aim at estimating the mass that will fall back. 
This is straight forward to calculate in a very simple picture where
some mass is launched, follows a ballistic trajectory and will later return to a radius where
its available energy is dissipated. The situation at the end 
of a numerical relativity simulation, however, is more complicated in the
sense that the mass distribution is rather continuous so that
idealized objects like ``the disk'' are not straight forward to 
identify. To illustrate the mass distributions in our runs, we 
show in Fig.~\ref{fig:M_r} the (fractional) baryonic mass that 
is included in a given radius. Not too surprisingly, the non-spinning
cases are more compact and reach a large fraction of the total 
enclosed mass, say, 99\% already at $\sim 100$~km, while this 
takes substantially larger radii ($\sim 600$~km) for the $\chi=0.5$-cases.
To estimate the fallback mass, we only consider fluid parts with
densities below a threshold of $\rho_{\rm thresh}= 10^9$~g~cm$^{-3}$. This value has been chosen after careful inspection
of all the simulations (at $t= 22.9$~ms). Details may depend on
the exact value chosen, but none of the main results depends on 
the exact value.\\
To ensure the robustness of the estimate, we use two different
criteria. First, we use the general relativistic Bernoulli criterion, see Eq.~(\ref{eq:Bernoulli}). Since the specific 
enthalpy, $\mathcal{E}= 1 + u+ P/n$, tends to unity at infinity and the zero-component
of the four-velocity tends to the negative Lorentz factor, $-\Gamma_\infty$, the condition
$\mathcal{B}\equiv -\mathcal{E} U_t > 1$ means
that a matter portion still has a finite velocity at infinity, i.e. it is 
unbound\footnote{There are some subtleties involved, for a discussion of which we refer to \cite{foucart21b}.}. If instead $\mathcal{B} < 1$, the flow portion will return and it is considered as fallback. As a second criterion, we use
the Newtonian eccentricity of a particle $a$ (e.g. \cite{shapiro83})
\be
e_a= \sqrt{1 + \frac{2 E_a J_a^2}{G^2 \mu_a^3 M}},
\ee
where $E_a$, $J_a$ and $\mu_a$ are the particle's Newtonian orbital energy, angular momentum and reduced mass 
and $M$ the enclosed mass. For this criterion, we identify as ``fallback''
matter with $\rho < \rho_{\rm thresh}$ and with $e<1$. The fallback masses found for both criteria agree very well, even in the worst case the difference is only a few percent, see Table~\ref{tab:fallback_masses}. The amount of fallback matter is typically a factor of two
larger for the spinning cases. Such fallback could release energies $E_{\rm fb} \sim 2\times 10^{51}\,{\rm erg} \left(\frac{\epsilon}{0.1}\right) \left( \frac{M_{\rm fb}}{0.01\,{M}_\odot} \right)$.\\
We also estimate the fallback luminosities based on a simple analytical model
\citep{rosswog07a}.
To this end we assume a particle follows a Keplerian
orbit from its current position to apocentre and then back to its "circularisation radius", $R_{\rm circ}$, 
see e.g. \cite{frank02}. This is the radius of a circular orbit corresponding to a specific angular momentum value and it is an estimate of the size of a
forming accretion disk. We denote the time to reach this radius as $t_{\rm fb}$, it can be calculated analytically, see \cite{rosswog07a}. After falling back to $R_{\rm circ}$, it will still take approximately a viscous accretion disk time scale
to be accreted onto the central object. This time scale is given by \citep{shakura73,frank02}
\be
\tau_{\rm visc}= \frac{1}{\alpha \; \omega_{\rm K}} \left( \frac{R_{\rm circ}}{H}\right)^2.
\ee
Here $\alpha$ is the Shakura-Sunyaev dissipation parameter, $\omega_{\rm K}$, the Keplerian angular velocity and $H$ the disk scale height. So the time from the current position to being accreted is is estimated as $t_{\rm fb} + t_{\rm visc}$.
A similar approach had been taken before in \cite{rosswog13a}. For the
plot of the mass fallback rate, Fig.~\ref{fig:Mdot_t},
we use $\alpha= 0.01$ and $H= R_{\rm circ}/3$.
The spinning cases 
provide a substantially larger fallback luminosity 
than the non-spinning cases.
They in particular show a pronounced peak in the fallback rate at $t\sim 1$ s, where they are about an order of magnitude brighter than the non-spinning cases.
 The spinning cases also show a {\em longer} fallback time scale, at least in this
simple model.

\section{Discussion}
\label{sec:discussion}

\subsection{How likely are DNS mergers with a rapidly spinning NS?}
\label{sec:discussion_fraction}
Unlike the situation for double BH binaries, the fraction of DNS binaries (and mixed BH+NS binaries) formed dynamically in globular clusters has been suggested to be quite insignificant. \citet{yfk+20} estimate the merger-rate density in the local Universe to be $\sim 0.02\;{\rm Gpc}^{-3}\,{\rm yr}^{-1}$ for both DNS and BH+NS binaries in globular clusters, or a total of $\sim 0.04\;{\rm Gpc}^{-3}\,{\rm yr}^{-1}$ for both populations. In comparison, a conservative merger-rate density estimate based on simulated Galactic field (isolated) DNS and BH+NS systems combined is between 50 and $200\;{\rm Gpc}^{-3}\,{\rm yr}^{-1}$ \citep{ktl+18}, i.e. four orders of magnitude larger. There are, however, a number of circumstances pointing to a much higher fraction of DNS mergers with a rapidly spinning NS component, including: i) the fraction of DNS mergers with a rapidly spinning NS in the Milky Way is of order 0.04 (see below); and ii) triple systems may also contribute to the formation rate of rapidly spinning NSs in DNS systems \citep{ht19}.

Evaluating the fraction of DNS mergers with a high-spin NS component is non-trivial for several reasons. Besides the observed data, one has to factor in the lifetimes of MSPs versus less, or mildly, recycled NSs that are usually observed in DNS systems, beaming morphology and radio survey statistics \citep{lk12}. Note also that dynamical exchange encounter events have a dependence on e.g. orbital separation and cluster age. Thus, a detailed analysis is much beyond the scope of this paper.  
We will, nevertheless, try here to provide a rough estimate based on a few simple arguments. Four out of the total of 23~DNS systems discovered so far\footnote{{See ATNF Pulsar Catalogue \citep{mhth05} version 1.71: \url{https://www.atnf.csiro.au/research/pulsar/psrcat/}. We disregarded PSR J1755$-$2550 as a DNS candidate; nor did we include PSR~J0514$-$4002E which may have a BH companion \citep{bdf+24}.}} in our Galaxy are located within a globular cluster. Three of these systems (PSR~J0514$-$4002A, PSR~J0514$-$4002E and PSR~J1807-2459B), corresponding to $\sim 13\%$ of the full DNS sample, host a MSP, whereas the fourth system (B2127+11C), hosting a mildly recycled pulsar spinning at 30~ms, is the only one of those four systems that merges within a Hubble time (it will merge in 217~Myr). Fully recycled MSPs (required for a high-spin NS component investigated here) have, on average, weaker B-fields and are therefore known to have longer spin-down timescales than the mildly recycled NSs that are usually detected in DNS systems. Hence, to correct for this difference in detectability, the current set of empirical data of DNS systems containing an MSP has to be corrected by a smaller production rate (and hence smaller merger rate). 
From the current sample of DNS systems in the Galactic field, 25\% of the mildly recycled NSs in DNS systems have spin-down timescales of $\tau > 10\;{\rm Gyr}$, whereas 30\% of the 99 MSPs with spin periods, $P<6\;{\rm ms}$ in the Galactic field\footnote{{Measurements of intrinsic $\dot{P}$ values (and thus $\tau\equiv P/2\dot{P}$) of MSPs in globular clusters are unreliable due to acceleration in the cluster potential.}} have $\tau > 10\;{\rm Gyr}$. Thus, the difference in lifetimes as detectable radio pulsars is, perhaps, not as pronounced as one would immediately guess. However, the average and median values of $\tau$ for the 16 DNS systems in the Galactic field, in which a mildly recycled NS is detected, are 4.4~Gyr and 1.0~Gyr, respectively. For the 99 fully recycled MSPs, the values are 9.4~Gyr and 6.0~Gyr, respectively. These values reduce the fraction of DNS systems containing an MSP to 6.1\% and 2.2\%, respectively.

Among the full population of Galactic binary MSP systems, there is no correlation between the spin of the recycled pulsar and the orbital period up to about 200~days \citep{tlk12}. Assuming the same holds for globular cluster DNS systems, i.e. that the formation of DNS systems that will merge is independent of spin period, then, based on the above small number statistics, we may roughly expect a fraction of $4\pm2\%$ of all DNS mergers to contain an MSP.
We emphasize that a more detailed estimate must include e.g. survey selection effects etc. \citep{kkl03,pml+19}.
The construction of the full Square Kilometer Array is anticipated to increase the number of known Galactic DNS systems by almost an order of magnitude \citep{kbk+15}. Thus in the coming decade, empirical measures will constrain much better the ratio of DNS systems with one fast-spinning NS component.

\subsection{Nucleosynthesis}
The decompression of NS matter is very appealing as an r-process scenario,
since the extreme neutron-richness inside the original stars allows for an effortless
and robust production of heavy r-process ($A >130$) up to and  beyond the
``platinum peak'' at $A=195$ \citep{lattimer77,rosswog98b,rosswog99, freiburghaus99b,korobkin12a}. In the
last decade, however, increasingly more ejection channels were identified where the electron fraction becomes substantially increased compared to the pristine value inside
the original NSs, see e.g. \cite{wanajo14,perego14b,just15,martin15,wu16,siegel17a,siegel18,miller19, fujibayashi20a,fujibayashi20b}. The raised electron fractions can lead to a broad range of  r-process
elements and this is consistent
with the likely identification of strontium in the kilonova of the first detected NS
merger GW170817 \citep{watson19}. Extremely long-lived central remnants  may actually lead to such large electron fractions that the heavy r-process is underproduced.\\
DNS mergers containing a recycled, milli-second pulsar, however, are strongly
dominated by tidal ejecta at the original electron fraction and therefore major sources
of lanthanides and actinides, a property that they may share with unequal-mass DNS mergers. Since the central remnants are longer-lived due to the less efficient GW-emission, their neutrino emission and absorption by the surrounding debris
may lead to an additional, lighter r-process component. Some metal-poor (and therefore old) stars
show supersolar ratios of Th or U to Eu
\citep{roederer10,holmbeck19} and one may
wonder whether a DNS merger containing a ms-pulsar may be a candidate
for producing the heavy r-process early on in the Galactic history. The binary evolution leading to
the formation of such a binary, however, will take a long time with the bottleneck being the nuclear evolution timescale of the low-mass star that evolves to become the donor star in the LMXB system, see Fig.~\ref{fig:cartoon} in which the NS is recycled to ms spin period. Since the nuclear evolution time scale is at least 1.3~Gyr, we do not consider such mergers as good candidates for the very early enrichment of the Galaxy
with r-process elements.

\subsection{Electromagnetic emission}
The presence of a rapidly spinning NS can drastically 
increase the dynamic ejecta, in the studied cases typically by one order 
of magnitude. Although we do not model weak interactions in 
the current study, we expect that the vast majority of the 
dynamic ejecta will be very low in $Y_e$, since it never was shock-heated, and will thus produce 
a bright red kilonova. This is because a) the initial rotational 
flattening leads to large tidal ejecta amounts, b) the less violent
collision disfavours shock-heated ejecta and c) it also should lead, 
at least initially, to lower neutrino luminosities disfavoring 
neutrino captures (and therefore $Y_e$-changes) by the already escaping, far away 
ejecta matter. Also the secular ejecta are increased, but to a lower degree: we find a factor of $\sim 2$ for the softest EOS (SLy) and smaller values for the stiffer EOSs. This ejecta component predominently determines the kilonova properties so that the dynamic ejecta alone will be difficult to infer.\\
The less violent compression and therefore weaker re-bounce
in the spinning case also results in a smaller amount of high velocity ejecta. While
the latter are only a small fraction of the overall ejecta that is not well enough
resolved to draw firm, quantitative conclusions, it seems safe to state that, if
indeed they produce a blue transient $\sim1$~hour post-merger \citep{metzger15a},
this transient should be substantially weaker for the spinning
cases.
 The combination of a weak, blue precursor
together with a particularly bright red kilonova may thus be the 
tell-tale signature for DNS mergers that contain a fast-spinning NS.\\
The high quantity of the ejecta itself could also be a smoking-gun signature of a DNS merger with a fast-spinning NS. However, it is unclear whether a high-spin origin can be robustly estimated from observations, especially in light of other origins of large ejecta amounts such as unequal mass ratios, although these could in principle be verified through gravitational-wave data in the event of a multi-messenger discovery.\\
The kilonova afterglow could also be used to discern whether a merger had at least one spinning NS. We find that mergers with a spinning NS component will produce a brighter kilonova afterglow that peaks at later times, consistent with physical intuition, given the larger mass and kinetic energies. Whether this difference in brightness/peak timescale can be robustly verified given typical uncertainties in afterglow parameters is an open question.
\\
While modelling uncertainties in kilonovae and kilonova afterglows may prevent distinguishing a merger with a spinning component from a merger with two non-spinning components, the combination of a weak blue precursor, but brighter kilonova and kilonova afterglow may provide strong constraints on the spin of the NS. For multi-messenger events, such an independent constraint on the spin of the NS through electromagnetic emission, will help to break the degeneracy between component masses and spin and help provide stronger inferences about the mass of merging NSs. \\
We also find that mergers with a spinning component have longer period of fall-back activity. This may provide a more natural solution to explain the duration of a new class of long GRBs that appear to originate from mergers~\citep{rastinejad22, Levan2023} or power late-time X-ray flares such as the low significance flare from GW170817~\citep{Piro2019}.

\section{Summary}
\label{sec:summary}
In this paper, we have studied a scenario 
that by and large has 
probably not been taken seriously enough: the merger of two NSs where one of the components has a spin close to zero while
the other is a spun-up MSP. Such binary systems form
regularly in regions of large stellar density such as globular clusters and,
based on observed systems, we estimate
that such binaries may constitute a non-negligible fraction  ($\sim$ 4\%) of the merging NS binaries,
see Sec.~\ref{sec:discussion_fraction}.
\\
The merger of such NS binaries has a number of distinctive
signatures, that we briefly summarize below, for more details,
we refer to the corresponding sections of this paper. 
\bi
\i Morphologically, the spinning (equal mass) cases ressemble mergers with a mass ratio that is significantly different from unity, see Figs.~\ref{fig:dens_evol} and \ref{fig:volren_APR3}. This is because
the spinning stars are substantially more extended due to the 
rotational flattening, therefore they are more vulnerable to 
tidal forces and thus produce a single massive tidal tail.
\i This implies that the dynamic ejecta are enhanced by an order of magnitude, see Fig.~\ref{fig:mesc_t}, and consist
predominantly of low $Y_e$-material at the original NS electron fraction. Also the secular ejecta are enhanced, depending on the EOS, by up to a factor of 2, see Table~\ref{tab:ejecta_masses}. Spinning cases could --- depending on the currently not well determined secular ejecta composition --- potentially eject vary large amounts of lanthanide-/actinide-rich matter. They are, however, not good candidates for the source of "actinide boost stars", since the latter must have formed early in the galactic history, while the formation time of the spinning binary NS mergers are determined by the long
nuclear evolution timescale of the low-mass star that evolves to 
become the donor star in the LMXB system.
\i The resulting kilonovae are substantially brighter and peak later than in the non-spinning cases, see Sec.~\ref{sec:kilonova} and Figs.~\ref{fig:KN_bolometric_dynamic}, \ref{fig:KN_bolometric_all} and \ref{fig:knmagnitude}. Not too surprisingly, the overall kilonova appearance 
is determined by the secular ejecta properties, 
since they constitute the largest ejecta fraction.
\i Due to the larger tides of the rotationally flattened spinning stars, the collisions themselves are less violent, lead to smaller compressions of the remnants and to substantially reduced gravitational wave emission, see Table~\ref{tab:runs} and 
Fig.~\ref{fig:GWs_dEdJ}.
\i For the softer EOSs (SLy and APR3) the polar ejecta are
about 10\% faster than the equatorial ejecta, while for the
very stiff MS1b EOS the polar ejecta are actually slower,
probably because it is difficult to have shocks with such a stiff equation of state. These statements apply to both the spinning and non-spinning cases.
\i The smaller compression in the spinning cases also goes along with
smaller amounts of fast ejecta. Therefore, spinning cases produce weaker blue kilonova precursor emission.
However, since the ejecta carry more mass and kinetic energy, the kilonova afterglow is brighter and peaks substantially later than in the corresponding irrotational cases, see Fig.~\ref{fig:KNafterglow}.
\i Since more matter is launched into non-circular orbits, the fallback accretion is substantially brighter and is expected to last longer, see
Fig.~\ref{fig:Mdot_t}. Whether such systems could be behind GRBs that appear to have a bright kilonova, but are otherwise uncomfortably long for a NS merger \citep{rastinejad22,Levan2023}, given their shorter accretion timescales~\citep{narayan92}, needs to be explored in future studies.
\i As a corollary of the less efficient GW-emission, see Fig.~\ref{fig:GWs_dEdJ}, the merger remnant contains larger amounts of angular momentum, should therefore be longer-lived and could potentially power a GRB for a longer time scale. Since the collision is less violent, neutrino driven winds (at least initially) may be weaker, therefore providing an easier path to the launch of ultra-relativistic outflows.
\i Last, but not least, given that the rate of such mergers is non-negligible, caution should be applied in interpretation of GW-data and BNS merger populations where often the prior assumption of negligible spins is considered as the most likely case. 
\ei

\section*{Acknowledgements}
It is a great pleasure to acknowledge interesting discussions with Sam Tootle 
and we are very grateful for his generous help with the FUKA library.
SR has been supported by the Swedish Research Council (VR) under 
grant number 2020-05044, by the research environment grant
``Gravitational Radiation and Electromagnetic Astrophysical
Transients'' (GREAT) funded by the Swedish Research Council (VR) 
under Dnr 2016-06012, by the Knut and Alice Wallenberg Foundation
under grant Dnr. KAW 2019.0112,   by the Deutsche 
Forschungsgemeinschaft (DFG, German Research Foundation) under 
Germany's Excellence Strategy - EXC 2121 ``Quantum Universe'' - 390833306 
and by the European Research Council (ERC) Advanced 
Grant INSPIRATION under the European Union's Horizon 2020 research 
and innovation programme (Grant agreement No. 101053985). FT was supported
the "GREAT" research environment grant of VR.
NS is supported by a Nordita Fellowship. Nordita is funded in part by NordForsk.\\
The simulations for this paper have been performed on the facilities of
 North-German Supercomputing Alliance (HLRN), and at the SUNRISE 
 HPC facility supported by the Technical Division at the Department of 
 Physics, Stockholm University. Special thanks go to Holger Motzkau 
 and Mikica Kocic for their excellent support in upgrading and maintaining 
 SUNRISE. 
 
\section*{Data availability}
The data underlying this article will be shared on reasonable request to the corresponding author.

\appendix

\bibliographystyle{mn2e}
\bibliography{astro_SKR.bib,further_refs.bib}

\hyphenation{Post-Script Sprin-ger}
\begin{thebibliography}{}

\bibitem[\protect\citeauthoryear{{Abbott}, {Abbott}, {Abbott}, {Abernathy},
  {Acernese}, {Ackley}, {Adams}, {Adams}, {Addesso}, {Adhikari} \& et
  al.}{{Abbott} et~al.}{2016}]{abbott16a}
{Abbott} B.~P.,  {Abbott} R.,  {Abbott} T.~D.,  {Abernathy} M.~R.,  {Acernese}
  F.,  {Ackley} K.,  {Adams} C.,  {Adams} T.,  {Addesso} P.,  {Adhikari} R.~X.,
     et al. 2016, Physical Review Letters, 116, 061102

\bibitem[\protect\citeauthoryear{{Abbott}, {Abbott}, {Abbott}, {Abernathy},
  {Ackley} et~al.,}{{Abbott} et~al.}{2017}]{CosmicExplorer}
{Abbott} B.~P.,  {Abbott} R.,  {Abbott} T.~D.,  {Abernathy} M.~R.,  {Ackley}
  K.,    et~al., 2017, Classical and Quantum Gravity, 34, 044001

\bibitem[\protect\citeauthoryear{{Abbott}, {Abbott}, {Abbott}, {Acernese},
  {Ackley}, {Adams}, {Adams}, {Addesso}, {Adhikari}, {Adya} \& et al.}{{Abbott}
  et~al.}{2017a}]{abbott17a}
{Abbott} B.~P.,  {Abbott} R.,  {Abbott} T.~D.,  {Acernese} F.,  {Ackley} K.,
  {Adams} C.,  {Adams} T.,  {Addesso} P.,  {Adhikari} R.~X.,  {Adya} V.~B.,
  et al. 2017a, Nature, 551, 85

\bibitem[\protect\citeauthoryear{{Abbott}, {Abbott}, {Abbott}, {Acernese},
  {Ackley}, {Adams}, {Adams}, {Addesso}, {Adhikari}, {Adya} \& et al.}{{Abbott}
  et~al.}{2017b}]{abbott17b}
{Abbott} B.~P.,  {Abbott} R.,  {Abbott} T.~D.,  {Acernese} F.,  {Ackley} K.,
  {Adams} C.,  {Adams} T.,  {Addesso} P.,  {Adhikari} R.~X.,  {Adya} V.~B.,
  et al. 2017b, Physical Review Letters, 119, 161101

\bibitem[\protect\citeauthoryear{{Abbott}, {Abbott}, {Abbott}, {Acernese},
  {Ackley}, {Adams}, {Adams}, {Addesso}, {Adhikari}, {Adya} \& et al.}{{Abbott}
  et~al.}{2017c}]{abbott17c}
{Abbott} B.~P.,  {Abbott} R.,  {Abbott} T.~D.,  {Acernese} F.,  {Ackley} K.,
  {Adams} C.,  {Adams} T.,  {Addesso} P.,  {Adhikari} R.~X.,  {Adya} V.~B.,
  et al. 2017c, ApJL, 848, L12

\bibitem[\protect\citeauthoryear{{Abbott}, {Abbott}, {Abraham}, {Acernese},
  {Ackley}, {Adams}, {Adams}, {Adhikari}, {Adya}, {Affeldt}, {Agarwal},
  {Agathos}, {Agatsuma} \& many more}{{Abbott} et~al.}{2021}]{abbott21a}
{Abbott} R.,  {Abbott} T.~D.,  {Abraham} S.,  {Acernese} F.,  {Ackley} K.,
  {Adams} A.,  {Adams} C.,  {Adhikari} R.~X.,  {Adya} V.~B.,  {Affeldt} C.,
  {Agarwal} D.,  {Agathos} M.,  {Agatsuma} K.,    many more 2021, ApJL, 915, L5

\bibitem[\protect\citeauthoryear{{Abbott}, {Abbott}, {Abraham}, {Acernese},
  {Ackley}, {Adams}, {Adhikari}, {Adya}, {Affeldt}, {Agathos}, {Agatsuma},
  {Aggarwal} \& more}{{Abbott} et~al.}{2020}]{GW190814}
{Abbott} R.,  {Abbott} T.~D.,  {Abraham} S.,  {Acernese} F.,  {Ackley} K.,
  {Adams} C.,  {Adhikari} R.~X.,  {Adya} V.~B.,  {Affeldt} C.,  {Agathos} M.,
  {Agatsuma} K.,  {Aggarwal} N.,    more 2020, ApJL, 896, L44

\bibitem[\protect\citeauthoryear{{Abbott}, {Abbott}, {Acernese}, {Ackley},
  {Adams} \& more}{{Abbott} et~al.}{2023}]{GWTC-3}
{Abbott} R.,  {Abbott} T.~D.,  {Acernese} F.,  {Ackley} K.,  {Adams} C.,
  more 2023, Physical Review X, 13, 041039

\bibitem[\protect\citeauthoryear{{Ackley}, {Adya}, {Agrawal}, {Altin},
  {Ashton}, {Bailes}, {Baltinas}, {Barbuio} et~al.,}{{Ackley}
  et~al.}{2020}]{NEMO}
{Ackley} K.,  {Adya} V.~B.,  {Agrawal} P.,  {Altin} P.,  {Ashton} G.,  {Bailes}
  M.,  {Baltinas} E.,  {Barbuio} A.,    et~al., 2020, Publications of the Astronomical Society of Australia, 37, e047

\bibitem[\protect\citeauthoryear{{Akmal}, {Pandharipande} \&
  {Ravenhall}}{{Akmal} et~al.}{1998}]{akmal98}
{Akmal} A.,  {Pandharipande} V.~R.,    {Ravenhall} D.~G.,  1998, Phys. Rev. C,
  58, 1804

\bibitem[\protect\citeauthoryear{{Alcubierre}}{{Alcubierre}}{2008}]{alcubierre08}
{Alcubierre} M.,  2008, {Introduction to 3+1 Numerical Relativity}.
Oxford University Press

\bibitem[\protect\citeauthoryear{Alcubierre, Bruegmann, Diener, Koppitz,
  Pollney, Seidel \& Takahashi}{Alcubierre et~al.}{2003}]{alcubierre02}
Alcubierre M.,  Bruegmann B.,  Diener P.,  Koppitz M.,  Pollney D.,  Seidel E.,
     Takahashi R.,  2003, Phys. Rev. D, 67, 084023

\bibitem[\protect\citeauthoryear{Alpar, Cheng, Ruderman \& Shaham}{Alpar
  et~al.}{1982}]{acrs82}
Alpar M.~A.,  Cheng A.~F.,  Ruderman M.~A.,    Shaham J.,  1982, \nat, 300, 728

\bibitem[\protect\citeauthoryear{{Antoniadis}, {Freire}, {Wex}, {Tauris},
  {Verbiest} \& {Whelan}}{{Antoniadis} et~al.}{2013}]{antoniadis13}
{Antoniadis} J.,  {Freire} P.~C.~C.,  {Wex} N.,  {Tauris} T.~M.,  {Verbiest}
  J.~P.~W.,    {Whelan} D.~G.,  2013, Science, 340, 448

\bibitem[\protect\citeauthoryear{{Arnett}}{{Arnett}}{1980}]{arnett80}
{Arnett} W.~D.,  1980, ApJ, 237, 541

\bibitem[\protect\citeauthoryear{{Arnett}}{{Arnett}}{1982}]{arnett82}
{Arnett} W.~D.,  1982, ApJ, 253, 785

\bibitem[\protect\citeauthoryear{{Barnes}, {Kasen}, {Wu} \&
  {Martinez-Pinedo}}{{Barnes} et~al.}{2016}]{barnes16a}
{Barnes} J.,  {Kasen} D.,  {Wu} M.-R.,    {Martinez-Pinedo} G.,  2016, ApJ,
  829, 110

\bibitem[\protect\citeauthoryear{{Barr}, {Dutta}, {Freire}, {Cadelano},
  {Gautam}, {Kramer}, {Pallanca}, {Ransom}, {Ridolfi}, {Stappers}, {Tauris} \&
  {et~al.}}{{Barr} et~al.}{2024}]{bdf+24}
{Barr} E.~D.,  {Dutta} A.,  {Freire} P. C.~C.,  {Cadelano} M.,  {Gautam} T.,
  {Kramer} M.,  {Pallanca} C.,  {Ransom} S.~M.,  {Ridolfi} A.,  {Stappers}
  B.~W.,  {Tauris} T.~M.,    {et~al.} 2024, arXiv e-prints, p. arXiv:2401.09872

\bibitem[\protect\citeauthoryear{{Bartos}, {Rosswog}, {Gayathri}, {Miller},
  {Veske} \& {Marka}}{{Bartos} et~al.}{2023a}]{bartos23}
{Bartos} I.,  {Rosswog} S.,  {Gayathri} V.,  {Miller} M.~C.,  {Veske} D.,
  {Marka} S.,  2023a, arXiv e-prints, p. arXiv:2302.10350

\bibitem[\protect\citeauthoryear{{Bartos}, {Rosswog}, {Gayathri}, {Miller},
  {Veske} \& {Marka}}{{Bartos} et~al.}{2023b}]{brg+23}
{Bartos} I.,  {Rosswog} S.,  {Gayathri} V.,  {Miller} M.~C.,  {Veske} D.,
  {Marka} S.,  2023b, arXiv e-prints, p. arXiv:2302.10350

\bibitem[\protect\citeauthoryear{{Baumgarte} \& {Shapiro}}{{Baumgarte} \&
  {Shapiro}}{1999}]{baumgarte99}
{Baumgarte} T.~W.,  {Shapiro} S.~L.,  1999, Phys. Rev. D, 59, 024007

\bibitem[\protect\citeauthoryear{{Baumgarte} \& {Shapiro}}{{Baumgarte} \&
  {Shapiro}}{2010}]{baumgarte10}
{Baumgarte} T.~W.,  {Shapiro} S.~L.,  2010, {Numerical Relativity: Solving
  Einstein's Equations on the Computer}.
Cambridge University Press, Cambridge

\bibitem[\protect\citeauthoryear{{Bauswein}, {Bastian}, {Blaschke},
  {Chatziioannou} \& et al.}{{Bauswein} et~al.}{2019}]{Bauswein2019}
{Bauswein} A.,  {Bastian} N.-U.~F.,  {Blaschke} D.~B.,  {Chatziioannou} K.,
  et al. 2019, Phys. Rev. Lett., 122, 061102

\bibitem[\protect\citeauthoryear{{Bauswein}, {Just}, {Janka} \&
  {Stergioulas}}{{Bauswein} et~al.}{2017}]{bauswein17}
{Bauswein} A.,  {Just} O.,  {Janka} H.-T.,    {Stergioulas} N.,  2017, ApJL,
  850, L34

\bibitem[\protect\citeauthoryear{{Bernuzzi}, {Dietrich}, {Tichy} \&
  {Br{\"u}gmann}}{{Bernuzzi} et~al.}{2014}]{bernuzzi14a}
{Bernuzzi} S.,  {Dietrich} T.,  {Tichy} W.,    {Br{\"u}gmann} B.,  2014, Phys.
  Rev. D, 89, 104021

\bibitem[\protect\citeauthoryear{{Bildsten} \& {Cutler}}{{Bildsten} \&
  {Cutler}}{1992}]{bildsten92}
{Bildsten} L.,  {Cutler} C.,  1992, ApJ, 400, 175

\bibitem[\protect\citeauthoryear{Biswas}{Biswas}{2022}]{Biswas:2021pvm}
Biswas B.,  2022, Astrophys. J., 926, 75

\bibitem[\protect\citeauthoryear{{Biswas} \& {Datta}}{{Biswas} \&
  {Datta}}{2021}]{biswas22}
{Biswas} B.,  {Datta} S.,  2021, arXiv e-prints, p. arXiv:2112.10824

\bibitem[\protect\citeauthoryear{{Bozzola}}{{Bozzola}}{2021}]{kuibit}
{Bozzola} G.,  2021, The Journal of Open Source Software, 6, 3099

\bibitem[\protect\citeauthoryear{{Cabezon}, {Garcia-Senz} \&
  {Relano}}{{Cabezon} et~al.}{2008}]{cabezon08}
{Cabezon} R.~M.,  {Garcia-Senz} D.,    {Relano} A.,  2008, Journal of
  Computational Physics, 227, 8523

\bibitem[\protect\citeauthoryear{{Cowperthwaite}, {Berger}, {Villar} \&
  {Metzger}}{{Cowperthwaite} et~al.}{2017}]{cowperthwaite17}
{Cowperthwaite} P.~S.,  {Berger} E.,  {Villar} V.~A.,    {Metzger} B.~D.,
  2017, ApJL, 848, L17

\bibitem[\protect\citeauthoryear{{Cromartie}, {Fonseca}, {Ransom}, {Demorest},
  {Arzoumanian}, {Blumer}, {Brook}, {DeCesar}, {Dolch}, {Ellis}, {Ferdman},
  {Ferrara}, {Garver-Daniels}, {Gentile}, {Jones}, {Lam}, {Lorimer}, {Lynch},
  {McLaughlin}, {Ng}, {Nice}, {Pennucci}, {Spiewak}, {Stairs}, {Stovall},
  {Swiggum} \& {Zhu}}{{Cromartie} et~al.}{2020}]{cromartie20}
{Cromartie} H.~T.,  {Fonseca} E.,  {Ransom} S.~M.,  {Demorest} P.~B.,
  {Arzoumanian} Z.,  {Blumer} H.,  {Brook} P.~R.,  {DeCesar} M.~E.,  {Dolch}
  T.,  {Ellis} J.~A.,  {Ferdman} R.~D.,  {Ferrara} E.~C.,  {Garver-Daniels} N.,
   {Gentile} P.~A.,  {Jones} M.~L.,  {Lam} M.~T.,  {Lorimer} D.~R.,  {Lynch}
  R.~S.,  {McLaughlin} M.~A.,  {Ng} C.,  {Nice} D.~J.,  {Pennucci} T.~T.,
  {Spiewak} R.,  {Stairs} I.~H.,  {Stovall} K.,  {Swiggum} J.~K.,    {Zhu}
  W.~W.,  2020, Nature Astronomy, 4, 72

\bibitem[\protect\citeauthoryear{{Diener}, {Rosswog} \& {Torsello}}{{Diener}
  et~al.}{2022}]{diener22a}
{Diener} P.,  {Rosswog} S.,    {Torsello} F.,  2022, European Physical Journal
  A, 58, 74

\bibitem[\protect\citeauthoryear{{Dietrich}, {Bernuzzi}, {Ujevic} \&
  {Tichy}}{{Dietrich} et~al.}{2017}]{dietrich17b}
{Dietrich} T.,  {Bernuzzi} S.,  {Ujevic} M.,    {Tichy} W.,  2017, Phys. Rev.
  D, 95, 044045

\bibitem[\protect\citeauthoryear{{Dietrich} \& {Ujevic}}{{Dietrich} \&
  {Ujevic}}{2017}]{dietrich17}
{Dietrich} T.,  {Ujevic} M.,  2017, Classical and Quantum Gravity, 34, 105014

\bibitem[\protect\citeauthoryear{{Douchin} \& {Haensel}}{{Douchin} \&
  {Haensel}}{2001}]{SLY_eos}
{Douchin} F.,  {Haensel} P.,  2001, A \& A, 380, 151

\bibitem[\protect\citeauthoryear{{Dudi}, {Dietrich}, {Rashti}, {Br{\"u}gmann},
  {Steinhoff} \& {Tichy}}{{Dudi} et~al.}{2022}]{dudi22}
{Dudi} R.,  {Dietrich} T.,  {Rashti} A.,  {Br{\"u}gmann} B.,  {Steinhoff} J.,
   {Tichy} W.,  2022, Phys. Rev. D, 105, 064050

\bibitem[\protect\citeauthoryear{{East}, {Paschalidis}, {Pretorius} \&
  {Shapiro}}{{East} et~al.}{2016}]{east16}
{East} W.~E.,  {Paschalidis} V.,  {Pretorius} F.,    {Shapiro} S.~L.,  2016,
  Phys. Rev. D, 93, 024011

\bibitem[\protect\citeauthoryear{{East}, {Paschalidis}, {Pretorius} \&
  {Tsokaros}}{{East} et~al.}{2019}]{east19}
{East} W.~E.,  {Paschalidis} V.,  {Pretorius} F.,    {Tsokaros} A.,  2019,
  Phys. Rev. D, 100, 124042

\bibitem[\protect\citeauthoryear{Eichler, Livio, Piran \& Schramm}{Eichler
  et~al.}{1989}]{eichler89}
Eichler D.,  Livio M.,  Piran T.,    Schramm D.~N.,  1989, Nature, 340, 126

\bibitem[\protect\citeauthoryear{{Einstein}}{{Einstein}}{1916}]{einstein16}
{Einstein} A.,  1916, Sitzungsberichte der K{\"o}niglich Preu{\ss}ischen
  Akademie der Wissenschaften (Berlin), Seite 688-696., pp 688--696

\bibitem[\protect\citeauthoryear{{Einstein Toolkit web page}}{{Einstein Toolkit
  web page}}{2020}]{ETK:web}
{Einstein Toolkit web page}, 2020, https://einsteintoolkit.org/

\bibitem[\protect\citeauthoryear{{Farouqi}, {Thielemann}, {Rosswog} \&
  {Kratz}}{{Farouqi} et~al.}{2021}]{farouqi22}
{Farouqi} K.,  {Thielemann} F.-K.,  {Rosswog} S.,    {Kratz} K.-L.,  2021,
  arXiv e-prints, p. arXiv:2107.03486

\bibitem[\protect\citeauthoryear{{Fernandez}, {Tchekhovskoy}, {Quataert},
  {Foucart} \& {Kasen}}{{Fernandez} et~al.}{2019}]{fernandez19}
{Fernandez} R.,  {Tchekhovskoy} A.,  {Quataert} E.,  {Foucart} F.,    {Kasen}
  D.,  2019, MNRAS, 482, 3373

\bibitem[\protect\citeauthoryear{{Fonseca}, {Cromartie}, {Pennucci}, {Ray},
  {Kirichenko}, {Ransom}, {Demorest}, {Stairs} \& more}{{Fonseca}
  et~al.}{2021}]{fonseca21}
{Fonseca} E.,  {Cromartie} H.~T.,  {Pennucci} T.~T.,  {Ray} P.~S.,
  {Kirichenko} A.~Y.,  {Ransom} S.~M.,  {Demorest} P.~B.,  {Stairs} I.~H.,
  more 2021, ApJL, 915, L12

\bibitem[\protect\citeauthoryear{{Foucart}, {M{\"o}sta}, {Ramirez}, {Wright},
  {Darbha} \& {Kasen}}{{Foucart} et~al.}{2021}]{foucart21b}
{Foucart} F.,  {M{\"o}sta} P.,  {Ramirez} T.,  {Wright} A.~J.,  {Darbha} S.,
  {Kasen} D.,  2021, \prd, 104, 123010

\bibitem[\protect\citeauthoryear{{Frank}, {King} \& {Raine}}{{Frank}
  et~al.}{2002}]{frank02}
{Frank} J.,  {King} A.,    {Raine} D.~J.,  2002, {Accretion Power in
  Astrophysics: Third Edition}.
Accretion Power in Astrophysics, by Juhan Frank and Andrew King and Derek
  Raine, pp.~398.~ISBN 0521620538.~Cambridge, UK: Cambridge University Press,
  February 2002.

\bibitem[\protect\citeauthoryear{{Frankfurt University/Kadath Initial Data
  solver}}{{Frankfurt University/Kadath Initial Data solver}}{2023}]{kadath}
{Frankfurt University/Kadath Initial Data solver}, 2023,
  https://kadath.obspm.fr/

\bibitem[\protect\citeauthoryear{Freiburghaus, Rosswog \&
  Thielemann}{Freiburghaus et~al.}{1999}]{freiburghaus99b}
Freiburghaus C.,  Rosswog S.,    Thielemann F.-K.,  1999, ApJ, 525, L121

\bibitem[\protect\citeauthoryear{{Fryer}, {Belczynski}, {Ramirez-Ruiz},
  {Rosswog}, {Shen} \& {Steiner}}{{Fryer} et~al.}{2015}]{fryer15}
{Fryer} C.~L.,  {Belczynski} K.,  {Ramirez-Ruiz} E.,  {Rosswog} S.,  {Shen} G.,
     {Steiner} A.~W.,  2015, ApJ, 812, 24

\bibitem[\protect\citeauthoryear{{Fujibayashi}, {Shibata}, {Wanajo}, {Kiuchi},
  {Kyutoku} \& {Sekiguchi}}{{Fujibayashi} et~al.}{2020a}]{fujibayashi20a}
{Fujibayashi} S.,  {Shibata} M.,  {Wanajo} S.,  {Kiuchi} K.,  {Kyutoku} K.,
  {Sekiguchi} Y.,  2020a, Phys. Rev. D, 101, 083029

\bibitem[\protect\citeauthoryear{{Fujibayashi}, {Shibata}, {Wanajo}, {Kiuchi},
  {Kyutoku} \& {Sekiguchi}}{{Fujibayashi} et~al.}{2020b}]{fujibayashi20b}
{Fujibayashi} S.,  {Shibata} M.,  {Wanajo} S.,  {Kiuchi} K.,  {Kyutoku} K.,
  {Sekiguchi} Y.,  2020b, Phys. Rev. D, 102, 123014

\bibitem[\protect\citeauthoryear{{Godzieba}, {Radice} \& {Bernuzzi}}{{Godzieba}
  et~al.}{2021}]{grb21}
{Godzieba} D.~A.,  {Radice} D.,    {Bernuzzi} S.,  2021, \apj, 908, 122

\bibitem[\protect\citeauthoryear{{Goldstein}, {Veres}, {Burns}, {Briggs},
  {Hamburg}, {Kocevski}, {Wilson-Hodge}, {Preece}, {Poolakkil}, {Roberts} \&
  many more}{{Goldstein} et~al.}{2017}]{goldstein17}
{Goldstein} A.,  {Veres} P.,  {Burns} E.,  {Briggs} M.~S.,  {Hamburg} R.,
  {Kocevski} D.,  {Wilson-Hodge} C.~A.,  {Preece} R.~D.,  {Poolakkil} S.,
  {Roberts} O.~J.,    many more 2017, ApJL, 848, L14

\bibitem[\protect\citeauthoryear{{Gottlieb} \& {Shu}}{{Gottlieb} \&
  {Shu}}{1998}]{gottlieb98}
{Gottlieb} S.,  {Shu} C.~W.,  1998, Mathematics of Computation, 67, 73

\bibitem[\protect\citeauthoryear{Gourgoulhon, Grandcl{\'{e}}ment, Taniguchi,
  Marck \& Bonazzola}{Gourgoulhon et~al.}{2001}]{gourgoulhon01}
Gourgoulhon E.,  Grandcl{\'{e}}ment P.,  Taniguchi K.,  Marck J.-A.,
  Bonazzola S.,  2001, Physical Review D, 63

\bibitem[\protect\citeauthoryear{{Grossman}, {Korobkin}, {Rosswog} \&
  {Piran}}{{Grossman} et~al.}{2014}]{grossman14a}
{Grossman} D.,  {Korobkin} O.,  {Rosswog} S.,    {Piran} T.,  2014, MNRAS, 439,
  757

\bibitem[\protect\citeauthoryear{{Hajela}, {Margutti}, {Bright}, {Alexander},
  {Metzger}, {Nedora}, {Kathirgamaraju}, {Margalit}, {Radice}, {Guidorzi},
  {Berger}, {MacFadyen}, {Giannios}, {Chornock}, {Heywood}, {Sironi},
  {Gottlieb}, {Coppejans}, {Laskar}, {Cendes}, {Duran}, {Eftekhari}, {Fong},
  {McDowell}, {Nicholl}, {Xie}, {Zrake}, {Bernuzzi}, {Broekgaarden},
  {Kilpatrick}, {Terreran}, {Villar}, {Blanchard}, {Gomez}, {Hosseinzadeh},
  {Matthews} \& {Rastinejad}}{{Hajela} et~al.}{2022}]{hajela22}
{Hajela} A.,  {Margutti} R.,  {Bright} J.~S.,  {Alexander} K.~D.,  {Metzger}
  B.~D.,  {Nedora} V.,  {Kathirgamaraju} A.,  {Margalit} B.,  {Radice} D.,
  {Guidorzi} C.,  {Berger} E.,  {MacFadyen} A.,  {Giannios} D.,  {Chornock} R.,
   {Heywood} I.,  {Sironi} L.,  {Gottlieb} O.,  {Coppejans} D.,  {Laskar} T.,
  {Cendes} Y.,  {Duran} R.~B.,  {Eftekhari} T.,  {Fong} W.,  {McDowell} A.,
  {Nicholl} M.,  {Xie} X.,  {Zrake} J.,  {Bernuzzi} S.,  {Broekgaarden} F.~S.,
  {Kilpatrick} C.~D.,  {Terreran} G.,  {Villar} V.~A.,  {Blanchard} P.~K.,
  {Gomez} S.,  {Hosseinzadeh} G.,  {Matthews} D.~J.,    {Rastinejad} J.~C.,
  2022, ApJL, 927, L17

\bibitem[\protect\citeauthoryear{{Hamers} \& {Thompson}}{{Hamers} \&
  {Thompson}}{2019}]{ht19}
{Hamers} A.~S.,  {Thompson} T.~A.,  2019, \apj, 883, 23

\bibitem[\protect\citeauthoryear{{Hessels}, {Ransom}, {Stairs}, {Freire},
  {Kaspi} \& {Camilo}}{{Hessels} et~al.}{2006}]{hessels06}
{Hessels} J. W.~T.,  {Ransom} S.~M.,  {Stairs} I.~H.,  {Freire} P. C.~C.,
  {Kaspi} V.~M.,    {Camilo} F.,  2006, Science, 311, 1901

\bibitem[\protect\citeauthoryear{{Holmbeck}, {Frebel}, {McLaughlin},
  {Mumpower}, {Sprouse} \& {Surman}}{{Holmbeck} et~al.}{2019}]{holmbeck19}
{Holmbeck} E.~M.,  {Frebel} A.,  {McLaughlin} G.~C.,  {Mumpower} M.~R.,
  {Sprouse} T.~M.,    {Surman} R.,  2019, ApJ, 881, 5

\bibitem[\protect\citeauthoryear{{Hotokezaka}, {Kiuchi}, {Shibata}, {Nakar} \&
  {Piran}}{{Hotokezaka} et~al.}{2018}]{hotokezaka18a}
{Hotokezaka} K.,  {Kiuchi} K.,  {Shibata} M.,  {Nakar} E.,    {Piran} T.,
  2018, ApJ, 867, 95

\bibitem[\protect\citeauthoryear{{Hotokezaka} \& {Nakar}}{{Hotokezaka} \&
  {Nakar}}{2020}]{hotokezaka20}
{Hotokezaka} K.,  {Nakar} E.,  2020, ApJ, 891, 152

\bibitem[\protect\citeauthoryear{{Hotokezaka}, {Piran} \& {Paul}}{{Hotokezaka}
  et~al.}{2015}]{hotokezaka15a}
{Hotokezaka} K.,  {Piran} T.,    {Paul} M.,  2015, Nature Physics, 11, 1042

\bibitem[\protect\citeauthoryear{{Hotokezaka}, {Sari} \& {Piran}}{{Hotokezaka}
  et~al.}{2017}]{hotokezaka17a}
{Hotokezaka} K.,  {Sari} R.,    {Piran} T.,  2017, MNRAS, 468, 91

\bibitem[\protect\citeauthoryear{{Just}, {Bauswein}, {Pulpillo}, {Goriely} \&
  {Janka}}{{Just} et~al.}{2015}]{just15}
{Just} O.,  {Bauswein} A.,  {Pulpillo} R.~A.,  {Goriely} S.,    {Janka} H.-T.,
  2015, MNRAS, 448, 541

\bibitem[\protect\citeauthoryear{{Kasen}, {Badnell} \& {Barnes}}{{Kasen}
  et~al.}{2013}]{kasen13a}
{Kasen} D.,  {Badnell} N.~R.,    {Barnes} J.,  2013, ApJ, 774, 25

\bibitem[\protect\citeauthoryear{{Kasen} \& {Bildsten}}{{Kasen} \&
  {Bildsten}}{2010}]{kasen10}
{Kasen} D.,  {Bildsten} L.,  2010, ApJ, 717, 245

\bibitem[\protect\citeauthoryear{{Kasen}, {Metzger}, {Barnes}, {Quataert} \&
  {Ramirez-Ruiz}}{{Kasen} et~al.}{2017}]{kasen17}
{Kasen} D.,  {Metzger} B.,  {Barnes} J.,  {Quataert} E.,    {Ramirez-Ruiz} E.,
  2017, Nature, 551, 80

\bibitem[\protect\citeauthoryear{{Kastaun}, {Galeazzi}, {Alic}, {Rezzolla} \&
  {Font}}{{Kastaun} et~al.}{2013}]{kastaun13}
{Kastaun} W.,  {Galeazzi} F.,  {Alic} D.,  {Rezzolla} L.,    {Font} J.~A.,
  2013, Phys. Rev. D, 88, 021501

\bibitem[\protect\citeauthoryear{{Keane}, {Bhattacharyya}, {Kramer},
  {Stappers}, {Keane}, {Bhattacharyya}, {Kramer}, {Stappers}, {Bates},
  {Burgay}, {Chatterjee}, {Champion}, {Eatough}, {Hessels}, {Janssen}, {Lee},
  {van Leeuwen}, {Margueron}, {Oertel}, {Possenti}, {Ransom}, {Theureau} \&
  {Torne}}{{Keane} et~al.}{2015}]{kbk+15}
{Keane} E.,  {Bhattacharyya} B.,  {Kramer} M.,  {Stappers} B.,  {Keane} E.~F.,
  {Bhattacharyya} B.,  {Kramer} M.,  {Stappers} B.~W.,  {Bates} S.~D.,
  {Burgay} M.,  {Chatterjee} S.,  {Champion} D.~J.,  {Eatough} R.~P.,
  {Hessels} J.~W.~T.,  {Janssen} G.,  {Lee} K.~J.,  {van Leeuwen} J.,
  {Margueron} J.,  {Oertel} M.,  {Possenti} A.,  {Ransom} S.,  {Theureau} G.,
   {Torne} P.,  2015, in Advancing Astrophysics with the Square Kilometre Array
  (AASKA14) {A Cosmic Census of Radio Pulsars with the SKA}.
p.~40

\bibitem[\protect\citeauthoryear{{Kim}, {Kalogera} \& {Lorimer}}{{Kim}
  et~al.}{2003}]{kkl03}
{Kim} C.,  {Kalogera} V.,    {Lorimer} D.~R.,  2003, \apj, 584, 985

\bibitem[\protect\citeauthoryear{{Kochanek}}{{Kochanek}}{1992}]{kochanek92}
{Kochanek} C.~S.,  1992, ApJ, 398, 234

\bibitem[\protect\citeauthoryear{{Korobkin}, {Rosswog}, {Arcones} \&
  {Winteler}}{{Korobkin} et~al.}{2012}]{korobkin12a}
{Korobkin} O.,  {Rosswog} S.,  {Arcones} A.,    {Winteler} C.,  2012, MNRAS,
  426, 1940

\bibitem[\protect\citeauthoryear{{Kruckow}, {Tauris}, {Langer}, {Kramer} \&
  {Izzard}}{{Kruckow} et~al.}{2018}]{ktl+18}
{Kruckow} M.~U.,  {Tauris} T.~M.,  {Langer} N.,  {Kramer} M.,    {Izzard}
  R.~G.,  2018, \mnras, 481, 1908

\bibitem[\protect\citeauthoryear{{Lattimer}, {Mackie}, {Ravenhall} \&
  {Schramm}}{{Lattimer} et~al.}{1977}]{lattimer77}
{Lattimer} J.~M.,  {Mackie} F.,  {Ravenhall} D.~G.,    {Schramm} D.~N.,  1977,
  ApJ, 213, 225

\bibitem[\protect\citeauthoryear{Lattimer \& Schramm}{Lattimer \&
  Schramm}{1974}]{lattimer74}
Lattimer J.~M.,  Schramm D.~N.,  1974, ApJ, (Letters), 192, L145

\bibitem[\protect\citeauthoryear{{Levan}, {Gompertz}, {Salafia}, {Bulla} \& et
  al.}{{Levan} et~al.}{2023}]{Levan2023}
{Levan} A.,  {Gompertz} B.~P.,  {Salafia} O.~S.,  {Bulla} M.,    et al. 2023,
  arXiv e-prints, p. arXiv:2307.02098

\bibitem[\protect\citeauthoryear{{Lippuner} \& {Roberts}}{{Lippuner} \&
  {Roberts}}{2015}]{lippuner15}
{Lippuner} J.,  {Roberts} L.~F.,  2015, ApJ, 815, 82

\bibitem[\protect\citeauthoryear{{Liu} \& {Lai}}{{Liu} \& {Lai}}{2021}]{ll21}
{Liu} B.,  {Lai} D.,  2021, \mnras, 502, 2049

\bibitem[\protect\citeauthoryear{{L{\"o}ffler}, {Faber}, {Bentivegna}, {Bode},
  {Diener}, {Haas}, {Hinder}, {Mundim}, {Ott}, {Schnetter}, {Allen},
  {Campanelli} \& {Laguna}}{{L{\"o}ffler} et~al.}{2012}]{loeffler12}
{L{\"o}ffler} F.,  {Faber} J.,  {Bentivegna} E.,  {Bode} T.,  {Diener} P.,
  {Haas} R.,  {Hinder} I.,  {Mundim} B.~C.,  {Ott} C.~D.,  {Schnetter} E.,
  {Allen} G.,  {Campanelli} M.,    {Laguna} P.,  2012, Classical and Quantum
  Gravity, 29, 115001

\bibitem[\protect\citeauthoryear{{LORENE}}{{LORENE}}{2001}]{lorene}
{LORENE}, 2001, http://lorene.obspm.fr

\bibitem[\protect\citeauthoryear{{Lorimer} \& {Kramer}}{{Lorimer} \&
  {Kramer}}{2012}]{lk12}
{Lorimer} D.~R.,  {Kramer} M.,  2012, {Handbook of Pulsar Astronomy}

\bibitem[\protect\citeauthoryear{{Lynch}, {Freire}, {Ransom} \&
  {Jacoby}}{{Lynch} et~al.}{2012}]{lfrj12}
{Lynch} R.~S.,  {Freire} P.~C.~C.,  {Ransom} S.~M.,    {Jacoby} B.~A.,  2012,
  \apj, 745, 109

\bibitem[\protect\citeauthoryear{Maggiore}{Maggiore}{2008}]{maggiore08}
Maggiore M.,  2008, Gravitational Waves.
Oxford University Press, Oxford

\bibitem[\protect\citeauthoryear{{Manchester}, {Hobbs}, {Teoh} \&
  {Hobbs}}{{Manchester} et~al.}{2005}]{mhth05}
{Manchester} R.~N.,  {Hobbs} G.~B.,  {Teoh} A.,    {Hobbs} M.,  2005, Astronomical Journal, 129,
  1993

\bibitem[\protect\citeauthoryear{{Margalit} \& {Metzger}}{{Margalit} \&
  {Metzger}}{2017}]{margalit17}
{Margalit} B.,  {Metzger} B.~D.,  2017, \apjl, 850, L19

\bibitem[\protect\citeauthoryear{{Martin}, {Perego}, {Arcones}, {Thielemann},
  {Korobkin} \& {Rosswog}}{{Martin} et~al.}{2015}]{martin15}
{Martin} D.,  {Perego} A.,  {Arcones} A.,  {Thielemann} F.-K.,  {Korobkin} O.,
    {Rosswog} S.,  2015, ApJ, 813, 2

\bibitem[\protect\citeauthoryear{{Martinez}, {Stovall}, {Freire}, {Deneva},
  {Jenet}, {McLaughlin}, {Bagchi}, {Bates} \& {Ridolfi}}{{Martinez}
  et~al.}{2015}]{martinez15}
{Martinez} J.~G.,  {Stovall} K.,  {Freire} P.~C.~C.,  {Deneva} J.~S.,  {Jenet}
  F.~A.,  {McLaughlin} M.~A.,  {Bagchi} M.,  {Bates} S.~D.,    {Ridolfi} A.,
  2015, ApJ, 812, 143

\bibitem[\protect\citeauthoryear{{Metzger}}{{Metzger}}{2019}]{metzger19a}
{Metzger} B.~D.,  2019, Living Reviews in Relativity, 23, 1

\bibitem[\protect\citeauthoryear{{Metzger}, {Arcones}, {Quataert} \&
  {Martinez-Pinedo}}{{Metzger} et~al.}{2010}]{metzger10a}
{Metzger} B.~D.,  {Arcones} A.,  {Quataert} E.,    {Martinez-Pinedo} G.,  2010,
  MNRAS, 402, 2771

\bibitem[\protect\citeauthoryear{{Metzger}, {Bauswein}, {Goriely} \&
  {Kasen}}{{Metzger} et~al.}{2015}]{metzger15a}
{Metzger} B.~D.,  {Bauswein} A.,  {Goriely} S.,    {Kasen} D.,  2015, MNRAS,
  446, 1115

\bibitem[\protect\citeauthoryear{{Metzger}, {Martinez-Pinedo}, {Darbha},
  {Quataert}, {Arcones}, {Kasen}, {Thomas}, {Nugent}, {Panov} \&
  {Zinner}}{{Metzger} et~al.}{2010}]{metzger10b}
{Metzger} B.~D.,  {Martinez-Pinedo} G.,  {Darbha} S.,  {Quataert} E.,
  {Arcones} A.,  {Kasen} D.,  {Thomas} R.,  {Nugent} P.,  {Panov} I.~V.,
  {Zinner} N.~T.,  2010, MNRAS, 406, 2650

\bibitem[\protect\citeauthoryear{{Miller}, {Ryan}, {Dolence}, {Burrows},
  {Fontes}, {Fryer}, {Korobkin}, {Lippuner}, {Mumpower} \&
  {Wollaeger}}{{Miller} et~al.}{2019}]{miller19}
{Miller} J.~M.,  {Ryan} B.~R.,  {Dolence} J.~C.,  {Burrows} A.,  {Fontes}
  C.~J.,  {Fryer} C.~L.,  {Korobkin} O.,  {Lippuner} J.,  {Mumpower} M.~R.,
  {Wollaeger} R.~T.,  2019, \prd, 100, 023008

\bibitem[\protect\citeauthoryear{{Monaghan}}{{Monaghan}}{2005}]{monaghan05}
{Monaghan} J.~J.,  2005, Reports on Progress in Physics, 68, 1703

\bibitem[\protect\citeauthoryear{{Most}, {Papenfort}, {Tootle} \&
  {Rezzolla}}{{Most} et~al.}{2021}]{most21}
{Most} E.~R.,  {Papenfort} L.~J.,  {Tootle} S.~D.,    {Rezzolla} L.,  2021,
  \apj, 912, 80

\bibitem[\protect\citeauthoryear{{Most}, {Papenfort}, {Tsokaros} \&
  {Rezzolla}}{{Most} et~al.}{2019}]{most19}
{Most} E.~R.,  {Papenfort} L.~J.,  {Tsokaros} A.,    {Rezzolla} L.,  2019,
  \apj, 884, 40

\bibitem[\protect\citeauthoryear{{M{\"u}ller} \& {Serot}}{{M{\"u}ller} \&
  {Serot}}{1996}]{MS1_EOS}
{M{\"u}ller} H.,  {Serot} B.~D.,  1996, Nuc. Phys. A, 606, 508

\bibitem[\protect\citeauthoryear{{Nakar} \& {Piran}}{{Nakar} \&
  {Piran}}{2011}]{nakar11a}
{Nakar} E.,  {Piran} T.,  2011, Nature, 478, 82

\bibitem[\protect\citeauthoryear{{Narayan}, {Paczynski} \& {Piran}}{{Narayan}
  et~al.}{1992}]{narayan92}
{Narayan} R.,  {Paczynski} B.,    {Piran} T.,  1992, ApJ, 395, L83

\bibitem[\protect\citeauthoryear{{Neuweiler}, {Dietrich}, {Bulla}, {Chaurasia},
  {Rosswog} \& {Ujevic}}{{Neuweiler} et~al.}{2023}]{neuweiler23}
{Neuweiler} A.,  {Dietrich} T.,  {Bulla} M.,  {Chaurasia} S.~V.,  {Rosswog} S.,
     {Ujevic} M.,  2023, Phys. Rev. D, 107, 023016

\bibitem[\protect\citeauthoryear{{Pacilio}, {Maselli}, {Fasano} \&
  {Pani}}{{Pacilio} et~al.}{2022}]{pacilio22}
{Pacilio} C.,  {Maselli} A.,  {Fasano} M.,    {Pani} P.,  2022, Phys. Rev.
  Lett., 128, 101101

\bibitem[\protect\citeauthoryear{{Papenfort}, {Most}, {Tootle} \&
  {Rezzolla}}{{Papenfort} et~al.}{2022}]{papenfort22}
{Papenfort} L.~J.,  {Most} E.~R.,  {Tootle} S.,    {Rezzolla} L.,  2022, MNRAS,
  513, 3646

\bibitem[\protect\citeauthoryear{{Papenfort}, {Tootle}, {Grandcl{\'e}ment},
  {Most} \& {Rezzolla}}{{Papenfort} et~al.}{2021}]{papenfort21}
{Papenfort} L.~J.,  {Tootle} S.~D.,  {Grandcl{\'e}ment} P.,  {Most} E.~R.,
  {Rezzolla} L.,  2021, Phys. Rev. D, 104, 024057

\bibitem[\protect\citeauthoryear{{Perego}, {Rosswog}, {Cabez{\'o}n},
  {Korobkin}, {K{\"a}ppeli}, {Arcones} \& {Liebend{\"o}rfer}}{{Perego}
  et~al.}{2014}]{perego14b}
{Perego} A.,  {Rosswog} S.,  {Cabez{\'o}n} R.~M.,  {Korobkin} O.,
  {K{\"a}ppeli} R.,  {Arcones} A.,    {Liebend{\"o}rfer} M.,  2014, MNRAS, 443,
  3134

\bibitem[\protect\citeauthoryear{Peters \& Mathews}{Peters \&
  Mathews}{1964}]{peters64}
Peters P.,  Mathews J.,  1964, Phys. Review, B136, 1224

\bibitem[\protect\citeauthoryear{{Peters}}{{Peters}}{1964}]{pet64}
{Peters} P.~C.,  1964, Physical Review, 136, 1224

\bibitem[\protect\citeauthoryear{{Peters} \& {Mathews}}{{Peters} \&
  {Mathews}}{1963}]{peters63}
{Peters} P.~C.,  {Mathews} J.,  1963, Physical Review, 131, 435

\bibitem[\protect\citeauthoryear{{Pinto} \& {Eastman}}{{Pinto} \&
  {Eastman}}{2000}]{pinto00}
{Pinto} P.~A.,  {Eastman} R.~G.,  2000, ApJ, 530, 757

\bibitem[\protect\citeauthoryear{{Piran}, {Nakar} \& {Rosswog}}{{Piran}
  et~al.}{2013}]{piran13a}
{Piran} T.,  {Nakar} E.,    {Rosswog} S.,  2013, MNRAS, 430, 2121

\bibitem[\protect\citeauthoryear{{Piro}, {Troja}, {Zhang}, {Ryan} \& et
  al.}{{Piro} et~al.}{2019}]{Piro2019}
{Piro} L.,  {Troja} E.,  {Zhang} B.,  {Ryan} G.,    et al. 2019, \mnras, 483,
  1912

\bibitem[\protect\citeauthoryear{{Pol}, {McLaughlin} \& {Lorimer}}{{Pol}
  et~al.}{2019}]{pml+19}
{Pol} N.,  {McLaughlin} M.,    {Lorimer} D.~R.,  2019, \apj, 870, 71

\bibitem[\protect\citeauthoryear{{Price}}{{Price}}{2012}]{price12a}
{Price} D.~J.,  2012, Journal of Computational Physics, 231, 759

\bibitem[\protect\citeauthoryear{{Punturo}, {Abernathy}, {Acernese}, {Allen},
  {Andersson} et~al.,}{{Punturo} et~al.}{2010}]{EinsteinTelescope}
{Punturo} M.,  {Abernathy} M.,  {Acernese} F.,  {Allen} B.,  {Andersson} N.,
  et~al., 2010, Class. Quant. Grav., 27, 194002

\bibitem[\protect\citeauthoryear{{Radhakrishnan} \&
  {Srinivasan}}{{Radhakrishnan} \& {Srinivasan}}{1982}]{rs82}
{Radhakrishnan} V.,  {Srinivasan} G.,  1982, Current Science, 51, 1096

\bibitem[\protect\citeauthoryear{{Ransom}, {Stairs}, {Archibald}, {Hessels},
  {Kaplan}, {van Kerkwijk}, {Boyles}, {Deller}, {Chatterjee},
  {Schechtman-Rook}, {Berndsen}, {Lynch}, {Lorimer}, {Karako-Argaman}, {Kaspi},
  {Kondratiev}, {McLaughlin}, {van Leeuwen}, {Rosen}, {Roberts} \&
  {Stovall}}{{Ransom} et~al.}{2014}]{sra+14}
{Ransom} S.~M.,  {Stairs} I.~H.,  {Archibald} A.~M.,  {Hessels} J.~W.~T.,
  {Kaplan} D.~L.,  {van Kerkwijk} M.~H.,  {Boyles} J.,  {Deller} A.~T.,
  {Chatterjee} S.,  {Schechtman-Rook} A.,  {Berndsen} A.,  {Lynch} R.~S.,
  {Lorimer} D.~R.,  {Karako-Argaman} C.,  {Kaspi} V.~M.,  {Kondratiev} V.~I.,
  {McLaughlin} M.~A.,  {van Leeuwen} J.,  {Rosen} R.,  {Roberts} M.~S.~E.,
  {Stovall} K.,  2014, \nat, 505, 520

\bibitem[\protect\citeauthoryear{{Rastinejad}, {Gompertz}, {Levan}, {Fong},
  {Nicholl}, {Lamb}, {Malesani}, {Nugent} \& more}{{Rastinejad}
  et~al.}{2022}]{rastinejad22}
{Rastinejad} J.~C.,  {Gompertz} B.~P.,  {Levan} A.~J.,  {Fong} W.-f.,
  {Nicholl} M.,  {Lamb} G.~P.,  {Malesani} D.~B.,  {Nugent} A.~E.,    more
  2022, Nature, 612, 223

\bibitem[\protect\citeauthoryear{{Read}, {Lackey}, {Owen} \& {Friedman}}{{Read}
  et~al.}{2009}]{read09}
{Read} J.~S.,  {Lackey} B.~D.,  {Owen} B.~J.,    {Friedman} J.~L.,  2009, Phys.
  Rev. D, 79, 124032

\bibitem[\protect\citeauthoryear{{Rezzolla}, {Most} \& {Weih}}{{Rezzolla}
  et~al.}{2018}]{rezzolla18}
{Rezzolla} L.,  {Most} E.~R.,    {Weih} L.~R.,  2018, \apjl, 852, L25

\bibitem[\protect\citeauthoryear{{Rezzolla} \& {Zanotti}}{{Rezzolla} \&
  {Zanotti}}{2013}]{rezzolla13a}
{Rezzolla} L.,  {Zanotti} O.,  2013, {Relativistic Hydrodynamics}

\bibitem[\protect\citeauthoryear{{Rodriguez}, {Zevin}, {Pankow}, {Kalogera} \&
  {Rasio}}{{Rodriguez} et~al.}{2016}]{rzp+16}
{Rodriguez} C.~L.,  {Zevin} M.,  {Pankow} C.,  {Kalogera} V.,    {Rasio} F.~A.,
   2016, \apjl, 832, L2

\bibitem[\protect\citeauthoryear{{Roederer}, {Cowan}, {Karakas}, {Kratz},
  {Lugaro}, {Simmerer}, {Farouqi} \& {Sneden}}{{Roederer}
  et~al.}{2010}]{roederer10}
{Roederer} I.~U.,  {Cowan} J.~J.,  {Karakas} A.~I.,  {Kratz} K.-L.,  {Lugaro}
  M.,  {Simmerer} J.,  {Farouqi} K.,    {Sneden} C.,  2010, ApJ, 724, 975

\bibitem[\protect\citeauthoryear{Rosswog}{Rosswog}{2007}]{rosswog07a}
Rosswog S.,  2007, MNRAS, 376, L48

\bibitem[\protect\citeauthoryear{Rosswog}{Rosswog}{2009}]{rosswog09b}
Rosswog S.,  2009, New Astronomy Reviews, 53, 78

\bibitem[\protect\citeauthoryear{{Rosswog}}{{Rosswog}}{2010}]{rosswog10a}
{Rosswog} S.,  2010, Classical and Quantum Gravity, 27, 114108

\bibitem[\protect\citeauthoryear{{Rosswog}}{{Rosswog}}{2015a}]{rosswog15b}
{Rosswog} S.,  2015a, MNRAS, 448, 3628

\bibitem[\protect\citeauthoryear{{Rosswog}}{{Rosswog}}{2015b}]{rosswog15c}
{Rosswog} S.,  2015b, Living Reviews of Computational Astrophysics (2015), 1

\bibitem[\protect\citeauthoryear{{Rosswog}}{{Rosswog}}{2020a}]{rosswog20b}
{Rosswog} S.,  2020a, ApJ, 898, 60

\bibitem[\protect\citeauthoryear{{Rosswog}}{{Rosswog}}{2020b}]{rosswog20a}
{Rosswog} S.,  2020b, MNRAS, 498, 4230

\bibitem[\protect\citeauthoryear{{Rosswog}, {Davies}, {Thielemann} \&
  {Piran}}{{Rosswog} et~al.}{2000}]{rosswog00}
{Rosswog} S.,  {Davies} M.~B.,  {Thielemann} F.-K.,    {Piran} T.,  2000, A\&A,
  360, 171

\bibitem[\protect\citeauthoryear{{Rosswog} \& {Diener}}{{Rosswog} \&
  {Diener}}{2021}]{rosswog21a}
{Rosswog} S.,  {Diener} P.,  2021, Classical and Quantum Gravity, 38, 115002

\bibitem[\protect\citeauthoryear{{Rosswog}, {Diener} \& {Torsello}}{{Rosswog}
  et~al.}{2022}]{rosswog22b}
{Rosswog} S.,  {Diener} P.,    {Torsello} F.,  2022, Symmetry, 14, 1280

\bibitem[\protect\citeauthoryear{{Rosswog} \& {Korobkin}}{{Rosswog} \&
  {Korobkin}}{2022}]{rosswog22c}
{Rosswog} S.,  {Korobkin} O.,  2022, Annalen der Physik

\bibitem[\protect\citeauthoryear{{Rosswog}, {Korobkin}, {Arcones}, {Thielemann}
  \& {Piran}}{{Rosswog} et~al.}{2014}]{rosswog14a}
{Rosswog} S.,  {Korobkin} O.,  {Arcones} A.,  {Thielemann} F.-K.,    {Piran}
  T.,  2014, MNRAS, 439, 744

\bibitem[\protect\citeauthoryear{Rosswog, Liebend\"orfer, Thielemann, Davies,
  Benz \& Piran}{Rosswog et~al.}{1999}]{rosswog99}
Rosswog S.,  Liebend\"orfer M.,  Thielemann F.-K.,  Davies M.,  Benz W.,
  Piran T.,  1999, A \&\ A, 341, 499

\bibitem[\protect\citeauthoryear{{Rosswog}, {Piran} \& {Nakar}}{{Rosswog}
  et~al.}{2013}]{rosswog13a}
{Rosswog} S.,  {Piran} T.,    {Nakar} E.,  2013, MNRAS, 430, 2585

\bibitem[\protect\citeauthoryear{{Rosswog}, {Sollerman}, {Feindt}, {Goobar},
  {Korobkin}, {Wollaeger}, {Fremling} \& {Kasliwal}}{{Rosswog}
  et~al.}{2018}]{rosswog18a}
{Rosswog} S.,  {Sollerman} J.,  {Feindt} U.,  {Goobar} A.,  {Korobkin} O.,
  {Wollaeger} R.,  {Fremling} C.,    {Kasliwal} M.~M.,  2018, A\&A, 615, A132

\bibitem[\protect\citeauthoryear{{Rosswog}, {Thielemann}, {Davies}, {Benz} \&
  {Piran}}{{Rosswog} et~al.}{1998}]{rosswog98b}
{Rosswog} S.,  {Thielemann} F.~K.,  {Davies} M.~B.,  {Benz} W.,    {Piran} T.,
  1998, in {Hillebrandt} W.,  {M\"uller} E.,  eds, Nuclear Astrophysics
  {Coalescing Neutron Stars: a Solution to the R-Process Problem?}.
Max-Planck-Institut f\"ur Physik und Astrophysik, Garching b. M\"unchen, p.~103

\bibitem[\protect\citeauthoryear{{Rosswog}, {Torsello} \& {Diener}}{{Rosswog}
  et~al.}{2023}]{rosswog23a}
{Rosswog} S.,  {Torsello} F.,    {Diener} P.,  2023, arXiv e-prints, p.
  arXiv:2306.06226

\bibitem[\protect\citeauthoryear{{Ruffert}, {Janka} \& {Schaefer}}{{Ruffert}
  et~al.}{1996}]{ruffert96}
{Ruffert} M.,  {Janka} H.,    {Schaefer} G.,  1996, A \& A, 311, 532

\bibitem[\protect\citeauthoryear{{Ruffert}, {Janka}, {Takahashi} \&
  {Schaefer}}{{Ruffert} et~al.}{1997}]{ruffert97a}
{Ruffert} M.,  {Janka} H.,  {Takahashi} K.,    {Schaefer} G.,  1997, A \& A,
  319, 122

\bibitem[\protect\citeauthoryear{{Ruiz}, {Tsokaros}, {Paschalidis} \&
  {Shapiro}}{{Ruiz} et~al.}{2019}]{ruiz19}
{Ruiz} M.,  {Tsokaros} A.,  {Paschalidis} V.,    {Shapiro} S.~L.,  2019, Phys.
  Rev. D, 99, 084032

\bibitem[\protect\citeauthoryear{{Sadeh}, {Guttman} \& {Waxman}}{{Sadeh}
  et~al.}{2023}]{Sadeh2023}
{Sadeh} G.,  {Guttman} O.,    {Waxman} E.,  2023, \mnras, 518, 2102

\bibitem[\protect\citeauthoryear{{Sarin}, {H{\"u}bner}, {Omand}, {Setzer} \& et
  al.}{{Sarin} et~al.}{2023}]{sarin23_redback}
{Sarin} N.,  {H{\"u}bner} M.,  {Omand} C. M.~B.,  {Setzer} C.~N.,    et al.
  2023, arXiv e-prints, p. arXiv:2308.12806

\bibitem[\protect\citeauthoryear{{Sarin}, {Lasky} \& {Ashton}}{{Sarin}
  et~al.}{2020}]{sarin20a}
{Sarin} N.,  {Lasky} P.~D.,    {Ashton} G.,  2020, \prd, 101, 063021

\bibitem[\protect\citeauthoryear{{Sarin}, {Omand}, {Margalit} \&
  {Jones}}{{Sarin} et~al.}{2022}]{sarin22_kne}
{Sarin} N.,  {Omand} C. M.~B.,  {Margalit} B.,    {Jones} D.~I.,  2022, \mnras,
  516, 4949

\bibitem[\protect\citeauthoryear{{Savchenko}, {Ferrigno}, {Kuulkers},
  {Bazzano}, {Bozzo}, {Brandt}, {Chenevez}, {Courvoisier}, {Diehl}, {Domingo}
  \& et al.}{{Savchenko} et~al.}{2017}]{savchenko17}
{Savchenko} V.,  {Ferrigno} C.,  {Kuulkers} E.,  {Bazzano} A.,  {Bozzo} E.,
  {Brandt} S.,  {Chenevez} J.,  {Courvoisier} T.~J.~L.,  {Diehl} R.,  {Domingo}
  A.,    et al. 2017, ApJL, 848, L15

\bibitem[\protect\citeauthoryear{{Shakura} \& {Sunyaev}}{{Shakura} \&
  {Sunyaev}}{1973}]{shakura73}
{Shakura} N.~I.,  {Sunyaev} R.~A.,  1973, A \& A, 24, 337

\bibitem[\protect\citeauthoryear{Shapiro \& Teukolsky}{Shapiro \&
  Teukolsky}{1983}]{shapiro83}
Shapiro S.,  Teukolsky S.~A.,  1983, {B}lack {H}oles, {W}hite {D}warfs and
  {N}eutron {S}tars.
Wiley \& Sons, New York

\bibitem[\protect\citeauthoryear{{Shibata}, {Fujibayashi}, {Hotokezaka},
  {Kiuchi}, {Kyutoku}, {Sekiguchi} \& {Tanaka}}{{Shibata}
  et~al.}{2017}]{shibata17c}
{Shibata} M.,  {Fujibayashi} S.,  {Hotokezaka} K.,  {Kiuchi} K.,  {Kyutoku} K.,
   {Sekiguchi} Y.,    {Tanaka} M.,  2017, Phys. Rev. D, 96, 123012

\bibitem[\protect\citeauthoryear{{Shibata} \& {Nakamura}}{{Shibata} \&
  {Nakamura}}{1995}]{shibata95}
{Shibata} M.,  {Nakamura} T.,  1995, Phys. Rev. D, 52, 5428

\bibitem[\protect\citeauthoryear{{Siegel} \& {Metzger}}{{Siegel} \&
  {Metzger}}{2017}]{siegel17a}
{Siegel} D.~M.,  {Metzger} B.~D.,  2017, Physical Review Letters, 119, 231102

\bibitem[\protect\citeauthoryear{{Siegel} \& {Metzger}}{{Siegel} \&
  {Metzger}}{2018}]{siegel18}
{Siegel} D.~M.,  {Metzger} B.~D.,  2018, ApJ, 858, 52

\bibitem[\protect\citeauthoryear{{Sironi} \& {Giannios}}{{Sironi} \&
  {Giannios}}{2013}]{Sironi2013}
{Sironi} L.,  {Giannios} D.,  2013, \apj, 778, 107

\bibitem[\protect\citeauthoryear{{Smartt}, {Chen}, {Jerkstrand}, {Coughlin},
  {Kankare}, {Sim}, {Fraser}, {Inserra}, {Maguire}, {Chambers}, {Huber} \& et
  al.}{{Smartt} et~al.}{2017}]{smartt17}
{Smartt} S.~J.,  {Chen} T.-W.,  {Jerkstrand} A.,  {Coughlin} M.,  {Kankare} E.,
   {Sim} S.~A.,  {Fraser} M.,  {Inserra} C.,  {Maguire} K.,  {Chambers} K.~C.,
  {Huber} M.~E.,    et al. 2017, Nature, 551, 75

\bibitem[\protect\citeauthoryear{{Symbalisty} \& {Schramm}}{{Symbalisty} \&
  {Schramm}}{1982}]{symbalisty82}
{Symbalisty} E.,  {Schramm} D.~N.,  1982, Astrophys. Lett., 22, 143

\bibitem[\protect\citeauthoryear{{Tanaka}, {Kato}, {Gaigalas} \&
  {Kawaguchi}}{{Tanaka} et~al.}{2020}]{tanaka20a}
{Tanaka} M.,  {Kato} D.,  {Gaigalas} G.,    {Kawaguchi} K.,  2020, MNRAS, 496,
  1369

\bibitem[\protect\citeauthoryear{{Tanvir}, {Levan}, {Gonz{a}lez-Fern{a}ndez},
  {Korobkin}, {Mandel}, {Rosswog}, {Hjorth}, {D'Avanzo} \& more}{{Tanvir}
  et~al.}{2017}]{tanvir17}
{Tanvir} N.~R.,  {Levan} A.~J.,  {Gonz{a}lez-Fern{a}ndez} C.,  {Korobkin} O.,
  {Mandel} I.,  {Rosswog} S.,  {Hjorth} J.,  {D'Avanzo} P.,    more 2017, ApJL,
  848, L27

\bibitem[\protect\citeauthoryear{{Tauris}, {Kramer}, {Freire}, {Wex}, {Janka},
  {Langer}, {Podsiadlowski}, {Bozzo}, {Chaty}, {Kruckow}, {van den Heuvel},
  {Antoniadis}, {Breton} \& {Champion}}{{Tauris} et~al.}{2017}]{tkf+17}
{Tauris} T.~M.,  {Kramer} M.,  {Freire} P.~C.~C.,  {Wex} N.,  {Janka} H.-T.,
  {Langer} N.,  {Podsiadlowski} P.,  {Bozzo} E.,  {Chaty} S.,  {Kruckow} M.~U.,
   {van den Heuvel} E.~P.~J.,  {Antoniadis} J.,  {Breton} R.~P.,    {Champion}
  D.~J.,  2017, \apj, 846, 170

\bibitem[\protect\citeauthoryear{{Tauris}, {Langer} \& {Kramer}}{{Tauris}
  et~al.}{2012}]{tlk12}
{Tauris} T.~M.,  {Langer} N.,    {Kramer} M.,  2012, \mnras, 425, 1601

\bibitem[\protect\citeauthoryear{{Tauris}, {Langer} \&
  {Podsiadlowski}}{{Tauris} et~al.}{2015}]{tlp15}
{Tauris} T.~M.,  {Langer} N.,    {Podsiadlowski} P.,  2015, \mnras, 451, 2123

\bibitem[\protect\citeauthoryear{{Tauris} \& {Savonije}}{{Tauris} \&
  {Savonije}}{1999}]{ts99}
{Tauris} T.~M.,  {Savonije} G.~J.,  1999, \aap, 350, 928

\bibitem[\protect\citeauthoryear{{Tauris} \& {van den Heuvel}}{{Tauris} \& {van
  den Heuvel}}{2023}]{tv23}
{Tauris} T.~M.,  {van den Heuvel} E. P.~J.,  2023, {Physics of Binary Star
  Evolution. From Stars to X-ray Binaries and Gravitational Wave Sources}.
Princeton University Press

\bibitem[\protect\citeauthoryear{{Thielemann}, {Arcones}, {K{\"a}ppeli},
  {Liebend{\"o}rfer}, {Rauscher}, {Winteler}, {Fr{\"o}hlich}, {Dillmann},
  {Fischer}, {Martinez-Pinedo}, {Langanke}, {Farouqi}, {Kratz}, {Panov} \&
  {Korneev}}{{Thielemann} et~al.}{2011}]{thielemann11}
{Thielemann} F.-K.,  {Arcones} A.,  {K{\"a}ppeli} R.,  {Liebend{\"o}rfer} M.,
  {Rauscher} T.,  {Winteler} C.,  {Fr{\"o}hlich} C.,  {Dillmann} I.,  {Fischer}
  T.,  {Martinez-Pinedo} G.,  {Langanke} K.,  {Farouqi} K.,  {Kratz} K.-L.,
  {Panov} I.,    {Korneev} I.~K.,  2011, Progress in Particle and Nuclear
  Physics, 66, 346

\bibitem[\protect\citeauthoryear{{Timmes} \& {Swesty}}{{Timmes} \&
  {Swesty}}{2000}]{timmes00a}
{Timmes} F.~X.,  {Swesty} F.~D.,  2000, ApJS, 126, 501

\bibitem[\protect\citeauthoryear{{Troja}, {van Eerten}, {Zhang}, {Ryan},
  {Piro}, {Ricci}, {O'Connor}, {Wieringa}, {Cenko} \& {Sakamoto}}{{Troja}
  et~al.}{2020}]{troja22}
{Troja} E.,  {van Eerten} H.,  {Zhang} B.,  {Ryan} G.,  {Piro} L.,  {Ricci} R.,
   {O'Connor} B.,  {Wieringa} M.~H.,  {Cenko} S.~B.,    {Sakamoto} T.,  2020,
  \mnras, 498, 5643

\bibitem[\protect\citeauthoryear{{Tsokaros}, {Ruiz}, {Paschalidis}, {Shapiro}
  \& {Ury{\={u}}}}{{Tsokaros} et~al.}{2019}]{tsokaros19}
{Tsokaros} A.,  {Ruiz} M.,  {Paschalidis} V.,  {Shapiro} S.~L.,    {Ury{\={u}}}
  K.,  2019, Phys. Rev. D, 100, 024061

\bibitem[\protect\citeauthoryear{{van den Heuvel} \& {De Loore}}{{van den
  Heuvel} \& {De Loore}}{1973}]{vhdl73}
{van den Heuvel} E.~P.~J.,  {De Loore} C.,  1973, \aap, 25, 387

\bibitem[\protect\citeauthoryear{{Wanajo}, {Sekiguchi}, {Nishimura}, {Kiuchi},
  {Kyutoku} \& {Shibata}}{{Wanajo} et~al.}{2014}]{wanajo14}
{Wanajo} S.,  {Sekiguchi} Y.,  {Nishimura} N.,  {Kiuchi} K.,  {Kyutoku} K.,
  {Shibata} M.,  2014, ApJL, 789, L39

\bibitem[\protect\citeauthoryear{{Watson}, {Hansen}, {Selsing}, {Koch},
  {Malesani}, {Andersen}, {Fynbo}, {Arcones}, {Bauswein}, {Covino}, {Grado},
  {Heintz}, {Hunt}, {Kouveliotou}, {Leloudas}, {Levan}, {Mazzali} \&
  {Pian}}{{Watson} et~al.}{2019}]{watson19}
{Watson} D.,  {Hansen} C.~J.,  {Selsing} J.,  {Koch} A.,  {Malesani} D.~B.,
  {Andersen} A.~C.,  {Fynbo} J. P.~U.,  {Arcones} A.,  {Bauswein} A.,  {Covino}
  S.,  {Grado} A.,  {Heintz} K.~E.,  {Hunt} L.,  {Kouveliotou} C.,  {Leloudas}
  G.,  {Levan} A.~J.,  {Mazzali} P.,    {Pian} E.,  2019, Nature, 574, 497

\bibitem[\protect\citeauthoryear{{W}endland}{{W}endland}{1995}]{wendland95}
{W}endland H.,  1995, Advances in Computational Mathematics, 4, 389

\bibitem[\protect\citeauthoryear{{Winteler}}{{Winteler}}{2012}]{winteler12}
{Winteler} C.,  2012, PhD thesis, University Basel, CH

\bibitem[\protect\citeauthoryear{{Wollaeger}, {Korobkin}, {Fontes}, {Rosswog},
  {Even}, {Fryer}, {Sollerman}, {Hungerford}, {van Rossum} \&
  {Wollaber}}{{Wollaeger} et~al.}{2018}]{wollaeger18a}
{Wollaeger} R.~T.,  {Korobkin} O.,  {Fontes} C.~J.,  {Rosswog} S.~K.,  {Even}
  W.~P.,  {Fryer} C.~L.,  {Sollerman} J.,  {Hungerford} A.~L.,  {van Rossum}
  D.~R.,    {Wollaber} A.~B.,  2018, MNRAS, 478, 3298

\bibitem[\protect\citeauthoryear{{Wu}, {Fernandez}, {Martinez-Pinedo} \&
  {Metzger}}{{Wu} et~al.}{2016}]{wu16}
{Wu} M.-R.,  {Fernandez} R.,  {Martinez-Pinedo} G.,    {Metzger} B.~D.,  2016,
  MNRAS

\bibitem[\protect\citeauthoryear{{Ye}, {Fong}, {Kremer}, {Rodriguez},
  {Chatterjee}, {Fragione} \& {Rasio}}{{Ye} et~al.}{2020}]{yfk+20}
{Ye} C.~S.,  {Fong} W.-f.,  {Kremer} K.,  {Rodriguez} C.~L.,  {Chatterjee} S.,
  {Fragione} G.,    {Rasio} F.~A.,  2020, \apjl, 888, L10

\bibitem[\protect\citeauthoryear{{Zhuge}, {Centrella} \& {McMillan}}{{Zhuge}
  et~al.}{1996}]{zhuge96}
{Zhuge} X.,  {Centrella} J.~M.,    {McMillan} S.~L.~W.,  1996, \prd, 54, 7261

\end{thebibliography}
\end{document}